\documentclass[11pt, a4paper]{article}

\usepackage[margin=1in]{geometry} 
\usepackage[utf8]{inputenc}
\usepackage[T1]{fontenc}
\usepackage{mathptmx}

\usepackage[affil-it]{authblk}

\usepackage{graphicx}
\usepackage{epstopdf, epsfig}
\usepackage{amsmath}
\usepackage{amssymb}
\usepackage{bm}
\usepackage{physics}
\usepackage{booktabs}
\usepackage{enumitem}
\usepackage{subcaption}
\usepackage{hyperref}
\usepackage{natbib} 
\usepackage{float}

\newcommand{\pderiv}[3]{\left(\frac{\partial #1}{\partial #2}\right)_{#3}}
\newcommand{\totalderiv}[2]{\frac{\mathrm{d} #1}{\mathrm{d} #2}}

\hypersetup{
    colorlinks=true,
    linkcolor=blue,
    citecolor=blue,
    urlcolor=blue
}

\title{Flash evaporation Riemann Problem: Formulation and its Exact Solution}

\author[1,2]{Haotong Bai}
\author[3]{Ping Yi}
\author[1,2]{Yixin Yang\thanks{Corresponding author: yangyixin@nudt.edu.cn}}
\author[1,2]{Guoyan Zhao\thanks{Corresponding author: zhaoguoyan09@nudt.edu.cn}}
\author[2]{Wenjia Xie}
\author[1,2]{Mingbo Sun\thanks{Corresponding author: sunmingbo@nudt.edu.cn}}

\affil[1]{Hypersonic Technology Laboratory, National University of Defense Technology, Kaifu District, Changsha, Hunan 41073, China}
\affil[2]{College of Aerospace Science and Engineering, National University of Defense Technology, Kaifu District, Changsha, Hunan 41073, China}
\affil[3]{Institute of Power Plants and Automation, Shanghai Jiao Tong University, Shanghai, 200240, China}

\date{}
\begin{document}

\maketitle

\begin{abstract}
Flash evaporation, a liquid-to-gas phase transition phenomenon in real fluids, is prevalent in aerospace propulsion systems. To elucidate the physical mechanisms of such complex flows and provide theoretical benchmarks for Computational Fluid Dynamics simulations, this paper formalizes the Flash evaporation Riemann problem (FeRP) characterized by the expansion branch crossing the saturation line, within the framework of Homogeneous Equilibrium and Vapor-Liquid Equilibrium assumptions. An exact solution framework that analytically resolves all thermodynamic derivatives of equilibrium two-phase fluids is established for arbitrary two-parameter equations of state. By evaluating the Landau fundamental derivative, the non-classical wave structures arising in the FeRP are analyzed, for which a stable iterative solution strategy incorporating the Chapman-Jouguet condition as an outer constraint is proposed. Furthermore, an exact solution for the FeRP based on Wood's mechanical equilibrium speed of sound is developed, enabling a comprehensive evaluation of its thermodynamic implications. Results indicate that Wood's model alters the definition of the two-phase mixture entropy in the Euler equations, introducing an isentropic path characterized by a "density lag" effect and non-physical entropy decrease. Comparative analysis of the FeRP under typical scramjet fuel injection conditions reveals that, although Wood's model captures the general trend of the Riemann solution curve, it significantly underestimates intermediate pressure, velocity, and the extent of vaporization relative to the complete equilibrium model.
\end{abstract}

\vspace{1em}

\section{Introduction}
\label{sec:intro}
The Riemann problem (RP) concerns the evolution of fluid states initially separated by a discontinuity \citep{Riemann1860}. In the context of classical ideal gases, the Euler equations naturally possess hyperbolicity \citep{Godunov1959, Toro2009}, yielding self-similar solutions comprising shock waves, rarefaction waves, and contact discontinuities \citep{Dafermos2016}. Conversely, for real fluids governed by complex thermodynamics, analyzing the RP presents more issues. Notably, within rocket engines \citep{Popp2004} and scramjets \citep{Li2023a}, liquid fuels commonly experience a violent expansion-driven phase transition, known as "flash evaporation" (or simply "flash"). Currently, accurate physical modeling of flash remains an open question. The RP specifically characterizing the flash phenomenon has yet to be formally defined, and its exact solution warrants a comprehensive investigation.

\begin{figure}[htbp]
    \centering
    \begin{minipage}[b]{0.60\textwidth}
        \centering
        \includegraphics[width=\linewidth]{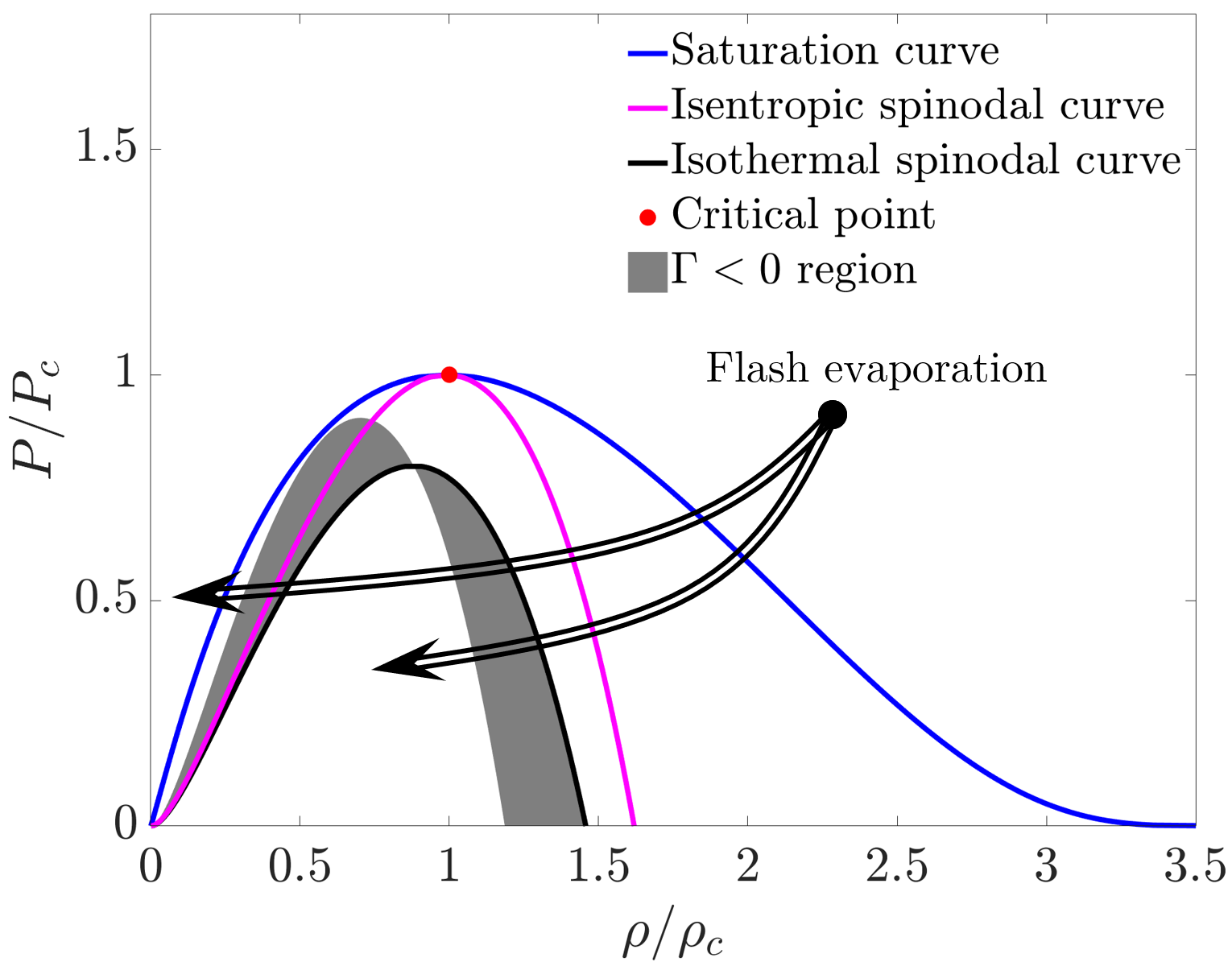}
        \caption{Phase diagram of n-dodecane}
        \label{fig1}
    \end{minipage}%
    \hfill
    \begin{minipage}[b]{0.36\textwidth}
        \centering
        \captionsetup{width=\linewidth} 
        
        \begin{subfigure}{\linewidth}
            \centering
            \includegraphics[width=0.95\linewidth]{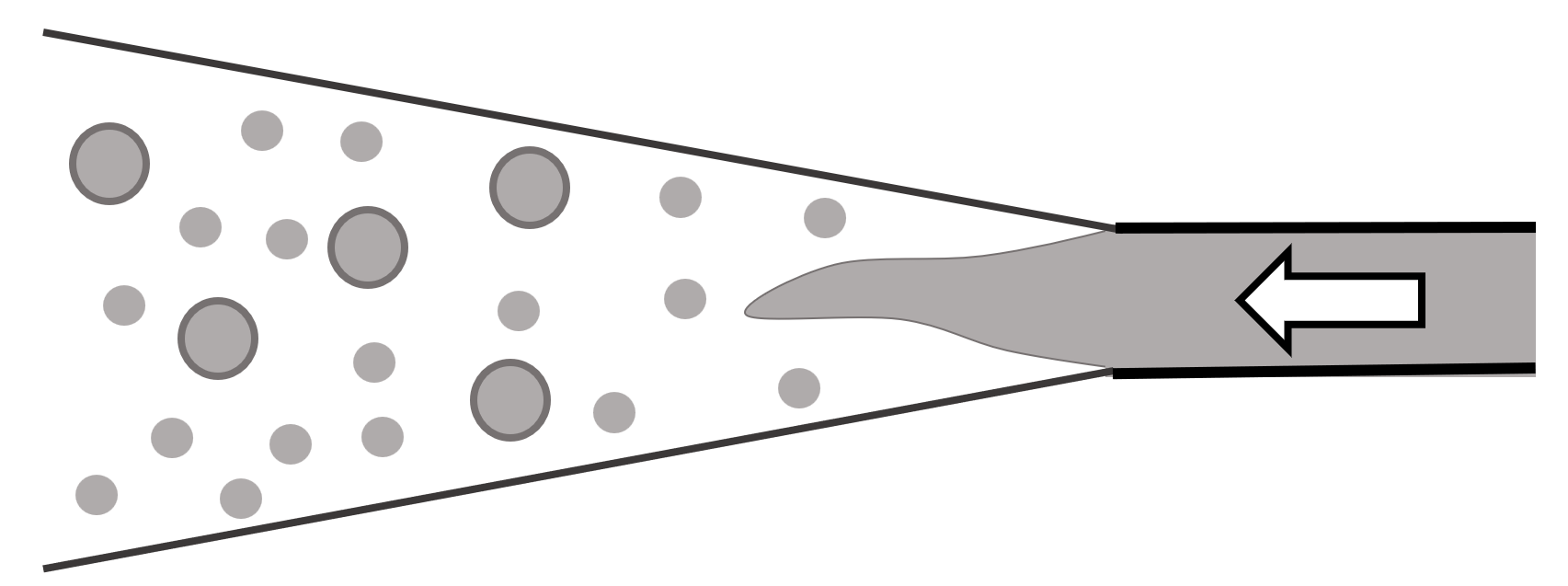}
            \caption{Spray flow}
            \label{fig:spray}
        \end{subfigure}

        \begin{subfigure}{\linewidth}
            \centering
            \includegraphics[width=0.95\linewidth]{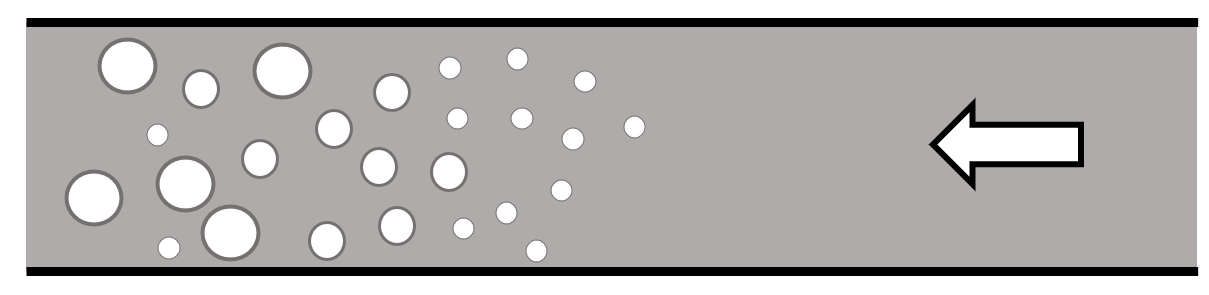}
            \caption{Frictional pipe flow}
            \label{fig:friction}
        \end{subfigure}

        \begin{subfigure}{\linewidth}
            \centering
            \includegraphics[width=0.95\linewidth]{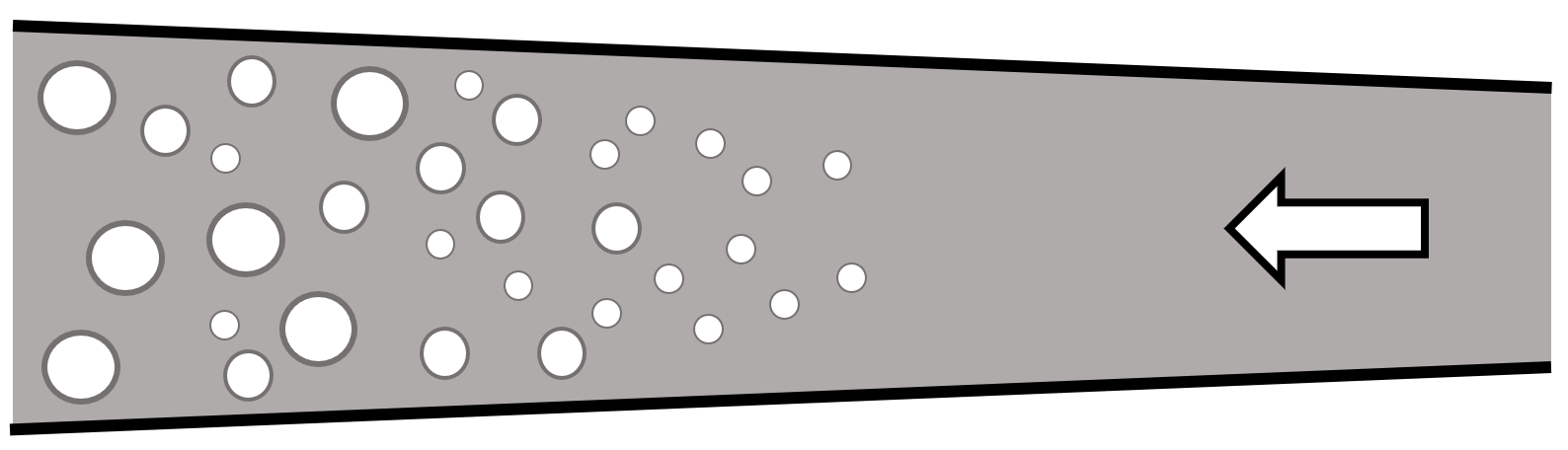}
            \caption{Expanding pipe flow}
            \label{fig:expanding}
        \end{subfigure}
        
        \caption{Flash evaporation in propulsion system}
        \label{fig2}
    \end{minipage}
\end{figure}

The description of liquid-vapor phase transitions constitutes a longstanding challenge in thermodynamics. Mathematically, there currently exists no single equation of state (EoS) capable of describing the continuous process of fluid phase transitions. Considering the phase diagram of n-dodecane, as shown in figure \ref{fig1}, cubic EoS predict a thermodynamic instability region \citep{LandauLifshitz1980} (bounded by the isothermal spinodal line) and a mechanical instability region (bounded by the isentropic spinodal line) within the saturation envelope. When a state point enters the mechanical instability region, the square of the speed of sound $c^2 = (\partial P / \partial \rho)_s$ becomes negative, rendering the Euler equations non-hyperbolic \citep{MenikoffPlohr1989}. Physically, the liquid-gas phase transition is inherently a non-equilibrium process, accompanied by nucleation and phase growth dynamics \citep{LifshitzPitaevskii1981}. These processes introduce a relaxation time or internal length scale into the system. The existence of this physical scale violates the scale invariance of the Euler equations, invalidating classical RP solutions that depends on the self-similar variable ($x/t$).

However, for the specific cases of phase transitions in aerospace propulsion systems, the RP remains amenable to self-similar solutions upon adopting appropriate physical assumptions. As illustrated in figure \ref{fig2}, such flows are ubiquitous in propulsion systems, manifested as spray flows (e.g., transverse jet injection in scramjets \citep{Li2023b}), frictional pipe flows (e.g., in cooling channels), and expanding pipe flows (e.g., in turbopump nozzles \citep{Caze2024}). These flows share a common mechanism: a depressurization-induced phase transition from liquid to vapor or a two-phase mixture, known as "flash evaporation." Flash flows typically involve intense turbulent mixing or extremely high pressure gradients. This strong thermodynamic driving force causes heterogeneous nucleation to occur extensively within extremely short temporal scales and small spatial scales. Consequently, interphase heat and mass transfer proceeds so rapidly that the non-equilibrium relaxation time is negligible compared to the macroscopic flow time scale.

Therefore, the specific characteristics of the flash phenomenon provides us with the physical basis for constructing its RP: the Homogeneous Equilibrium Model (HEM) \citep{Brennen2005} and the Vapor-Liquid Equilibrium (VLE) model \citep{Yi2019, Yang2020}. The HEM postulates instantaneous thermodynamic and dynamic equilibrium, where the liquid and vapor phases share identical pressure $P$, temperature $T = T_{\text{sat}}(P)$, and velocity $u$ \citep{Brennen2005}. Furthermore, the phase transition is assumed to initiate strictly on the saturation line, neglecting metastable states, and the VLE model is introduced via the equality of fugacity. Combined, the HEM and VLE construct a stable thermodynamic path, ensuring the equilibrium speed of sound is defined globally and restoring the hyperbolicity of the Euler equations, while simultaneously eliminating the influence of the system's additional characteristic time scales.

In summary, this paper formalizes the Flash evaporation Riemann problem (FeRP) as: \textbf{a Riemann problem based on homogeneous equilibrium and vapor-liquid equilibrium assumptions, where the expansion branch crosses the saturation line and induces a liquid-vapor phase transition}. This formulation physically captures the flash phenomenon of fluids caused by depressurization, while mathematically ensuring the self-similarity of the flow and the existence of exact solutions.

The flash process typically involves broad variations in temperature and pressure, thereby invalidating standard ideal gas or liquid assumptions. Consequently, establishing an exact solution for the FeRP requires a preliminary investigation into solutions under thermodynamically stable, single-phase conditions. \citet{ColellaGlaz1985} pioneered numerical methods for the RP in real fluid flows. Subsequently, \citet{Saurel1994} developed an exact Riemann solution for high-temperature real gases and constructed an approximate solution for expansion waves based on the isentropic expansion index. \citet{Quartapelle2003} provided a framework for classical real fluids (where ${{{\partial }^{2}}e\left( v,s \right)}/{\partial {{v}^{2}}}\;\ne 0$) by enforcing velocity and pressure continuity across the contact discontinuity. More recently, Jeremy C. H. Wang \citep{ WangHickey2020, Wang2022} constructed an exact RP solution for real fluids that resolves expansion waves based on the concept of generalized Riemann invariants \citep{Kouremenos1986}, applying it to transcritical nitrogen flows. In our previous work \citep{Bai2025}, we derived a Newton iteration method for the classical RP of two-parameter fluids, grounded in fundamental thermodynamic principles.

A classical solution implies that the wave structure consists solely of expansion rarefaction waves, compression shock waves, and contact discontinuities. In Bethe-Zel’dovich-Thompson (BZT) fluids \citep{LambrakisThompson1972, Thompson1972}, more complex non-classical wave phenomena may appear, such as expansion shocks and compression fans. The emergence of non-classical waves is determined by the Landau derivative (also called fundamental derivative ) $\Gamma$ \citep{LandauLifshitz1959}; the solution is classical when $\Gamma>0$, and non-classical otherwise. As shown in figure 1, a region where $\Gamma<0$ exists in the supercooled vapor region of n-dodecane (excluding the region inside the spinodal line, which lacks physical meaning). Fortunately, under the VLE model, propulsion fluids of interest such as the kerosene surrogate n-dodecane and the coolant CO$_2$ satisfy $\Gamma>0$ in both two-phase and single-phase regions; indeed, $\Gamma > 0$ everywhere except at the saturation line is an implicit prerequisite of the FeRP, distinguishing it from BZT flows. Nevertheless, traversing the saturation line inevitably leads to rarefaction wave splitting and the formation of composite waves containing expansion shocks, which will be discussed in Section 3. These non-classical waves are physically real; although they violate the Lax entropy condition \citep{Lax1973}, they satisfy the more generalized Liu entropy condition \citep{Liu1976}.

\citet{MenikoffPlohr1989} analyzed in detail the causes of non-classical waves and the conditions for rarefaction wave splitting and composite waves in phase-change fluids. \citet{MullerVoss2006, Dahmen2005} and \citet{Voss2005} analyzed the construction forms of various non-classical waves in BZT fluids and provided a basic solution framework based on wave curves, which is universal for various real fluids. Alternatively, other researchers have investigated specific simplified configurations of the RP involving phase transitions. Saurel et al., based on the stiff EoS and the Noble Abel Stiffened Gas EoS \citep{Saurel2016}, studied the exact solutions \citep{Saurel2008} and modeling methods \citep{LeMetayerSaurel2016} of phase-change fluids, successfully capturing the phenomenon of rarefaction wave splitting. \citet{Dreyer2013} provided the exact solution of RP for phase-change flows in isothermal processes.

In summary, extensive research has been conducted on the RP for real fluids with phase change, ranging from unified construction methods to those focused on simplified configurations. However, for the specific FeRP, a comprehensive exact solution framework is still lacking. On one hand, the non-classical wave structure of the FeRP is deterministic and does not require complex screening via the Liu entropy condition, unlike BZT fluids. On the other hand, flash evaporation often involves substantial state variations. This renders simple models like the stiff EoS insufficient, necessitating more complex EoS such as Peng-Robinson \citep{PengRobinson1976} or Redlich-Kwong \citep{RedlichKwong1949}. Therefore, addressing the FeRP, this paper establishes an exact solution framework based on Newton's method. In Section 2, starting from a general EoS $P=P(\rho,T)$ and specific heat function $C_v= C_v(\rho,T)$, we derive all thermodynamic derivatives for single-phase and equilibrium two-phase fluids, ensuring universality and computational convergence. In Section 3, we calculate the Landau derivative $\Gamma$ for common working fluids like n-dodecane and carbon dioxide, analyze their non-classical gas dynamic behaviors during the flash process, and provide exact solution methods for non-classical waves. Finally, we propose an exact solution to the RP based on Wood's mechanical equilibrium speed of sound \citep{Wood1946} in section 4. While this model is widely adopted in computational schemes, we scrutinize its intrinsic thermodynamic consistency within the FeRP, elucidating the "density lag" effect and non-physical entropy decrease relative to the complete equilibrium model.

\section{Exact Solution of the Flash evaporation Riemann Problem}
\label{sec:flash RP}

The Euler equations contains three types of discontinuities: shock waves, contact discontinuities, and rarefaction waves (the latter being weak discontinuities). This section begins by formulating the thermodynamic relations for single-phase and two-phase regimes, followed by the governing equations for these three discontinuities. Subsequently, we then establish a Newton iteration framework to obtain the exact solution of the FeRP.

\begin{figure}[htpb]
    \centering
    \begin{subfigure}{0.49\textwidth}
        \includegraphics[width=\linewidth]{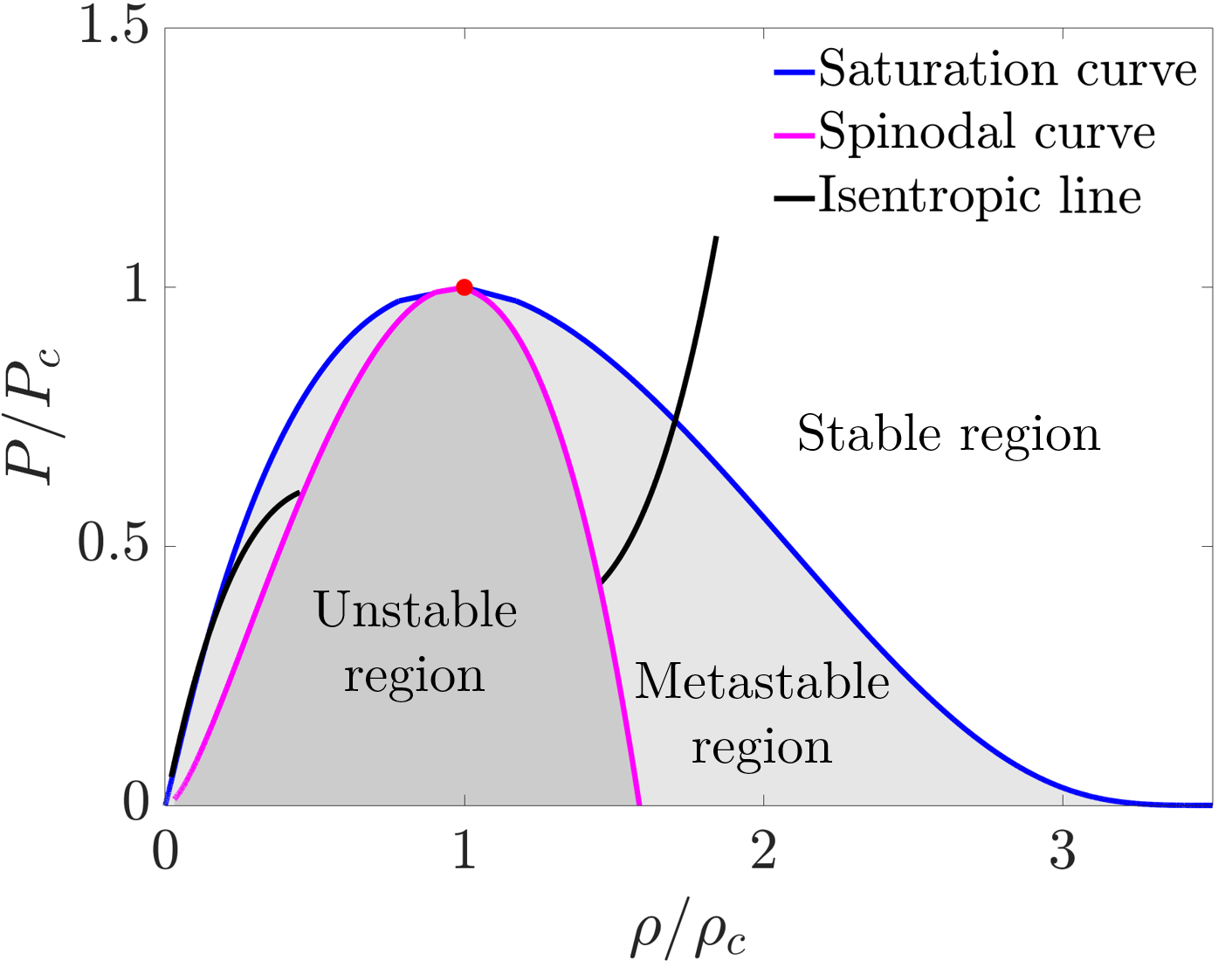}
        \caption{Thermodynamic boundary}
    \end{subfigure}
    \hfill
    \begin{subfigure}{0.49\textwidth}
        \includegraphics[width=\linewidth]{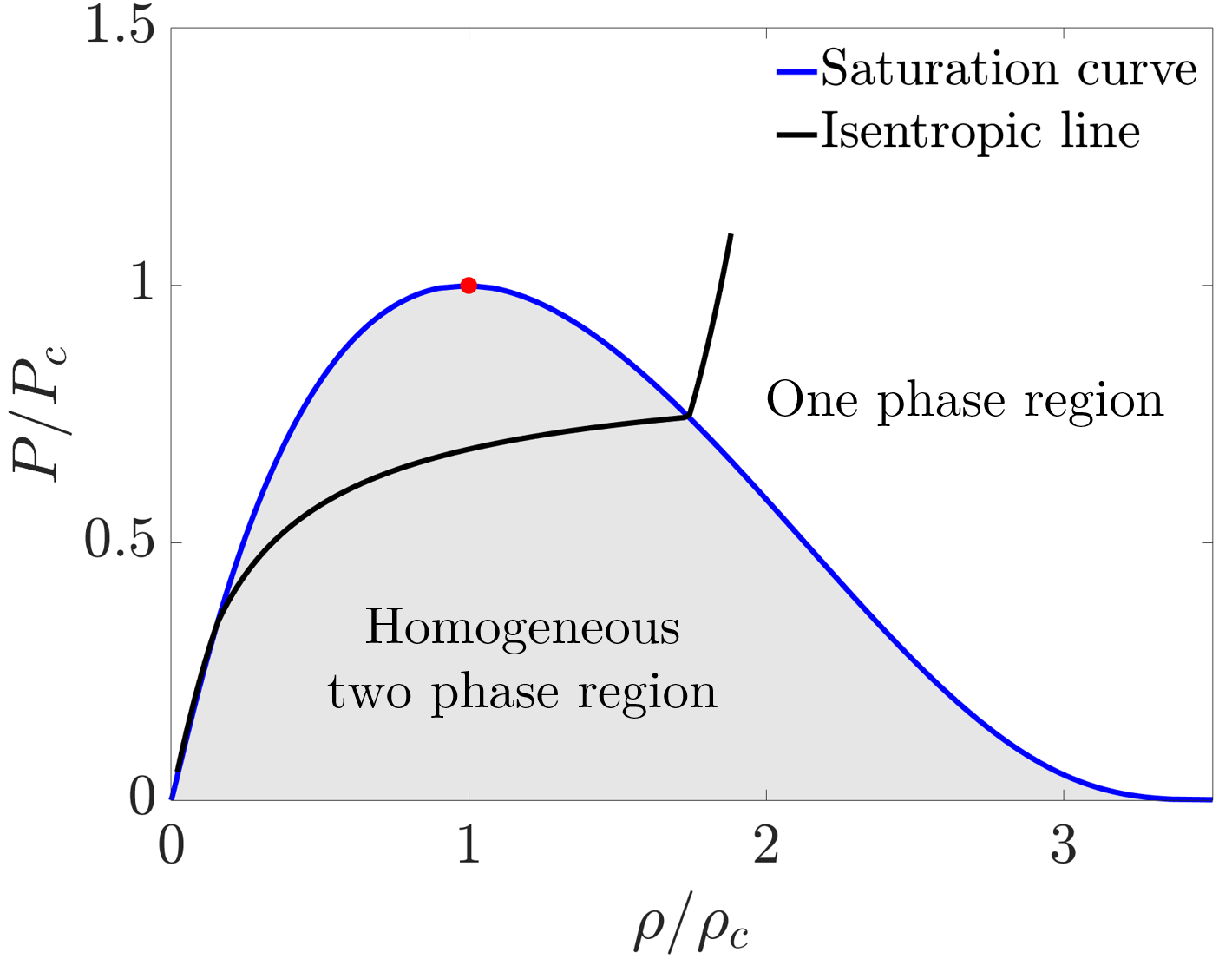}
        \caption{VLE model}
    \end{subfigure}
    \caption{Thermodynamic Boundary and Phase Model of n-dodecane on the $P-\rho$ Plane}
    \label{fig3}
\end{figure}

Prior to deriving the specific governing equations, we first clarify the definitions of "single-phase," "two-phase," and the aforementioned "thermodynamic equilibrium" adopted in this paper. As shown in Figure \ref{fig3}, in thermodynamics, fluid states are classified into thermodynamically stable regions, metastable regions (bounded by the saturation line), and unstable regions (bounded by the spinodal line)\citep{LandauLifshitz1980}. As shown in Figure \ref{fig3}(a), isentropic lines can cross the saturation line into the metastable region but have no physical meaning inside the spinodal line. Physical phase transitions may occur in the metastable region, where the fluid enters a state of subcooled vapor or superheated liquid after crossing the saturation line, requiring the overcoming of a nucleation barrier to trigger the transition. For instance, previous studies \citep{Saurel2008, LeMetayerSaurel2016} have modeled phase transitions within the metastable region. Theoretically, one can prescribe a phase transition initiation boundary (such as the Wilson line \citep{Yellott1934}) within the metastable region to solve the RP. However, this method presents significant challenges, as it requires not only specifying the location of the phase transition boundary, but also providing high-order derivatives on the boundary to ensure the proper construction of the wave structures. Furthermore, such an artificial boundary limits the generality of the method.

Accordingly, based on the VLE model, we assume that the flash phase transition initiates strictly at the saturation line, thereby neglecting the delay effects of metastability. As shown in Figure \ref{fig3}(b), throughout this paper, the "two-phase region" (or mixture region) refers to the area enclosed by the saturation line, while the "single-phase region" denotes the area outside the saturation line, corresponding to the "thermodynamically stable region." Through the condition of equal fugacity, the VLE model constructs a continuous equilibrium trajectory across the two-phase region, permitting isentropes to cross the saturation lines continuously.

\subsection{Governing Equations and Basic Thermodynamic Parameters}
\label{sec:euler equ}

The governing equations for the Riemann problem are the one-dimensional Euler equations:
\begin{equation}
 \frac{\partial \mathbf{U}}{\partial t}+\frac{\partial \mathbf{F}(\mathbf{U})}{\partial x}=0,
 \label{equ1}
\end{equation}
where $\mathbf{U}$ is the vector of conserved variables and $\mathbf{F}$ is the flux vector. Their specific expressions are:
\begin{equation}
\mathbf{U}=\left( \begin{matrix}
   \rho   \\
   \rho u  \\
   \rho (e+\frac{1}{2}{{u}^{2}})  \\
\end{matrix} \right),\quad \mathbf{F}=\left(\begin{matrix}
   \rho u  \\
   \rho {{u}^{2}}+P  \\
   \rho u(e+\frac{1}{2}{{u}^{2}})+Pu  \\
\end{matrix} \right).
\label{equ2}
\end{equation}

For the single-phase fluid, the EoS adopts $\rho$ and $T$ as independent variables, defining $P$ as:
\begin{equation}
    P=P\left( \rho ,T \right).
\label{equ:Peos}
\end{equation}
Here, $P$ is a single-valued function of $\rho$ and $T$, and $T$ is also uniquely determined by $P$ and $\rho$ (note that $\rho$ is not necessarily a single-valued function of $P$ and $T$). For the two-phase mixture, $P$ takes $T$ as the unique independent variable. The $P-T$ curve is referred to as the saturation line:

\begin{equation}
    P=P_\text{sat}\left(T \right).
\label{equ:Psat}
\end{equation}
Each point $(P,T)$ on the saturation line corresponds to two density values: the saturated vapor density $\rho_v$ and the saturated liquid density $\rho_l$. The two-phase equilibrium mixture density $\rho$ and the saturation densities satisfy the relation (also known as the lever rule):
\begin{equation}
    \rho=\alpha \rho_v +(1-\alpha)\rho_l,
    \label{equ:mix_rho}
\end{equation}
where $\alpha$ is the vapor volume fraction. Equation \ref{equ:mix_rho} is equivalent to:
\begin{equation}
    \alpha(\rho,T)=\frac{\rho-\rho_l(T)}{\rho_v(T) - \rho_l(T)}.
\label{equ:alpha_mix}
\end{equation}
\par
For the single-phase fluid, the internal energy is expressed as:
\begin{equation}
    e(\rho,T)=\int_{{{T}_{0}}}^{T}{{{C}_{v}}\mathrm{d}T}-\int_{{{\rho }_{0}}}^{\rho }{\frac{1}{{{\rho }^{2}}}\left( T\pderiv{P}{T}{\rho}-P \right)}\mathrm{d}\rho +e_0,
    \label{equ:e}
\end{equation}
where $C_v$ is the specific heat at constant volume. $C_v$ is also a function of $\rho$ and $T$, which often termed the caloric equation of state:
\begin{equation}
    {C_v}={C_v}(\rho ,T).
\label{equ:Cv}
\end{equation}

The internal energy expression for the two-phase mixture is:
\begin{equation}
    e(\rho,T)=\frac{\alpha \rho_v(T) e_v(T) +(1-\alpha)\rho_l(T) e_l(T)}{\rho},
    \label{equ:mix_e}
\end{equation}
where $e_v$ and $e_l$ are the internal energies on the vapor and liquid saturation lines at $(P,T)$, respectively; clearly, these depend solely on $T$ (or $P$).

The entropy expression for the single-phase fluid is:
\begin{equation}
    s(\rho,T)=\int_{{{T}_{0}}}^{T}{\frac{{{C}_{v}}}{T}\mathrm{d}T}-\int_{{{\rho }_{0}}}^{\rho }{\frac{1}{{{\rho }^{2}}}{\pderiv{P}{T}{\rho}}\mathrm{d}\rho }+s_0.
\label{equ:s}
\end{equation}
The entropy expression for the two-phase mixture is analogous to internal energy. Defining $s_v$ and $s_l$ as the entropy at point $(P,T)$ on the saturation lines:
\begin{equation}
    s(\rho,T)=\frac{\alpha \rho_v(T) s_v(T) +(1-\alpha)\rho_l(T) s_l(T)}{\rho}.
\label{equ:mix_s}
\end{equation}

The speed of sound for the single-phase fluid is:
\begin{equation}
    {c^2}= \pderiv{P}{\rho}{s}=\pderiv{P}{\rho}{T}+\frac{T}{{{C}_{v}}{{\rho }^{2}}}\pderiv{P}{T}{\rho}^{2}.
\label{equ:c}
\end{equation}\par
The complete thermodynamic equilibrium speed of sound for the two-phase mixture is calculated from the ratio of the derivatives of mixture entropy with respect to density and pressure. This simplifies to:
\begin{equation}
    \frac{1}{\rho c_{eq}^2}=\frac{\alpha }{{\rho_v}c_v^2}+\frac{1-\alpha }{{\rho_l}c_l^2}+T\left( \frac{\alpha {\rho_v}}{{C_{P,v}}}{{\left( \frac{\mathrm{d}s_v}{\mathrm{d}P} \right)}^{2}}+\frac{(1-\alpha ){\rho_l}}{C_{P,l}}{{\left( \frac{\mathrm{d}s_l}{\mathrm{d}P} \right)}^{2}} \right),
    \label{equ:c_eq}
\end{equation}
where $c_v$, $c_l$ and $C_{P,v}$, $C_{P,l}$ are the speed of sound and specific heat at constant pressure at $(P,T)$ on the saturation line, respectively. Clearly, they are also single-valued functions of $P$ or $T$. In addition, $C_v$ and $C_P$ satisfy the thermodynamic relation \citep{Sedov1973}:
\begin{equation}
    {C_P}-{C_v}=\frac{T}{{\rho^2}}\frac{\left( {\partial P}/{\partial T} \right)_\rho^2}{{\partial P}/{\partial \rho}}.
\label{equ:CPCv}
\end{equation}

\subsection{Types of Discontinuities in the Flash evaporation Riemann Problem}
\label{sec:type discontinuity}

The governing equations for these three types of discontinuities are identical to those for non-phase-changing real fluids \citep{Bai2025}. Shock waves satisfy the Rankine-Hugoniot conditions, contact discontinuities satisfy velocity and pressure equilibrium conditions, and rarefaction waves follow isentropic relations, resolved via the method of characteristics. Denoting the pre-wave state by subscript "1" and the post-wave state by "2", the governing equations are as follows:

\subsubsection{Shock wave}
The upstream and downstream states of a shock correspond to the intersection of the Hugoniot curve with the Rayleigh line. The Hugoniot relation is expressed as:
\begin{equation}
    {e_1}-{e_2}+\frac{P_1}{\rho_1}-\frac{P_2}{\rho_2}+\frac{(P_2-P_1)(\rho_1+\rho_2)}{2\rho_1 \rho_2}=0.
\label{equ:hugoniot}
\end{equation}
The Rayleigh line is defined by:
\begin{equation}
    \frac{P_2 - P_1}{1/\rho_2 - 1/\rho_1} = - (\rho_1 u_1)^2 = - (\rho_2 u_2)^2 = - j^2,
    \label{equ:j2}
\end{equation}
where $j$ represents the mass flux.

\subsubsection{Contact Discontinuity}
The states on either side of a contact discontinuity satisfy the relation:
\begin{equation}
    {u_1}={u_2}; \quad {P_1}={P_2}.
\label{equ17}
\end{equation}

\subsubsection{Rarefaction Wave}
Rarefaction waves are classified as weak discontinuities characterized by continuous state variables but discontinuous derivatives. They are solved analytically via the method of characteristics. Along a characteristic line within the wave:
\begin{equation}
    \frac{\mathrm{d}x}{\mathrm{d}t}=u\pm c,
    \label{equ18}
\end{equation}
the Riemann invariants are:
\begin{equation}
    {{J}^{\pm}}=u\pm\int_{{{C}^{\pm}}}{\frac{1}{\rho c}\mathrm{d}P}=\text{const}.
\label{equ19}
\end{equation}
For a left-propagating rarefaction wave, the Riemann invariant $J^+$ (associated with the right-propagating characteristic $C^+$) remains constant across the wave. Consequently, the signs in Equations \ref{equ18} and \ref{equ19} both take "$+$". Conversely, for a right-propagating rarefaction wave, the Riemann invariant $J^-$ (associated with the left-propagating characteristic $C^-$) remains constant. Thus, the signs in Equations \ref{equ18} and \ref{equ19} both take "$-$".
\subsection{Types of Solution Curves for the Flash evaporation Riemann Problem}

\begin{figure}[h]
\centering 
\includegraphics[width=0.64\textwidth]{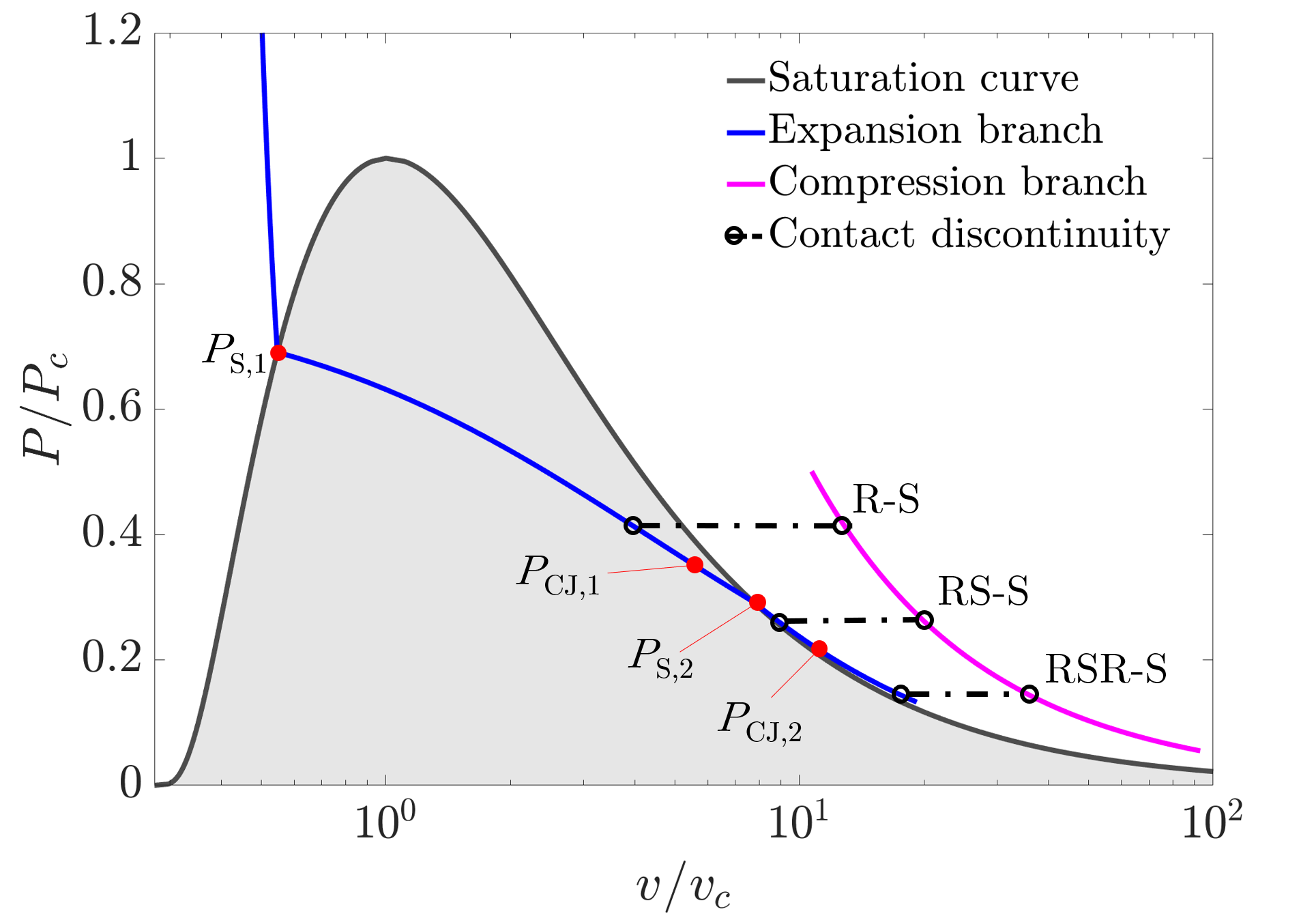} 
\caption{Three types of solution curves for the flash RP, using the n-dodecane phase diagram as an example}
\label{fig4}
\end{figure}

During the flash process, the liquid undergoes a phase transition due to expansion. Depending on fluid properties and equilibrium pressure, the liquid may expand into the two-phase region or traverse it entirely into the vapor phase. Meanwhile, the low-pressure gas on the other side is subjected to compression. \textbf{Without loss of generality, we specify the left state of the FeRP as a high-pressure liquid and the right state as a low-pressure gas.} Other cases, such as double rarefaction waves, follow similar principles and are omitted for brevity. Considering practical situations (e.g., fuel sprays in engines), we assume the low-pressure gas on the right side is sufficiently far from the saturation line and remains in the vapor phase state during compression. We analyze the solution curves of the RP on the $P-v$ diagram ($v=1/\rho$ denotes specific volume).Depending on the initial state and thermodynamic properties of the fluid, the FeRP yields three distinct solution types: the R-S (Rarefaction wave - Shock wave) solution, the RS-S (Rarefaction - Upstream Sonic Expansion Shock composite wave - Shock wave) solution, and the RSR-S (Rarefaction - Double Sonic Expansion Shock - Rarefaction composite wave - Shock wave) solution, as illustrated in Figure \ref{fig4}.

In Figure \ref{fig4}, the compression branch follows a Hugoniot curve in the vapor phase, whereas the expansion branch comprises either a pure isentrope or a composite path involving Hugoniot segments. We specify $P_\text{S,1}$ and $P_\text{S,2}$ as the intersections of the isentrope with the liquid and vapor saturation lines, respectively. Furthermore, $P_\text{CJ,1}$ and $P_\text{CJ,2}$ represent the pre-shock and post-shock pressures of the unique double sonic shock along a given isentrope. The resulting wave structure is determined by the intermediate pressure $P^*$ relative to these characteristic pressures:
\begin{itemize}
    \item $ P_\text{S,1} > P^* >P_\text{S,2}$: The expansion branch is a single rarefaction wave, denoted as an R wave. Although the wave splits at $P_\text{S,1}$ due to the discontinuity in the speed of sound, the integration of the Riemann invariants remains unaffected.
    \item $ P_\text{S,2} > P^* > P_\text{CJ,2}$: The expansion branch forms a composite wave comprising a rarefaction wave and an upstream sonic expansion shock (where the pre-shock state satisfies the Chapman-Jouguet (CJ) condition), denoted as an RS wave. The transition occurs at a CJ state point $a$, where $P_a \in (P_\text{CJ,1}, P_\text{S,2})$. The expansion shock connects $P_a$ to the post-shock pressure $P^*$.
    \item $ P_\text{CJ,2} > P^*$: The expansion branch is a composite structure consisting of a head rarefaction wave, a double sonic expansion shock (which achieves CJ states both upstream and downstream), and a tail rarefaction wave, denoted as an RSR wave. The double sonic shock bridges the two CJ states $P_\text{CJ,1}$ and $P_\text{CJ,2}$, connecting the head rarefaction to the tail rarefaction, which subsequently extends to the intermediate pressure $P^*$.
\end{itemize}

For a given left state, the characteristic pressures $P_\text{S,1}, P_\text{S,2}, P_\text{CJ,1}$, and $P_\text{CJ,2}$ are solely determined by the initial conditions, independent of the intermediate pressure $P^*$. In contrast, although the transition pressure $P_a$ varies with $P^*$, its determination relies on the same mathematical formulation as that derived for solving $P_\text{CJ}$. Given that the iterative solution for these values involves complex derivations, we assume in this section that these parameters are known, in order to maintain focus on the core algorithm of the exact Riemann solution. The specific derivations for $P_\text{CJ,1}, P_\text{CJ,2}$, and $P_a$, along with a detailed analysis of the expansion branch, are provided in the subsequent section.

\subsection{Method for Solving Intermediate Pressure in Flash evaporation Riemann Problem}
\label{sec:solve p*}

The primary objective of solving the RP is to determine the intermediate pressure $P^*$ and velocity $u^*$ such that pressure and velocity remain continuous across the contact discontinuity:
\begin{equation}
    u_L^* = u_R^* = u^* \quad \text{and} \quad P_L^* = P_R^* = P^*.
\end{equation}
We employ Newton's method with the intermediate pressure $P^*$ as the iteration variable. For a given estimate $P^*$, the residual function $U(P^*)$ is defined as the velocity difference at the contact discontinuity:
\begin{equation}
    U(P^*) = (u_L + \psi_L(P^*)) - (u_R - \varphi_R(P^*)) = 0,
\end{equation}
where $\psi_L$ represents the velocity variation across the expansion branch (R, RS, or RSR wave), referred to as the expansion $P-u$ function. Similarly, $\varphi_R$ represents the velocity variation across the compression branch (S wave), referred to as the compression $P-u$ function. The Newton iteration formula is:
\begin{equation}
    P_{k+1}^* = P_k^* - \frac{U(P_k^*)}{U'(P_k^*)}.
\end{equation}
Consequently, the key to solving for the intermediate pressure lies in deriving $U(P^*)$ and its derivative $U'(P^*)$.

\subsubsection{Compression Branch \texorpdfstring{$\varphi_R$}{varphiR}}
\label{sec:phiR- S}
As previously stated, the compression branch consists of a single shock wave. The function $\varphi_R(P^*)$ represents the shock $P-u$ relation:
\begin{equation}
    \varphi_{R}(P^*)={{j}_{R}}\left( \frac{1}{\rho _{R}^{*}}-\frac{1}{{\rho _R}} \right),
\end{equation}
where $j$ is the mass flux through the shock, determined by Equation \ref{equ:j2}, and $P^*$ and $\rho_R^*$ satisfy the Hugoniot relation \ref{equ:hugoniot}.\par
The derivative of $\varphi_R(P)$ with respect to $P$ is:
\begin{equation}
    \frac{\mathrm{d} {{\varphi }_{R}}}{\mathrm{d}{{P}^{*}}}=-\frac{1}{2{{j}_{R}}}-\frac{{{j}_{R}}}{2(\rho_R^*)^2}{\left( \frac{\mathrm{d}\rho }{\mathrm{d}P} \right)}_{H,R},
\end{equation}
where ${({\mathrm{d}\rho }/{\mathrm{d}P})}_{H,R}$ represents the derivative along the Hugoniot curve at the state $(P^*, \rho_R^*)$
\begin{equation}
    {\left( \frac{\mathrm{d}\rho }{\mathrm{d}P} \right)}_{H,R}=-\frac{(\partial{F_H}/\partial{P^*})_{\rho}}{(\partial{F_H}/\partial{\rho})_{P^*}} = -\frac{- \pderiv{e}{P}{\rho}+\frac{\rho - \rho_R}{2 \rho \rho_R}}{- \pderiv{e}{\rho}{P}+\frac{P^*+P_R}{2 \rho^2}}.
\end{equation}
To calculate $\varphi_R(P)$ and its derivative, a nested Newton iteration is required to determine the post-shock density $\rho_R^*$ corresponding to $P^*$ that satisfies the Hugoniot relation.\par
Let the residual function be the Hugoniot condition:
\begin{equation}
    F_H = e(\rho,P^*)-e(\rho_R,P_R)+\frac{P_R}{\rho_R}-\frac{P^*}{\rho}+\frac{(P^*-P_R)(\rho_R+\rho)}{2\rho_R \rho}=0.
\label{equ:F_H}
\end{equation}
Its derivative is:
\begin{equation}
    \pderiv{F_H}{\rho}{P^*} = - \pderiv{e}{\rho}{P}+\frac{P^*+P_R}{2 \rho^2}.
\end{equation}
Note that the internal energy depends on pressure and density via $e(\rho,P)=e(\rho,T(\rho,P))$. Computing $e$ via Equation \ref{equ:e} or \ref{equ:mix_e} requires an iterative inversion of the EoS \ref{equ:Peos} to obtain $T$ in the single-phase region; in the two-phase region, however, the temperature is explicitly determined by $T=T_\text{sat}(P)$. The algorithm for this iterative procedure can be found in \citet{Bai2025}. The analytical expressions for $({\partial e}/{\partial \rho})_{P}$ and $({\partial e}/{\partial P})_{\rho}$ are provided in Appendix \ref{sec:ds_de}.

\subsubsection{Expansion Branch \texorpdfstring{$\psi_L$}{psiL} - R Wave}
\label{sec:psiL- R}
When $ P_\text{S,1} > P^* >P_\text{S,2}$, the expansion branch is a single rarefaction wave. Although this rarefaction wave exhibits splitting behavior, the flow remains isentropic, and the definition of the Riemann invariant remains valid. This case represents the simplest configuration, where the expansion $P-u$ function is directly derived from the Riemann invariant:
\begin{equation}
    \psi_L(P^*) = \int_{P^*}^{P_L} \frac{1}{\rho c}\mathrm{d}P.
\end{equation}
This integral is evaluated using numerical integration algorithms \citep{Bai2025}. Along the integration path, $\rho$ and $P$ satisfy the isentropic relation, and $c = c_{eq}(\rho,P)$. The derivative of $\psi_L(P^*)$ is:
\begin{equation}
    \frac{\mathrm{d}\psi_L}{\mathrm{d}P^*} = - \frac{1}{\rho^*_L c^*_L}.
\end{equation}
To calculate $\psi(P)$ and its derivative, a nested Newton iteration is needed to determine $\rho, c$ corresponding to $P^*$ subject to the isentropic relation.
\par
Let the residual function be the isentropic condition:
\begin{equation}
     F_s = s(\rho_L,P_L) - s(\rho,P).
\label{equ:F_s}
\end{equation}

Note that entropy depends on density and pressure via $s(\rho,P)=s(\rho,T(\rho,P))$. The derivative of this residual function is:
\begin{equation}
     \pderiv{F_s}{\rho}{P} = -\pderiv{s}{\rho}{P},
\end{equation}
where the analytical expression for $({\partial s}/{\partial \rho})_{P}$ is provided in the Appendix \ref{sec:ds_de}.

\subsubsection{Expansion Branch \texorpdfstring{$\psi_L$}{psiL} - RS Wave}
\label{sec:psiL- RS}
When $ P_\text{S,2} > P^* > P_\text{CJ,2}$, the expansion branch is a composite wave consisting of a rarefaction wave and an upstream sonic expansion shock. We denote the transition point—where the rarefaction terminates and the shock initiates—as state $a$. At point $a$, the shock wave is co-moving with the tail of the rarefaction wave; specifically, the propagation speed of the shock equals the characteristic speed of the rarefaction wave tail. This is also referred to as the CJ condition, where the mass flux satisfies $j_a = \rho_a c_a$. The $P-u$ function of the RS wave is given by:
\begin{equation}
    \psi_L(P^*) = - \int_{P_L}^{P_a} \frac{1}{\rho c}\mathrm{d}P + \frac{P_a-P^*}{j_a}.
\end{equation}
The derivative of $\psi_L(P^*)$ is:
\begin{equation}
    \frac{\mathrm{d}\psi_L}{\mathrm{d}P^*} = - \frac{1}{j_a}\left(1+ \frac{P_a-P^*}{j_a}\frac{\mathrm{d}j_a}{\mathrm{d}P_a}\frac{\mathrm{d}P_a}{\mathrm{d}P^*} \right),
    \label{equ:dRS}
\end{equation}
where the transition pressure $P_a$ resides within the interval $(P_\text{CJ,1}, P_\text{S,2})$. Note that in deriving Equation \ref{equ:dRS}, the derivative terms associated with the variation of the transition point $P_a$ in the rarefaction and shock portions cancel out due to the CJ condition. The methodology for determining $P_a$, as well as the analytical expressions for ${\mathrm{d}j_a}/{\mathrm{d}P_a}$ and ${\mathrm{d}P_a}/{\mathrm{d}P^*}$, are detailed in the subsequent section after introducing the double sonic shock. The path from $P_L$ to $P_a$ represents a single rarefaction wave, the calculation of which is analogous to that described in the preceding subsection.

\subsubsection{Expansion Branch \texorpdfstring{$\psi_L$}{psiL} - RSR Wave}
\label{sec:psiL- RSR}
When $ P_\text{CJ,2} > P^*$, the expansion branch is a composite wave consisting of a head rarefaction, a double sonic expansion shock, and a tail rarefaction wave. We denote the upstream CJ state point {CJ,1} as the transition point where the head rarefaction terminates and the shock initiates, while the downstream CJ state point {CJ,2} marks the transition point from the shock to the tail rarefaction. In this configuration, the shock flux satisfies the dual CJ condition: $j_\text{CJ} = \rho_\text{CJ,1} c_\text{CJ,1} = \rho_\text{CJ,2} c_\text{CJ,2}$. The $P-u$ function of the RSR wave is given by:
\begin{equation}
    \psi_L(P^*) = - \int_{P_L}^{P_\text{CJ,1}} \frac{1}{\rho c}\mathrm{d}P + \frac{P_\text{CJ,1}-P_\text{CJ,2}}{j_\text{CJ}}- \int_{P_\text{CJ,2}}^{P^*} \frac{1}{\rho c}\mathrm{d}P.
\end{equation}
The derivative of $\psi_L(P^*)$ is:
\begin{equation}
     \frac{\mathrm{d}\psi_L}{\mathrm{d}P^*} = - \frac{1}{\rho^*_L c^*}.
\end{equation}
Note that the Riemann invariant for the tail rarefaction wave is integrated along the isentrope originating from $P_\text{CJ,2}$ to $P^*$. This isentrope is distinct from the integration isentrope of the head rarefaction wave (which starts from the initial left state), as the entropy increases across the expansion shock.

\subsection{State Solution Inside Split Rarefaction Waves}
\label{sec:solve raf}

The ultimate objective of solving the RP is to obtain the flow state at any characteristic coordinate $\xi = x/t$. After determining the intermediate pressure, it is necessary to consider the relative position of the wave structure and the coordinate $\xi$ to calculate the flux. The calculation of fluxes in the undisturbed and intermediate regions is straightforward and analogous to the ideal gas case, so it is omitted here. However, determining the flux within a rarefaction wave is more complex, specifically due to the wave splitting arising from the discontinuous drop in the speed of sound across the saturation boundary. This requires a solution strategy distinct from that for single-phase fluids \citep{Bai2025}. Based on the preceding discussion, we provide a Newton iteration method specifically tailored for this scenario. Considering a left-propagating R wave (also applicable to rarefaction segments within RS and RSR structures), the rarefaction fan spans the domain defined by:
\begin{equation}
    {\lambda_\text{head}}<\frac{x}{t}<{\lambda_\text{tail}},
\end{equation}
where $\lambda_\text{head} = u_L-c_L$ denotes the velocity of the rarefaction wave head, and $\lambda_\text{tail}$ denotes the wave tail velocity, defined as:
\begin{equation}
   \lambda_\text{tail} = \begin{cases}
   u^* - c^*, & \text{R wave} \\ 
   u_a - c_a, & \text{RS wave} \\ 
   u_{\text{CJ,1}} - c_{\text{CJ,1}}. & \text{RSR wave}
   \end{cases}
\end{equation}
However, traversing the liquid saturation boundary (point $S,1$ in Figure \ref{fig4}) induces a discontinuous drop in the speed of sound, leading to a discontinuity in the characteristic wave speed. The saturation point $S,1$ is characterized by two distinct sound speeds: the liquid speed $c=c_{l,\text{S}}$ and the equilibrium speed $c=c_{eq,\text{S}}$, defined as:
\[
c_{eq,\text{S}} = \left(\frac{1}{c^2_{l,\text{S}}} + \frac{\rho^2_{l,\text{S}}}{C_{P,l\text{S}}}{{\left( \frac{\mathrm{d}s_l}{\mathrm{d}P} \right)}^{2} }\right)^{-1/2}.
\] 
The characteristic wave speeds satisfy $\lambda_{l,\text{S}} = u_S - c_{l,\text{S}} < u_S - c_{eq,\text{S}} = \lambda_{eq,\text{S}}$. Since the rarefaction wave is left-propagating, this sudden increase in characteristic wave speed causes the wave to split at point $S,1$. The wave segment in the characteristic speed interval $(\lambda_\text{head}, \lambda_{l,\text{S}})$ travels faster in the negative $x$-direction than the segment in the interval $(\lambda_{eq,\text{S}}, \lambda_\text{tail})$, leading to the splitting of the rarefaction wave in the $x-t$ diagram. It is crucial to distinguish rarefaction wave splitting from a discontinuity. The split rarefaction wave remains continuous on the $P-v$ diagram (following the isentrope), but a constant state region manifests on the $x-t$ diagram. In contrast, a true discontinuity involves a jump in thermodynamic states on the $P-v$ diagram governed by the R-H relations. For instance, a double sonic shock in RSR wave may be considered as a "discontinuity" embedded within a rarefaction wave (though this analogy is imperfect as the two R components of an RSR wave do not share the same isentrope).

In summary, the rarefaction wave in the classical sense (characterized by continuous wave speed variation) exists only when the coordinate $\xi$ lies within the characteristic speed intervals $(\lambda_\text{head}, \lambda_{l,\text{S}})$ and $(\lambda_{eq,\text{S}}, \lambda_\text{tail})$. For coordinates within these two intervals, we use Newton's method to resolve the internal wave structure. Denoting the known pre-wave state by the subscript "1" and the unknown state to be resolved by "0", the residual function is written as:

\begin{equation}
    \begin{aligned}
         & u_1+\int_{P_1}^{P_1}{\frac{1}{\rho c}\mathrm{d}P}= u_0+\int_{P_0}^{P_1}{\frac{1}{\rho c}\mathrm{d}P} = c_0 +\frac{x}{t}-\int_{P_0}^{P_1}{\frac{1}{\rho c}\mathrm{d}P} \\ 
         & \Rightarrow F_R = c\left( \rho_0,P_0 \right)+\frac{x}{t}-u_1+\int_{P_0}^{P_1}{\frac{1}{\rho c}\mathrm{d}P} =0.
\label{equ34}
    \end{aligned}
\end{equation}
Differentiating $F_R$ with respect to $\rho$ and substituting the definition of the isentropic speed of sound ${c^2}= ({\partial P}/{\partial \rho})_{s}$, we obtain:
\begin{equation}
    \pderiv{F_R}{\rho_0}{P_0}= \left. {\pderiv{c}{\rho}{P}} \right|_{P=P_0} + \left. {\pderiv{c}{P}{\rho}} \right|_{\rho=\rho_0} c^2_0 + \frac{c_0}{\rho_0},
\end{equation}
where the analytical expressions for $({\partial c}/{\partial \rho})_{P}$ and $({\partial c}/{\partial P})_{\rho}$ are given in Appendix \ref{sec:dc_eq}. These expressions are also instrumental in the evaluation of the Landau derivative $\Gamma$ discussed later. When the coordinate $\xi$ lies within the interval $(\lambda_{l,\text{S}}, \lambda_{eq,\text{S}})$, the state remains constant in the $x-t$ plane (corresponding to point $S,1$ in the $P-v$ plane):
\begin{equation}
    u=u_S; \quad P=P_S; \quad \rho=\rho_s.
\end{equation}
These values can be determined from the initial state via isentropic relations. It is worth noting that for the Newton iteration algorithm designed for flash rarefaction waves, we select $\rho$ as the iteration variable rather than $P$, contrasting with the approach in \citet{Bai2025}. This choice is motivated by the fact that when the isentrope crosses the liquid saturation boundary into the two-phase region (particularly at low pressures), the speed of sound drops significantly, causing the isentrope to become very flat in the $P-\rho$ plane. If pressure were used as the iteration variable, small pressure perturbations would result in large density variations, making convergence difficult or impossible. Numerical experiments confirm that selecting $\rho$ as the independent variable confers robust convergence characteristics.
\par

\section{Non-classical Waves in the Flash evaporation Riemann Problem and Their Solution Methods}

As previously discussed, the FeRP admits three types of wave structures: rarefaction waves, RS composite waves, and RSR composite waves. The physical mechanisms underpinning non-classical waves in BZT and phase-changing fluids have been analyzed in detail by \citet{MenikoffPlohr1989} and \citet{Dahmen2005}. Building upon their findings, this section investigates the properties of these waves, specifically from the perspective of solution algorithm design.

\subsection{Landau Derivative of Classical Equilibrium Phase-Change Fluids}
\label{sec:Landau Gamma}

The sign and smoothness of the Landau derivative determine whether the fluid dynamics exhibit "classical" behavior. It is defined as:
\begin{equation}
    \Gamma = -\frac{v}{2}\frac{\partial ^2 P(v,s)/\partial v^2}{\partial P(v,s)/\partial v}.
\end{equation}
Through algebraic manipulation, $\Gamma$ can be reformulated in terms of the thermodynamic derivatives obtained previously:
\begin{equation}
    \Gamma = 1+\frac{\rho}{c} \left(c^2\pderiv{c}{P}{\rho} +\pderiv{c}{\rho}{P} \right),
     \label{eq:Gamma2}
\end{equation}
where $({\partial c}/{\partial \rho})_{P}$ and $({\partial c}/{\partial P})_{\rho}$ correspond to the terms used in the iteration for calculating rarefaction waves in Section \ref{sec:solve raf}. Clearly, the term in the brackets
\begin{equation}
    c^2\pderiv{c}{P}{\rho} +\pderiv{c}{\rho}{P} = \pderiv{c}{\rho}{s}
\end{equation}
represents the derivative of the speed of sound with respect to density along the isentrope, which also reflects the intrinsic link between rarefaction wave properties and the value of $\Gamma$. Fundamentally, the value of $\Gamma$ characterizes the convexity of the isentrope. When $\Gamma > 0$ and strict smoothness holds globally, the isentrope is smooth and globally convex, and the wave dynamic behavior of the fluid is classical; otherwise, non-classical waves appear.

In this paper, we consistently employ the PR EoS to evaluate the Landau derivative $\Gamma$, resolve phase equilibrium, and compute the Riemann solution. The PR EoS is widely adopted in the chemical engineering field and demonstrates high predictive accuracy for propulsion working fluids such as $\text{CO}_2$, $\text{CH}_4$, n-dodecane, and other hydrocarbon gases over a wide pressure range. For the calculation of internal energy and entropy, we use exact integration based on the PR EoS (see \citet{GuardoneArgrow2005} for details), with the corresponding coefficients calibrated against NIST data \citep{NIST2023}. The expressions for the PR EoS, its thermodynamic derivatives, and the values of related coefficients are provided in Appendices \ref{sec:PR_EOS} and \ref{sec:thermo_dodecane}.
\par
\begin{figure}[htpb]
    \centering
    \begin{subfigure}{0.49\textwidth}
        \includegraphics[width=\linewidth]{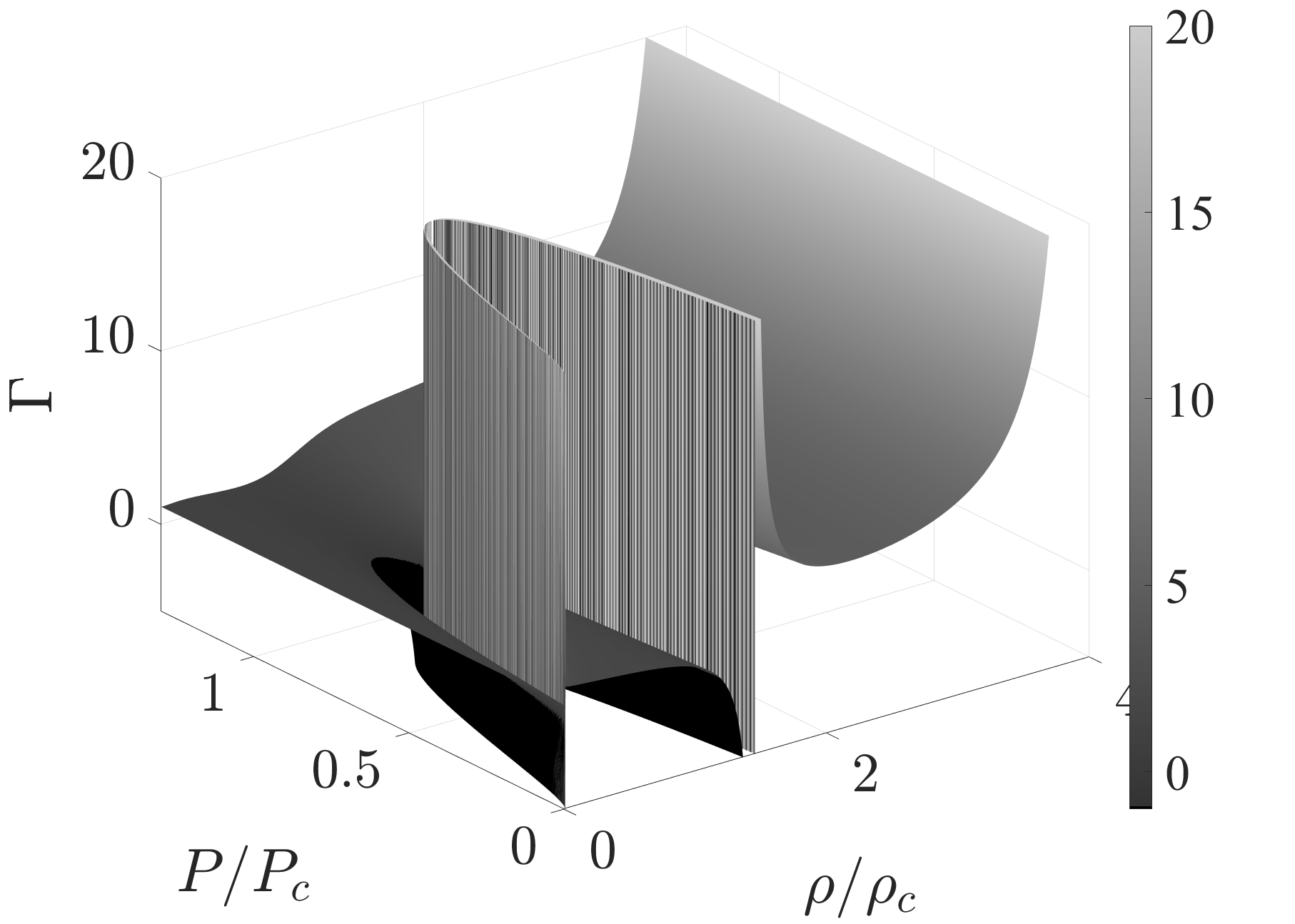}
        \caption{Without VLE}
        \label{fig5a}
    \end{subfigure}
    \hfill
    \begin{subfigure}{0.49\textwidth}
        \includegraphics[width=\linewidth]{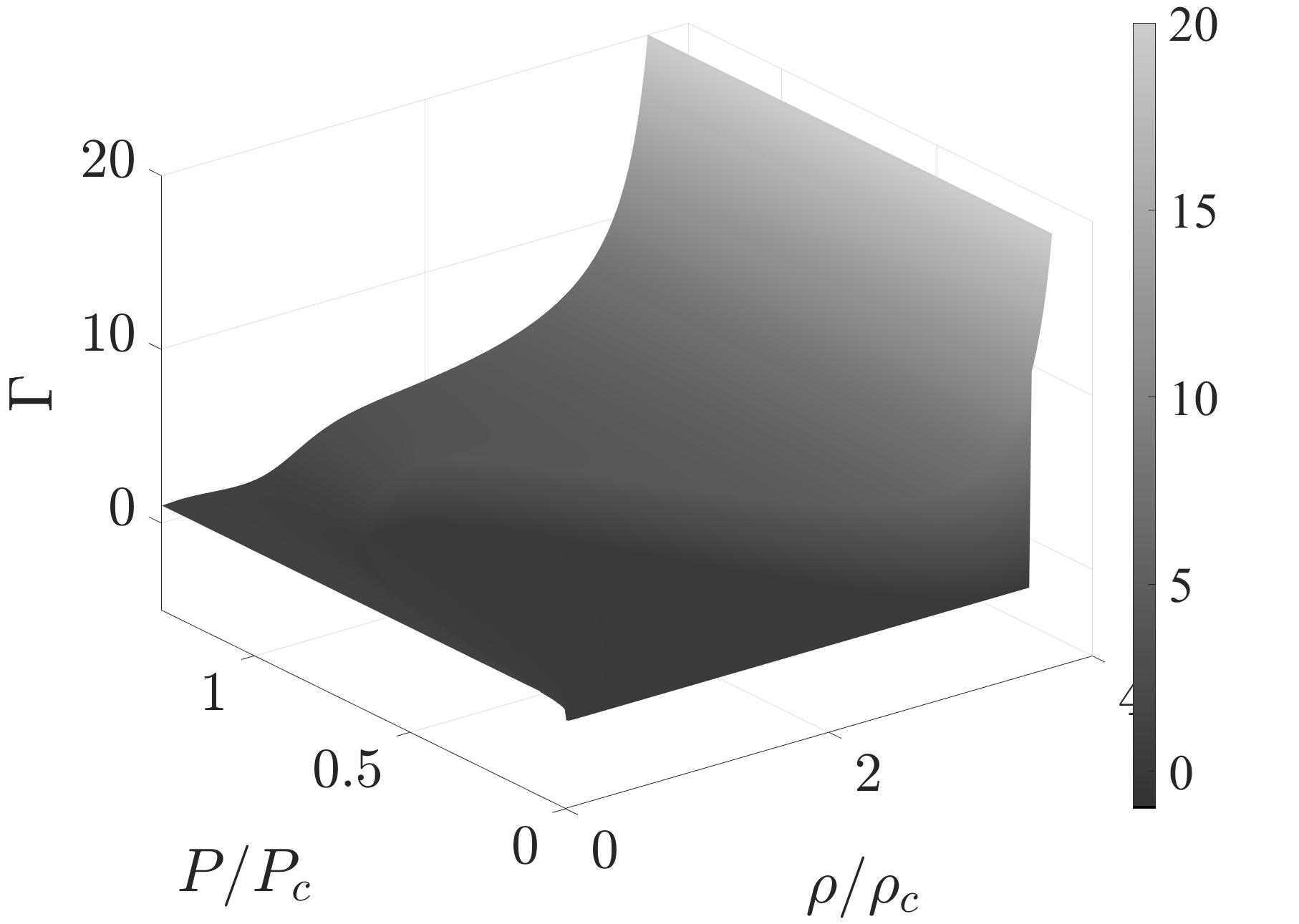}
        \caption{VLE model}
        \label{fig5b}
    \end{subfigure}
    \caption{Landau derivative $\Gamma$ of n-dodecane}
    \label{fig5}
\end{figure}

Figure \ref{fig5} illustrates the Landau derivative $\Gamma$ for n-dodecane calculated via the PR EoS. Specifically, Figure \ref{fig5a} depicts the distribution of $\Gamma$ computed directly without incorporating the VLE model. The derivative $\Gamma$ diverges to infinity at the isentropic spinodal line ($c^2=0$), which is evident from the denominator in Equation \ref{eq:Gamma2}. For the majority of fluids (such as $\text{CO}_2$, $\text{N}_2$, and $\text{O}_2$), the domain where $\Gamma<0$ is totally enclosed by the isothermal spinodal line; consequently, these regions are excluded from consideration as they lack physical meaning. What is of real concern is the scenario where the $\Gamma<0$ region extends outside the spinodal line, as observed in the supercooled vapor of n-dodecane (see Figure \ref{fig1}). However, upon incorporating the VLE model, we observe that $\Gamma$ for n-dodecane remains positive and smooth throughout the entire domain, except at the saturation line, as demonstrated in Figure \ref{fig5b}. Common propellants of interest, such as $\text{CO}_2$, $\text{N}_2$, $\text{O}_2$, and $\text{CH}_4$, share this characteristic. Consequently, the analysis of the wave structure in the FeRP reduces to considering non-classical waves solely at the phase boundaries.

\begin{figure}[htpb]
    \centering
    \begin{subfigure}{0.47\textwidth}
       \includegraphics[width=\linewidth]{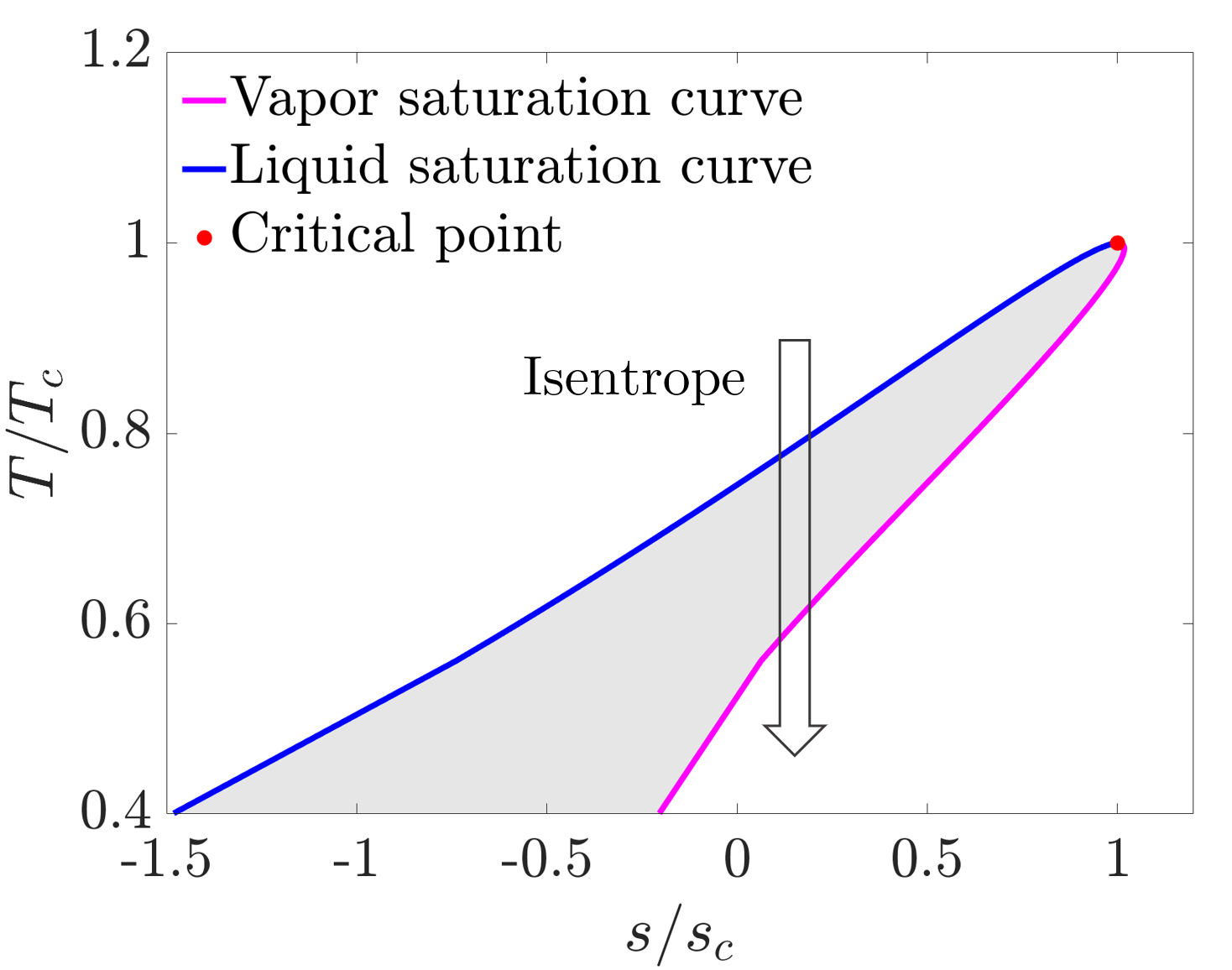}
        \caption{n-dodecane}
        \label{fig6a}
    \end{subfigure}
    \hfill
    \begin{subfigure}{0.47\textwidth}
        \includegraphics[width=\linewidth]{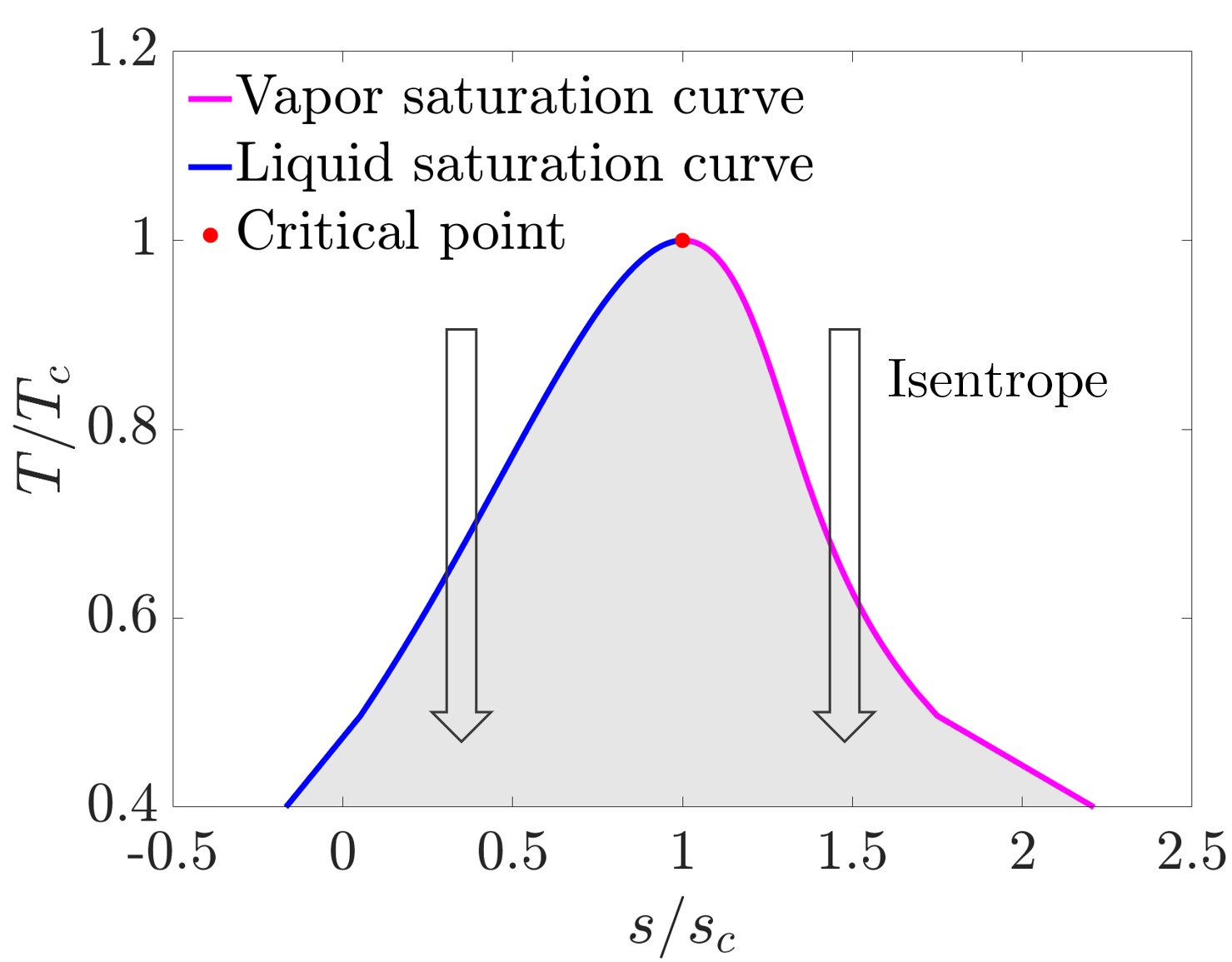}
        \caption{$\text{CO}_2$}
        \label{fig6b}
    \end{subfigure}
    \caption{Phase diagram of a low heat capacity fluid. Isentropic lines cannot enter the Vapor region from the two-phase region}
    \label{fig6}
\end{figure}

There is a theoretical detail in Figure \ref{fig5} that cannot be captured by numerical calculations. Revisiting the definition of $\Gamma$ in Equation \ref{eq:Gamma2}, the discontinuous jump in the speed of sound at the phase boundary induces a divergence in $({\partial c}/{\partial \rho})_{P}$ and $({\partial c}/{\partial P})_{\rho}$, which manifests in $\Gamma$ as a local Dirac "$\delta$-function" singularity on the saturation line. The impact of this singularity on the wave structure varies according to the fluid properties:
\begin{itemize}
    \item When the isentrope crosses the liquid saturation line, the characteristic speed always exhibits a positive jump $\lambda_{l} < \lambda_{eq}$. Consequently, $\Gamma$ exhibits a positive $\delta$-function singularity. The condition $\Gamma>0$ still holds globally, and a split rarefaction wave exists without non-classical waves.
    \item For the vapor saturation line, the direction in which the isentrope crosses varies for different fluids. For fluids with high specific heat capacity (such as n-dodecane), the isentrope crosses from the two-phase region into the gas phase region, which corresponds to a retrograde crossing direction, as shown in Figure \ref{fig6a}. In this configuration, the characteristic speed exhibits a negative jump $\lambda_{eq} \Rightarrow \lambda_{v}$, with $\lambda_{eq} > \lambda_{v}$. Consequently, $\Gamma$ exhibits a negative $\delta$-function singularity, locally violating the condition $\Gamma>0$ and giving rise to RS or RSR composite waves.
    \item For general low heat capacity, low molecular weight fluids (such as $\text{CO}_2$, $\text{CH}_4$, $\text{O}_2$), expansion drives the isentrope deeper into the two-phase region but does not cross the vapor saturation line retrogradely. Therefore, the isentrope only crosses from the vapor phase region into the two-phase region, as shown in Figure \ref{fig6b}. In this configuration, the characteristic speed exhibits a positive jump $\lambda_{v} \Rightarrow \lambda_{eq}$. Consequently, $\Gamma$ exhibits a positive $\delta$-function singularity, and only a split rarefaction wave exists.
\end{itemize}

In summary, the dynamic condition for the emergence of non-classical waves is a retrograde jump in characteristic speed — thermodynamically equivalent to the manifestation of a negative $\delta$-function singularity in $\Gamma$ — which consequently gives rise to RS and RSR composite waves. This phenomenon occurs during the flash evaporation of high heat capacity fluids like n-dodecane. Conversely, for fluids like $\text{CO}_2$, $\text{CH}_4$, and $\text{O}_2$, non-classical waves are absent, and only rarefaction wave splitting occurs. In addition, when gaseous n-dodecane is compressed into the two-phase region, non-classical waves may also appear, specifically manifesting as Shock-Compression Fan (SC) composite waves or Shock-Compression Fan-Shock (SCS) composite waves—a behavior that is the opposite of the expansion case. However, such scenarios are rare in practice, as they occur only when the initial state of gaseous n-dodecane is located in close proximity to the vapor saturation line. In most cases, the compression branch of gaseous n-dodecane in the RP is a simple vapor-phase shock, as shown in Figure \ref{fig4}. The methodology for constructing SCS waves is analogous to the method for RSR waves detailed subsequently. Given the specific focus on the "flash evaporation" problem, we will not further discuss the cases involving SRS waves.

\subsection{Properties of RSR Double Sonic Shock and RS Upstream Sonic Shock}
\label{sec:RSR_property}

\begin{figure}[htpb]
    \centering
    \begin{subfigure}{0.45\textwidth}
        \includegraphics[width=\linewidth]{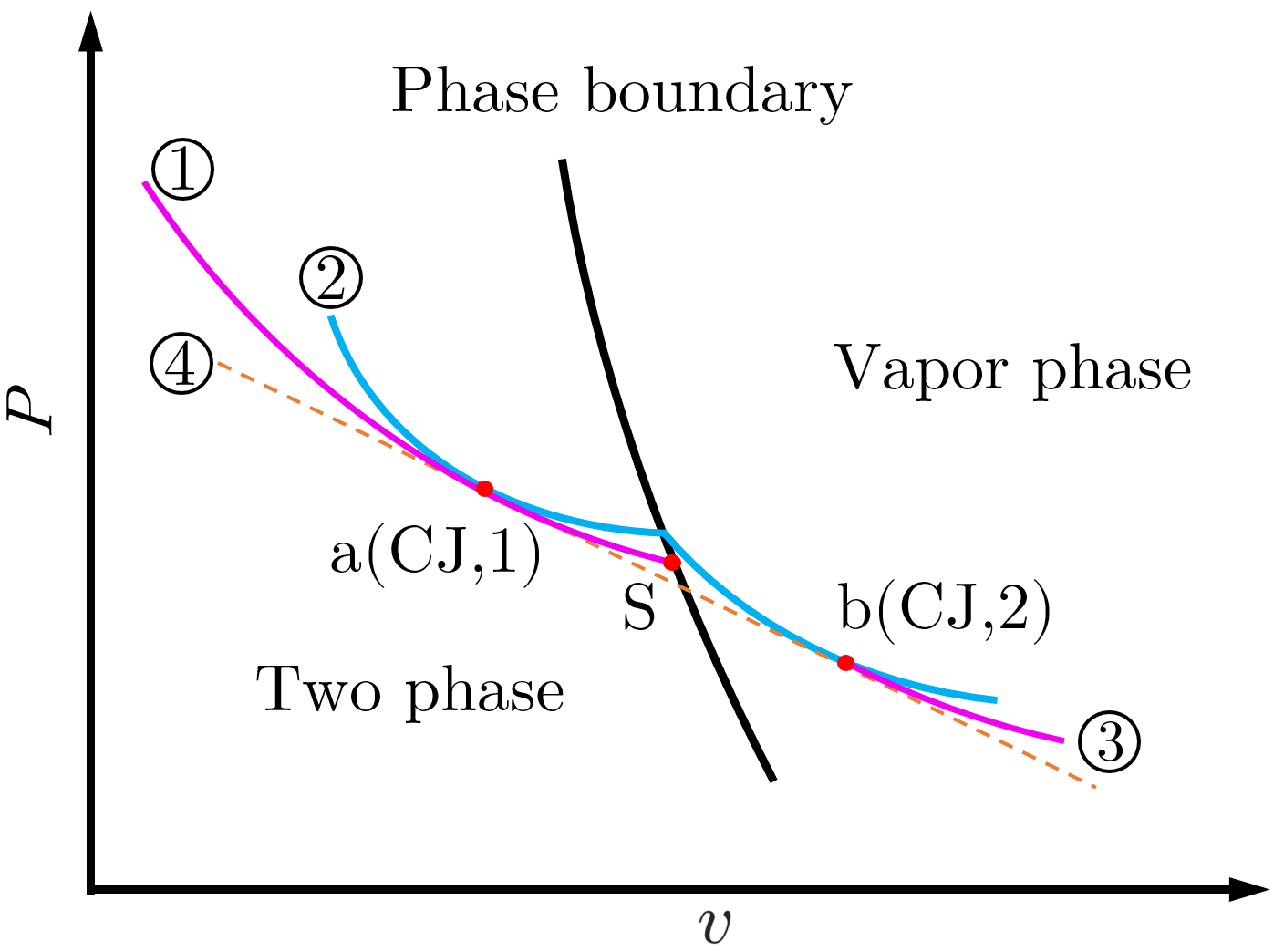}
        \caption{RSR wave}
        \label{fig7a}
    \end{subfigure}
    \hfill
    \begin{subfigure}{0.45\textwidth}
        \includegraphics[width=\linewidth]{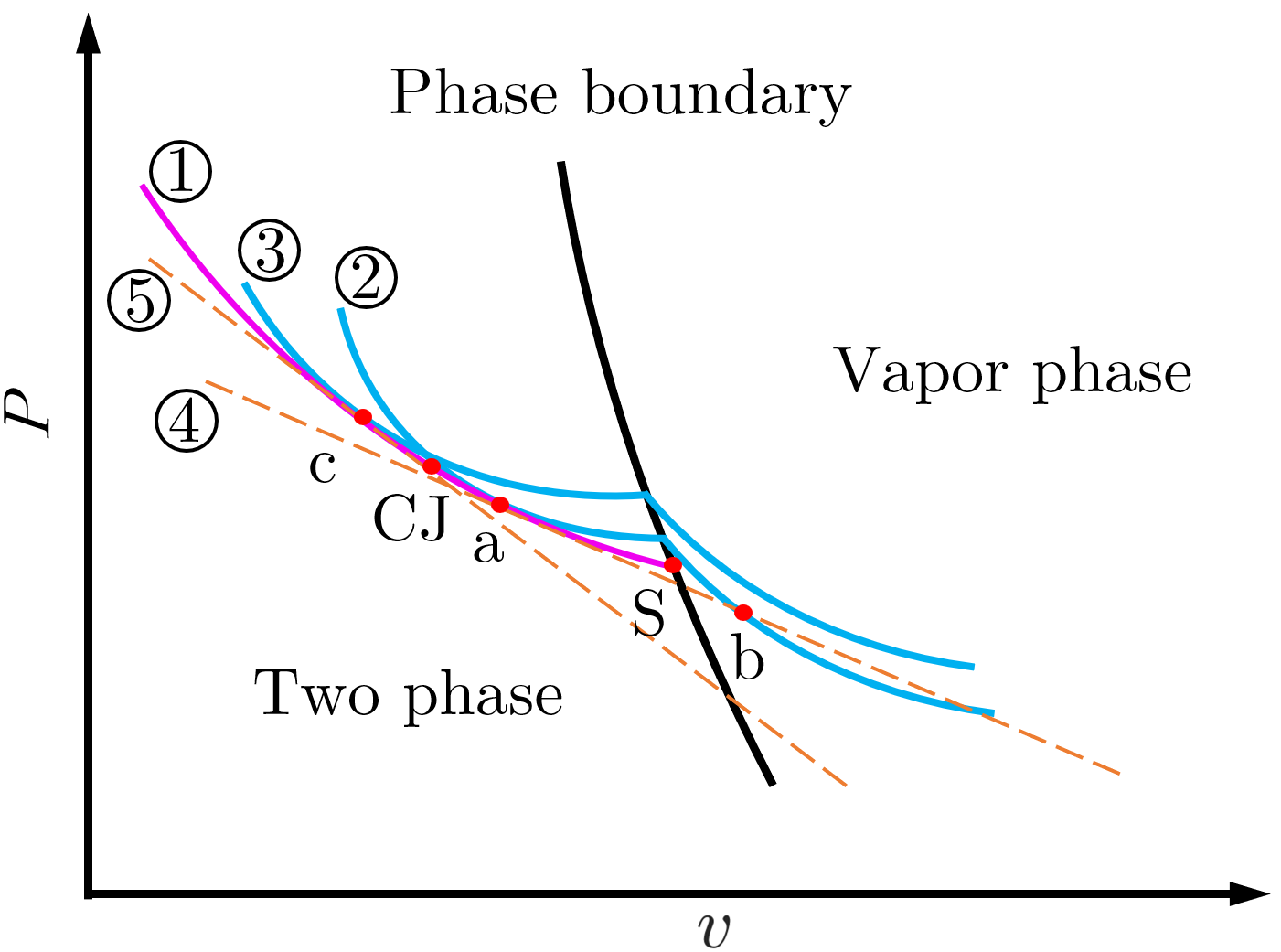}
        \caption{RS wave}
        \label{fig7b}
    \end{subfigure}
    \caption{Relationship between isentropic lines, Hugoniot curves, and Rayleigh lines in nonclassical waves}
    \label{fig7}
\end{figure}

We first examine the physical properties of the RSR wave on the $P-v$ diagram. As shown in Figure \ref{fig7a}, the double sonic shock is an expansion shock connecting state $a$ $(P_a, \rho_a, e_a, v_a, c_a)$ to state $b$ $(P_b, \rho_b, e_b, v_b, c_b)$, subject to the following conditions:\\
\begin{enumerate}
    \item[\textbf{C1:}] \textbf{Isentropic}: State $a$ lies on the initial isentrope, i.e., the isentrope \textcircled{1} intersects with the Hugoniot curve \textcircled{2}.
    \begin{equation}
        F_s = s(P_a, \rho_a) - s_L = 0.
    \end{equation}
    
    \item[\textbf{C2:}] \textbf{R-H}: States $a$ and $b$ must satisfy the Rankine-Hugoniot energy conservation, i.e., $a$ and $b$ are on the same Hugoniot curve \textcircled{2}.
    \begin{equation} \label{eq:C2_Frh_rsr}
        F_H = e_a - e_b + \frac{P_a}{\rho_a} - \frac{P_b}{\rho_b} + \frac{(P_b - P_a)(\rho_a + \rho_b)}{2\rho_a\rho_b} = 0.
    \end{equation}
    
    \item[\textbf{C3:}] \textbf{Pre-wave CJ}: The shock mass flux $j$ must equal the pre-wave acoustic impedance. Geometrically, the isentrope \textcircled{1}, Hugoniot curve \textcircled{2}, and Rayleigh line \textcircled{4} are tangent at point $a$.
    \begin{equation}
        f_1 = j^2 - (\rho_a c_a)^2 = 0.
    \end{equation}
    
    \item[\textbf{C4:}] \textbf{Post-wave CJ}: The shock mass flux $j$ must equal the post-wave acoustic impedance. Geometrically, the isentrope \textcircled{3}, Hugoniot curve \textcircled{2}, and Rayleigh line \textcircled{4} are tangent at point $b$.
    \begin{equation}
        f_2 = j^2 - (\rho_b c_b)^2 = 0.
    \end{equation}
\end{enumerate}

The combination of these four conditions uniquely determines the pre-wave and post-wave states of the double sonic shock in an RSR wave. For the upstream sonic shock of RS wave, conditions \textbf{C1, C2} and \textbf{C3} are retained, whereas condition \textbf{C4} is relaxed. Consequently, to determine an RS wave, the post-wave pressure $P_b$ must be specified. This dependence explains why the starting state point $a$ of the RS wave varies as a function of the end pressure $P_b$. Furthermore, state $b$ implies an implicit condition: the shock mass flux should be less than the post-wave acoustic impedance, $j^2 < (\rho_b c_b)^2$ (i.e., the post-wave subsonic condition). In Figure \ref{fig7b}, this indicates that the Rayleigh line \textcircled{4} must not only be tangent to the isentrope \textcircled{1} and Hugoniot curve \textcircled{2} at point $a$, but also intersect the Hugoniot curve\textcircled{2} at a distinct second point $b$. This condition is equivalent to requiring point $a$ in Figure \ref{fig7b} to be located downstream of the CJ point of the RSR wave (point $a$ in Figure \ref{fig7a}). If the shock of the RS wave starts upstream of this CJ point—like point $c$ in Figure \ref{fig7b}—the solution satisfying conditions \textbf{C1, C2, C3} collapses to a trivial solution coinciding with $c$. In other words, the Rayleigh line \textcircled{5} fails to intersect the Hugoniot curve \textcircled{3} at any point downstream of $c$. The existence of a trivial solution for the RS wave imposes specific requirements on the design of the Newton iteration method for RSR and RS waves. Since four conditions must be satisfied simultaneously, a nested solution strategy is required. It is worth noting that we cannot use the R-H condition as the governing condition for the outermost iteration, because $P_a$ tends to converge to the trivial solution as it approaches $P_\text{CJ}$. Based on this consideration, we employ the CJ condition as the constraint for the outer-loop Newton iteration. This design ensures that a non-trivial shock wave always exists throughout the iteration process.

\subsection{Newton's Method for the Double Sonic Shock in RSR Waves}
\label{sec:solve RSR}

The objective of solving for the double sonic shock is to determine the pre-wave pressure $P_a$ and post-wave pressure $P_b$. To achieve this, we formulate a $2\times2$ matrix Newton iteration method. The vector of unknowns $\bm{x}$ is defined as:

\begin{equation}
    \bm{x} = \begin{bmatrix} P_a \\ P_b \end{bmatrix}.
\end{equation}
The residual function vector $\bm{F}(\bm{x}) = \bm{0}$ is governed by conditions \textbf{C3, C4} in Section \ref{sec:RSR_property}:
\begin{equation}
    \bm{F_\text{RSR}}(\bm{x}) = \begin{bmatrix} f_1 \\ f_2 \end{bmatrix} = \begin{bmatrix} j^2 - (\rho_a c_a)^2 \\ j^2 - (\rho_b c_b)^2 \end{bmatrix} = \begin{bmatrix} 0 \\ 0 \end{bmatrix}.
\end{equation}
Conditions \textbf{C1, C2} are satisfied implicitly within the inner loop via the Newton iterations for the R-H relation (Equation \ref{equ:F_H}) and the isentropic relation (Equation \ref{equ:F_s}) presented in Section \ref{sec:solve p*}. Consequently, conditions \textbf{C1, C2} are automatically satisfied as constraints throughout the outer iteration. The Newton iteration formula is:
\begin{equation}
    \bm{x}^{(k+1)} = \bm{x}^{(k)} - \bm{J}^{-1} \bm{F_\text{RSR}}^{(k)},
\end{equation}
where $\bm{J}$ denotes the Jacobian matrix:
\begin{equation}
    \bm{J} = \begin{bmatrix}
    \pdv{f_1}{P_a} & \pdv{f_1}{P_b} \\
    \pdv{f_2}{P_a} & \pdv{f_2}{P_b}
\end{bmatrix}.
\end{equation}

To evaluate $\bm{J}$, we first introduce three key implicit derivatives, which are the derivatives of conditions \textbf{C1, C2} with respect to the inputs:
\begin{align}
    D_a &=\totalderiv{\rho_a}{P_a}\Big|_{s} = \frac{1}{c_a^2}, \label{equ:Da_rsr} \\
    D_{b,P_b} & =\totalderiv{\rho_b}{P_b}\Big|_{H, a} = - \frac{(\partial{F_H}/ \partial{P_b})_{\rho_b}}{(\partial{F_H}/ \partial{\rho_b})_{P_b}}, 
    \label{equ:DbPb_rsr} \\
    D_{b,P_a} &=\frac{\partial{\rho_b}}{\partial{P_a}}\Big|_{H, P_b} = - \frac{\mathrm{d}{F_H}/\mathrm{d}{P_a}}{(\partial{F_H}/ \partial{\rho_b})_{P_b}}.
\label{equ:DbPa_rsr}
\end{align}
Here, Equation \ref{equ:Da_rsr} is the derivative of condition \textbf{C1} with respect to $P_a$, i.e., the slope of the isentrope at point $a$. Equation \ref{equ:DbPb_rsr} is the derivative of condition \textbf{C2} with respect to $P_b$, i.e., the slope of the Hugoniot curve at point $b$ for a fixed state $a$. In Section \ref{sec:phiR- S}, we have already obtained:
\begin{align}
    \pderiv{F_H}{P_b}{\rho_b} &= - \pderiv{e}{P}{\rho}\Big|_{P=P_b} +\frac{\rho_b - \rho_a}{2 \rho_b \rho_a}, \\
     \pderiv{F_H}{\rho_b}{P_b} &= - \pderiv{e}{\rho}{P}\Big|_{\rho=\rho_b}+\frac{P_a+P_b}{2 \rho_b^2}.
\end{align}    
Equation \ref{equ:DbPa_rsr} describes the sensitivity of condition \textbf{C2} to $P_a$, specifically the density variation at point $b$ induced by shifting the Hugoniot origin $a$ while holding $P_b$ constant. Its numerator $\mathrm{d}{F_H}/\mathrm{d}{P_a}$ represents the variation when state $b$ is fixed and state $a$ moves along the isentrope:

\begin{equation}
    \begin{aligned}
      \totalderiv{F_H}{P_a}= &\frac{\partial F_H}{\partial P_a} + \frac{\partial F_H}{\partial \rho_a}\totalderiv{\rho_a}{P_a}\Big|_{s} + \frac{\partial F_H}{\partial e_a}\frac{\mathrm{d}{e_a}}{\mathrm{d}{P_a} }
    \Big|_{s}\\
      = &\frac{\rho_b -\rho_a}{2\rho_a\rho_b}  - \frac{P_b-P_a}{2\rho_a^2 c_a^2}.
\end{aligned}
\end{equation}
Here, the isentropic relations $\mathrm{d}{\rho_a}/\mathrm{d}{P_a}|_{s} = 1/c_a^2$ and $\mathrm{d}{e_a}/\mathrm{d}{P_a}|_s = P_a / (\rho_a^2 c_a^2)$ are employed to simplify this expression. With these implicit derivatives, the components of the Jacobian matrix $\bm{J}$ are derived as follows:

\textbf{1. Component} $J_{11}$:
\begin{equation}
J_{11} = \pderiv{(j^2)}{P_a}{P_b}- \totalderiv{(\rho_a c_a)^2}{P_a}\Big|_{s},
\end{equation}
where $({\partial(j^2)}/{\partial P_a})_{P_b}$ is the partial derivative of $j^2(P_a, \rho_a(P_a), P_b, \rho_b(P_a, P_b))$ with respect to $P_a$:
\begin{align}
    \pderiv{(j^2)}{P_a}{P_b} &= \pderiv{(j^2)}{P_a}{\rho_a,\rho_b} + \frac{\partial(j^2)}{\partial(\rho_a)}\totalderiv{\rho_a}{P_a}\Big|_{s
    } + \frac{\partial(j^2)}{\partial \rho_b} \frac{\partial{\rho_b}}{\partial{P_a}}\Big|_{H}\notag \\
    &= \frac{-1}{v_a - v_b} + \frac{j^2}{(v_a-v_b)\rho_a^2 c_a^2} + \frac{j^2}{(v_a-v_b)\rho_b^2} D_{b,P_a}.
\label{equ:dj2dPa}
\end{align}
The second term ${\mathrm{d}(\rho_a c_a)^2}/\mathrm{d}{P_a}\Big|_{s}$ is the derivative along the isentrope:
\begin{align}
    \totalderiv{(\rho_a c_a)^2}{P_a} &= 2\rho_a c_a^2 \totalderiv{\rho_a}{P_a}\Big|_{s} + 2\rho_a^2 c_a \totalderiv{c_a}{P_a}\notag \\
    &= 2\rho_a + 2\rho_a^2 c_a\pderiv{c}{P}{\rho} + \frac{2\rho_a^2}{c_a} \pderiv{c}{\rho}{P}.
\label{equ:dja2dPa}
\end{align}

\textbf{2. Component} $J_{12}$:
\begin{equation}
    J_{12} = \pderiv{(j^2)}{P_b}{P_a}- \frac{\partial(\rho_a c_a)^2}{\partial P_b} = \pderiv{(j^2)}{P_b}{P_a}.
\end{equation}
Since $\rho_a c_a$ is independent of $P_b$, the second term vanishes. $({\partial(j^2)}/{\partial P_b})_{P_a}$ is the partial derivative of $j^2(P_a, \rho_a, P_b, \rho_b(P_b))$ with respect to $P_b$:
\begin{align}
    \pderiv{(j^2)}{P_b}{P_a} &= \pderiv{(j^2)}{P_b}{\rho_b} + \frac{\partial(j^2)}{\partial \rho_b}\totalderiv{\rho_b}{P_b}\Big|_{H} \notag\\
    &= \frac{1}{v_a - v_b} + \frac{j^2}{(v_a-v_b)\rho_b^2}D_{b,P_b}.
\label{equ:dj2dPb}
\end{align}    

\textbf{3. Component} $J_{21}$:
\begin{equation}
J_{21} = \pderiv{(j^2)}{P_a}{P_b}- \totalderiv{(\rho_b c_b)^2}{P_a}.
\end{equation}
Here, the first term $({\partial(j^2)}/{\partial P_a})_{P_b}$ is identical to Equation \ref{equ:dj2dPa} in $J_{11}$. $\mathrm{d}(\rho_b c_b)^2/\mathrm{d}{P_a}$ is the implicit derivative of $(\rho_b c_b)^2$ with respect to $P_a$ via the intermediate variable $\rho_b$:
\begin{align}
    \frac{\partial(\rho_b c_b)^2}{\partial P_a} &= \pderiv{(\rho_b c_b)^2}{\rho_b}{P_b} \cdot \frac{\partial\rho_b}{\partial P_a}\Big|_{H}\notag \\
    &= \left( 2\rho_b c_b^2 + 2\rho_b^2 c_b\pderiv{c_b}{\rho_b}{P_b} \right) \cdot D_{b,P_a}.
\end{align}

\textbf{4. Component} $J_{22}$:
\begin{equation}
J_{22} = \pderiv{(j^2)}{P_b}{P_a}- \totalderiv{(\rho_b c_b)^2}{P_b}\Big|_{H}.
\end{equation}
Here, the first term $({\partial(j^2)}/{\partial P_b})_{P_a}$ is identical to Equation \ref{equ:dj2dPb} in $J_{12}$. The second term ${\mathrm{d}(\rho_b c_b)^2}/\mathrm{d}{P_b}\Big|_{H}$ is the derivative along the Hugoniot curve:
\begin{align}
    \totalderiv{(\rho_b c_b)^2}{P_b}\Big|_{H} &= \pderiv{(\rho_b c_b)^2}{P_b}{\rho_b} + \pderiv{(\rho_b c_b)^2}{\rho_b}{P_b} \cdot \totalderiv{\rho_b}{P_b}\Big|_{H}\notag \\
    &= 2\rho_b^2 c_b \pderiv{c_b}{P_b}{\rho_b}  + \left( 2\rho_b c_b^2 + 2\rho_b^2 c_b\pderiv{c_b}{\rho_b}{P_b} \right) \cdot D_{b,P_b}.
\label{equ:djb2dPb}
\end{align}

\subsection{Newton's Method for the Upstream Sonic Shock in RS Waves}
\label{sec:solve RS}
The calculation of the upstream sonic shock resolves the pre-wave pressure $P_a$ given a post-wave pressure $P_b$. According to the analysis in Section \ref{sec:RSR_property}, we establish a Newton iteration with the pre-wave CJ condition \textbf{C3} as governing equation. With $P_a$ as the independent variable, the residual function is given by:
\begin{equation}
    F_\text{RS}(P_a) = j^2 - (\rho_a c_a)^2.
\end{equation}
Similarly, conditions \textbf{C1} and \textbf{C2} are satisfied implicitly within the inner loop via the Newton iterations for the R-H relation (Equation \ref{equ:F_H}) and the isentropic relation (Equation \ref{equ:F_s}).
\par
The derivative of the residual function is:
\begin{equation}
    \totalderiv{F_\text{RS}}{P_a} = \totalderiv{j^2 }{P_a}- \totalderiv{(\rho_a c_a)^2}{P_a}\Big|_{s},
\end{equation}
where the first term ${\mathrm{d}(j)^2}/\mathrm{d}{P_a}$ is identical to the term $({\partial(j^2)}/{\partial P_a})_{P_b}$ in $J_{11}$, with the distinction that $P_b$ is treated here as a known value, and thus we use the total derivative. Consequently:
\begin{equation*}
    \totalderiv{F_\text{RS}}{P_a} = J_{11}.
\end{equation*}

\subsection{Derivatives of the RS Wave \texorpdfstring{$P-u$}{P-u} function}     
\label{sec:RS derivative}

In Section \ref{sec:psiL- RS}, we deferred the analytical expressions for the derivatives ${\mathrm{d}j_a}/{\mathrm{d}P_a}$ and ${\mathrm{d}P_a}/{\mathrm{d}P^*}$ in the $P-u$ curve of the RS wave. Obviously, ${\mathrm{d}j_a}/{\mathrm{d}P_a}$ is already given by Equation \ref{equ:dja2dPa}. Regarding ${\mathrm{d}P_a}/{\mathrm{d}P^*}$, for consistency with previous derivations, we set $P_b = P^*$. By implicit differentiation of the condition $F_\text{RS}(P_a, P_b) = 0$::
\begin{equation}
    \totalderiv{P_a}{P_b} = -\frac{(\partial{F_\text{RS}}/ \partial{P_b})_{P_a}}{(\partial{F_\text{RS}}/ \partial{P_a})_{P_b}} = - \frac{J_{12}}{J_{11}}.
\end{equation}

At this point, we have established the Newton iteration method for the exact solution of the FeRP under the HEM and VLE assumptions. It should be noted that this exact solution is difficult to apply directly to the construction of first-order Godunov methods, as two-phase flow Godunov methods suffer from the well-known problem of pressure oscillations at contact discontinuities \citep{Abgrall1996}. We recommend employing approximate Riemann solvers combined with double-flux methods \citep{Ma2017, Zhang2024} for Computational Fluid Dynamics (CFD) simulations. The convergence, energy conservation errors, and other related properties of these methods can be validated against the exact solution established here. To facilitate implementation and reproducibility, we provide an open-source program for the algorithm described in this paper, as detailed in Appendix \ref{sec:matlab}.

\section{Thermodynamic Analysis and Exact Solution of the Flash evaporation Riemann Problem based on Wood's Model}
\subsection{Thermodynamic Properties of Wood's Speed of Sound}

Wood's speed of sound assumes mechanical equilibrium of pressure and velocity between two phases, but neglects thermodynamic equilibrium. Its definition is:
\begin{equation}
    \frac{1}{\rho c_{W}^2}=\frac{\alpha }{{\rho_v}c_v^2}+\frac{1-\alpha }{{\rho_l}c_l^2}.
\label{equ_c_wood}
\end{equation}
Defining Wood's compressibility:
\begin{equation}
    {K_W} = \rho  \left(\frac{\alpha}{\rho_v c_v^2} + \frac{1-\alpha}{\rho_l c_l^2}\right),
    \label{equ:Kw}
\end{equation}
and the acoustic compressibility associated with the change of two-phase mixture entropy:
\begin{equation}
    {K_H} =\rho T\left( \frac{\alpha {\rho_v}}{{C_{P,v}}}{{\left( \frac{\mathrm{d}s_v}{\mathrm{d}P} \right)}^{2}}+\frac{(1-\alpha ){\rho_l}}{C_{P,l}}{{\left( \frac{\mathrm{d}s_l}{\mathrm{d}P} \right)}^{2}} \right).
\end{equation}
Then we have:
\begin{equation}
   c_{eq} = (K_W + K_H)^{-1/2}; \quad  c_{W} = (K_W)^{-1/2}.
\end{equation}
Evidently, Wood's speed of sound $c_W$ differs from the complete equilibrium speed of sound $c_{eq}$ by the compressibility term $K_H$, which is related to entropy change. This indicates that $c_W$ represents the derivative along the isentrope of a non-equilibrium mixture entropy. When thermal equilibrium is considered, the entropies of the two phases will reach the same temperature through heat exchange. During this process, entropy changes occur in both phases, making the final two-phase mixture entropy consistent with Equation \ref{equ:mix_s}.
\par
\begin{figure}[htpb]
    \centering
    \begin{subfigure}{0.47\textwidth}
        \includegraphics[width=\linewidth]{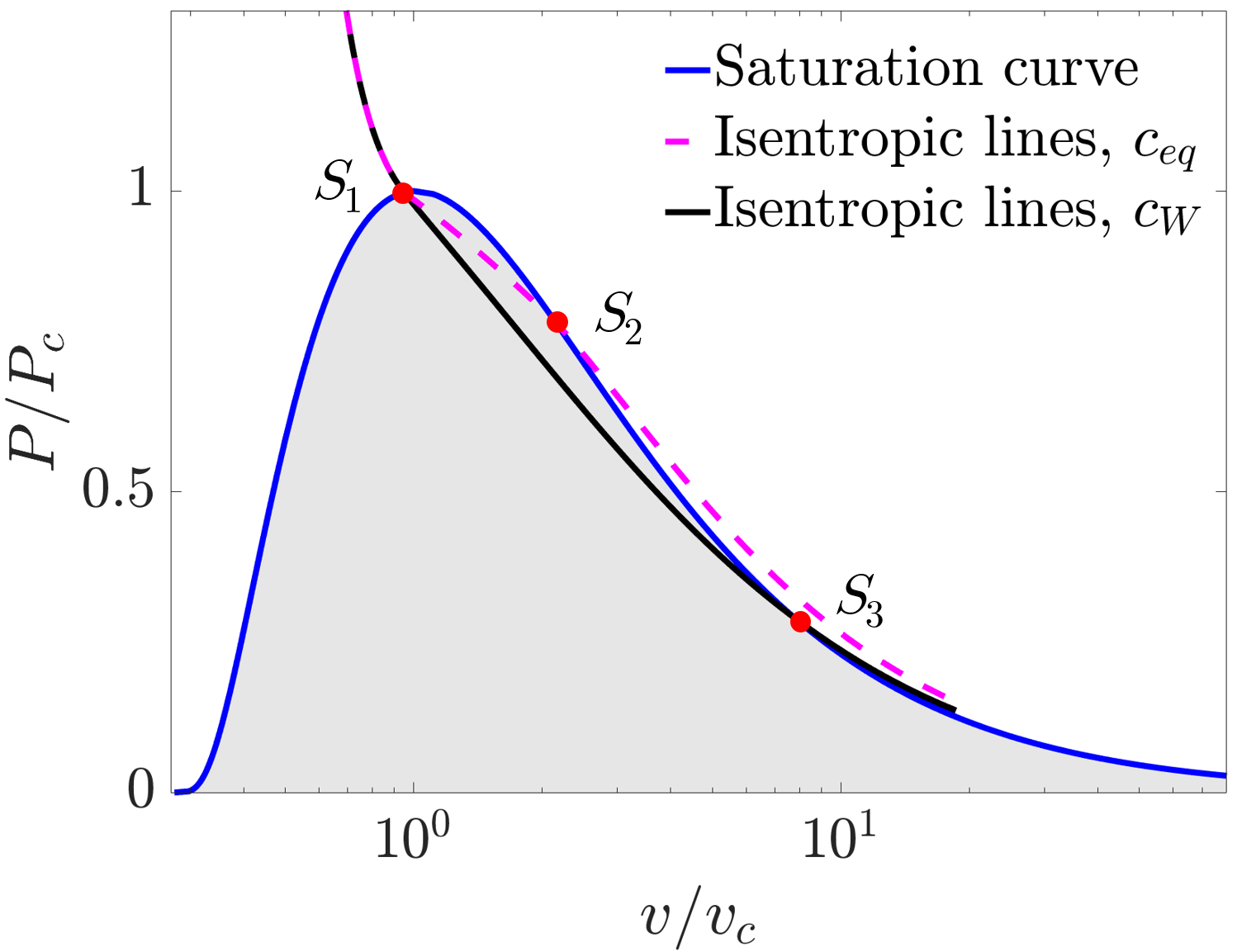}
        \caption{Isentropic lines on $P-v$ coordinates}
        \label{fig8a}
    \end{subfigure}
    \hfill
    \begin{subfigure}{0.52\textwidth}
        \includegraphics[width=\linewidth]{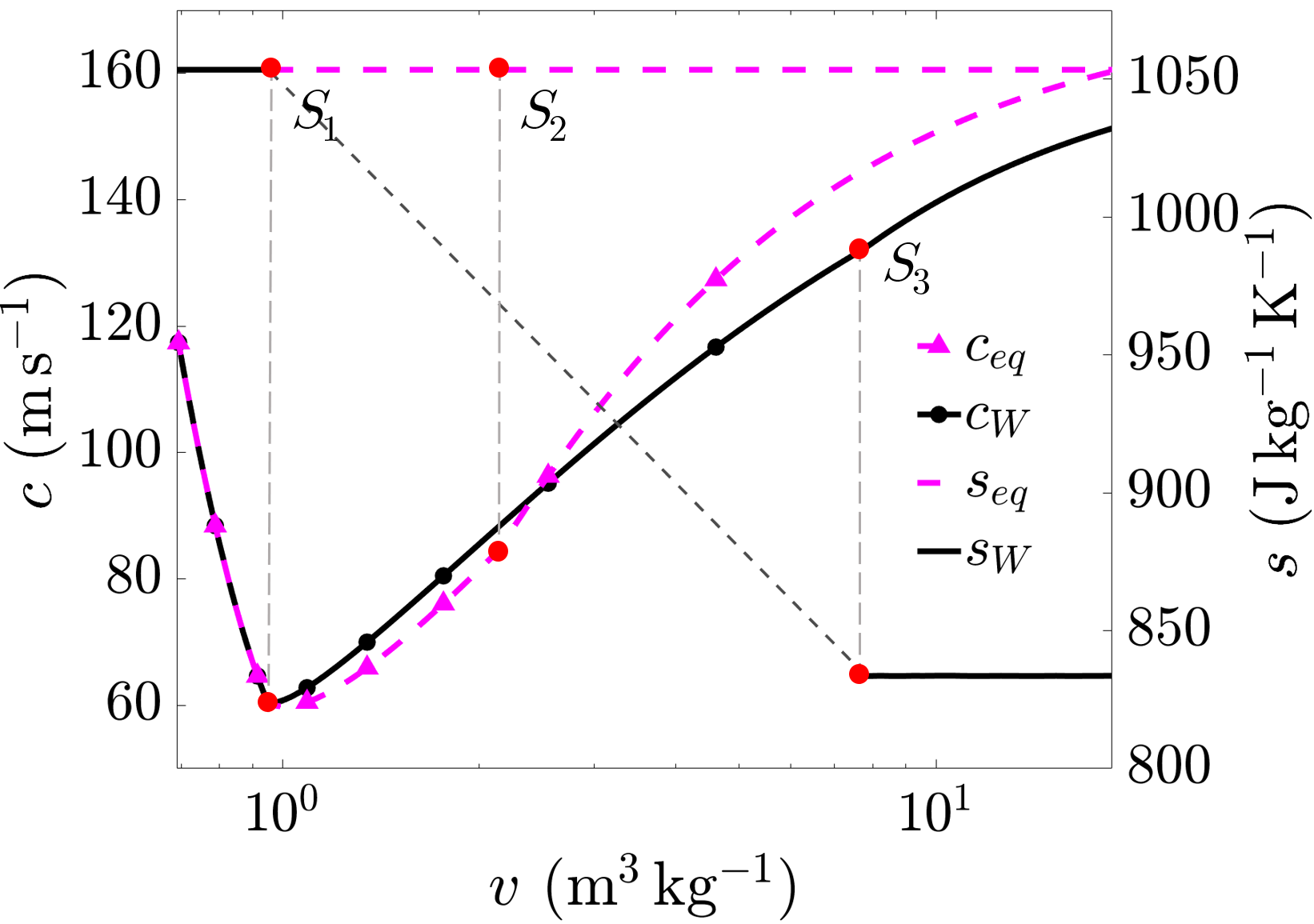}
        \caption{Speed of sound along isentropic lines}
        \label{fig8b}
    \end{subfigure}
    \caption{Isentropy paths of $c_{eq}$ and $c_W$}
    \label{fig8}
\end{figure}

However, we point out that applying Wood's speed of sound within the HEM and VLE framework is \textbf{thermodynamically inconsistent}. Specifically, we cannot simultaneously adopt the definition of homogeneous equilibrium mixing entropy (which inherently implies $c_{eq}$) and use $c_W$ to specify the characteristic propagation speed. Particularly in Godunov methods based on approximate Riemann solvers, once $c_W$ is specified as the eigenvalue, the isentropic path in the two-phase region is forced to use $c_W^2$ as the slope. This is because the Euler equations themselves do not "know" what entropy is; they only "know" what the slope of the isentrope is. If entropy is defined according to Equation \ref{equ:mix_s}, the isentrope is uniquely determined, which in turn determines the expression for the speed of sound as $c_{eq}$ (Equation \ref{equ:c_eq}). In reality, the use of Wood's speed of sound implicitly redefines the entropy of the two-phase region as a non-equilibrium mechanical mixture entropy. Denoting this mechanical mixture entropy as $s_W$, the slope of the isentrope becomes:

\begin{equation}
    \pderiv{P}{\rho}{s_p,Y_p} =-\frac{(\partial s_W/ \partial \rho)_P }{(\partial s_W/ \partial P)_\rho } =  c_W^2,
\end{equation}
where the subscript $(s_p, Y_p)$ signifies that the entropies of each phase and the mass fraction are invariant. Figure \ref{fig8} illustrates the isentropic path for n-dodecane starting from an initial pressure of $2.7\mathrm{MPa}$ and density of $240\mathrm{kg\cdot m^{-3}}$. Within the two-phase region, the isentropic path under Wood's model corresponds to a "mechanical equilibrium" process, characterized by frozen phase composition and phasic entropies during differential changes. This path diverges significantly from the integration path derived using the complete equilibrium speed of sound $c_{eq}$, accompanied by a large deviation in the sound speed values along the isentrope. Consequently, this inevitably introduces significant deviations in the integration of the Riemann invariants, thereby causing a difference in the resulting intermediate pressure state.
\par
Note that we specifically adopt the term "Wood's model" because Wood's speed of sound is not merely a kinematic velocity concept; more importantly, it introduces an underlying thermodynamic model based on mechanical equilibrium mixture entropy. Consequently, what stands in contrast to Wood's speed of sound is not simply the equilibrium speed $c_{eq}$, but rather the complete equilibrium model discussed throughout the previous chapters. Hereinafter, we refer to Wood's speed of sound model as Wood's model (or $c_W$ model), and the complete equilibrium model as the standard model (or $c_{eq}$ model).

Furthermore, we highlight the \textbf{non-physical nature} of calculating flash expansion using Wood's model. Since $c_W$ is continuous across the entire $P-\rho$ domain, the expansion branch can continuously traverse the two-phase region, transitioning from the liquid phase to the vapor phase along a single smooth trajectory (path $S_1 \Rightarrow S_3$ in Figure \ref{fig8}; note that this path only exists when $S_1$ is close to the critical point). In the single-phase region (upstream of start point $S_1$ and downstream of end point $S_3$), the thermodynamic state of the fluid is defined by the single-phase equilibrium entropy (Equation \ref{equ:s}). In the two-phase region, the mechanical mixture entropy does not correspond to any point on the standard equilibrium thermodynamic phase diagram; hence, no value is plotted in the figure. (In fact, if a value had to be assigned to $s_W$, it would equal the initial entropy, simply because the path is defined as "isentropic".) However, after expanding from the two-phase region into the vapor phase region (downstream of point $S_3$ in Figure \ref{fig8}), the definition of entropy reverts to the single-phase equilibrium entropy (Equation \ref{equ:s}). From Figure \ref{fig8b}, we observe that after a so-called "isentropic" expansion, the entropy value decreases. The mechanism for this non-physical result is as follows:

\begin{itemize}
  \item \textbf{Isentropic Path Separation}: Once the expansion path crosses into the two-phase region from liquid phase, Wood's model constrains the system to follow a non-equilibrium mechanical mixture path. This path exhibits a steeper gradient on the $P-v$ diagram compared to the complete equilibrium isentrope. Mathematically, the isentropic path represented by Wood's model actually evolves within a high-dimensional non-equilibrium phase space $\mathcal{S}_W = \{P, \rho, T_l, T_v, g_l, g_v \dots\}$ containing inter-phase temperature and chemical potential differences ($g$ denotes the Gibbs free energy). In contrast, the $c_{eq}$ model evolves within the subspace $\mathcal{S}_{eq} \subset \mathcal{S}_W$, defined as $\mathcal{S}_{eq} = \left\{ \mathbf{z} \in \mathcal{S}_W \mid T_l = T_v, g_l = g_v \right\}$, where $\mathbf{z}$ denotes the state vector.
  \item \textbf{Cumulative Deviation}: As expansion penetrates deeper into the two-phase region, the non-equilibrium effect accumulates continuously. Since $c_W$ is always larger than $c_{eq}$, the non-equilibrium system accumulates a density lag. For an expansion process from pressure $P_{S,1}$ to $P_{S,3}$, this density deviation $\delta \rho$ is expressed as:
    \begin{equation}
    \delta \rho = \int_{P_{S,1}}^{P_{S,3}} \left( \frac{1}{c_W^2} - \frac{1}{c_{eq}^2} \right) \mathrm{d}P > 0,
    \end{equation}
  where for expansion $\mathrm{d}P<0$. This implies that Wood's model density $\rho_{W}$ is significantly higher than the complete equilibrium density $\rho_{eq}$.
  \item \textbf{Equilibrium Restoration}: When the fluid fully evaporates into the vapor phase, the system is forced to transition from a state of mechanical equilibrium back to complete equilibrium. The governing equations undergo a sudden jump in thermodynamic definition, where the EoS recalculates entropy based on strict equilibrium conditions. However, as a \textbf{holonomic constraint} defining the subspace $\mathcal{S}_{eq}$, the EoS dictates that the system state is uniquely determined by the current thermodynamic coordinates ($P, \rho$) and contains no memory of the non-equilibrium history path. Consequently, the density lag is "translated" into a thermodynamic entropy decrease $\Delta s$:
    \begin{equation}
    \Delta s \approx \left( \frac{\partial s}{\partial \rho} \right)_P \cdot \delta \rho < 0,
    \end{equation}
    where $(\partial s/\partial \rho)_P = {C_p} (\partial T/\partial \rho)_P / {T} < 0$ holds for any single-phase vapor.
\end{itemize}  
\par
In summary, this non-physical entropy decrease is an inevitable consequence of the system being forcibly projected from a high-dimensional non-equilibrium phase space $\mathcal{S}_W$ onto a low-dimensional equilibrium subspace $\mathcal{S}_{eq}$ when employing Wood's model to describe an isentrope crossing the vapor saturation line retrogradely. Although the practice of using Wood's speed of sound combined with the HEM framework is thermodynamically inconsistent and even leads to non-physical entropy decrease, many well-known algorithms \citep{Yi2019, Saurel2016, Yang2020} adopt this model. On one hand, this is because solving for the complete equilibrium speed of sound is relatively difficult, especially for multi-component cases \citep{Zhang2024}. On the other hand, although the $c_{eq}$ model is rigorous in its thermodynamic definition, it represents an infinite-rate mass transfer limit which may not align with physical reality in certain regimes \citep{Benjelloun2021}. For instance, in flows characterized by high velocities relative to the slow rate of inter-phase heat exchange (such as two-phase flow in supersonic nozzles), the "frozen" behavior captured by Wood's speed of sound may actually provide a superior approximation to the experimental observations.

\subsection{Exact Solution Method for the Riemann Problem based on Wood's Model}

The exact solution framework based on Wood's model is fundamentally similar to the method for the standard model proposed in this paper, yet simpler. Since the speed of sound is continuous across the entire $P-\rho$ domain under Wood's model, non-classical waves are absent. Consequently, the compression branch is a simple shock wave, solved using the exact same method as in Section \ref{sec:phiR- S}, while the expansion branch is a simple rarefaction wave. It is worth noting that when using Wood's model to solve for Riemann invariants and rarefaction waves, the isentropic path is no longer governed by Equation \ref{equ:F_s}; instead, it must be integrated using $c_W^2$ as the slope of the isentrope. The Riemann invariant is:

\begin{equation}
    {{J}^{\pm}_W}=u\pm\int_{{{C_W}^{\pm}}}{\frac{1}{\rho c}\mathrm{d}P}=\text{const}.
    \label{equ:JW}
\end{equation}
On the characteristic line $C_W$, the $P-\rho$ relation satisfies the isentropic condition of mechanical mixture entropy:

\begin{equation}
    \begin{aligned}
    &P = P_W(\rho), \\
    \frac{\mathrm{d}P_W}{\mathrm{d}\rho} & =\pderiv{P}{\rho}{s_p,Y_p} = c_W^2.
\end{aligned}
    \label{equ:P_W}
\end{equation}
\par
We reiterate that employing Wood's model within the HEM framework is thermodynamically inconsistent, because we rely on the HEM model to make the fluid thermodynamic potential dependent on two state variables, thereby defining mixture density, mixture energy, and mixture entropy to achieve the closure of the thermodynamic system. However, Wood's model implies that the fluid in the two-phase region possesses two distinct temperatures at the same $P-\rho$ state point. Therefore, we must completely abandon the concept of equilibrium entropy (Equation \ref{equ:mix_s}) during the integration process and directly integrate $c_W^2$ to obtain the "frozen" isentrope $P_W(\rho)$. Here, we introduce a logarithmic integration method to enhance numerical stability.\par
Introducing the logarithmic variable $L = \ln \rho$, we have:

\begin{equation}
     \frac{\mathrm{d}L}{\mathrm{d}P} =\frac{\mathrm{d}\ln \rho}{\mathrm{d}P} = \frac{1}{\rho c_W^2(\rho, P)}.
\label{equ:dLdP}
\end{equation}
Following the numerical integration of Equation \ref{equ:dLdP}, the isentropic density at a given pressure is recovered via the transformation $\rho = e^L$. The numerical integration can be performed using standard 3rd or 4th order Runge-Kutta methods, the details of which are omitted here for brevity. In addition, solving for the internal parameters of the rarefaction wave requires the calculation of the derivatives $({\partial c_W}/{\partial \rho})_{P}$ and $({\partial c_W}/{\partial P})_{\rho}$. These derivatives are constituent components of the derivatives of $c_{eq}$; see Appendix \ref{sec:dc_w} for details.

\subsection{Comparison of Exact Solutions of the Flash evaporation Riemann Problem Based on Wood's Model and Standard Model}

We now employ the exact solution of the FeRP to design a numerical experiment: we simulate the n-dodecane FeRP employing Wood's model ($c_W$) and the standard model ($c_{eq}$) model to assess the influence of the two thermodynamic models on the resulting wave structures and solution profiles. The calculation conditions represent a practical scenario where high-pressure liquid n-dodecane is injected into low-pressure gaseous n-dodecane in an engine. The computational domain length is $1\,\mathrm{m}$, and the time is set to $0.8\,\mathrm{ms}$. For all cases, the right state (vapor phase) is fixed at $P_R=1\times 10^5 \mathrm{Pa}$, $\rho_R = 2 \,\mathrm{kg \,m^{-3}}$, and $u_R = 0 \,\mathrm{m \, s^{-1}}$. The varying left initial states and the corresponding expansion branch types are listed in Table \ref{tab:initial_values}:

\begin{table}
    \centering
    \caption{Initial Values for the Riemann Problem Calculation}
    \label{tab:initial_values}
    \begin{tabular}{cccccc}
        \toprule
       Case & Branch Type& $P_L$ (Pa) & $\rho_L$ ($\mathrm{kg \,m^{-3}}$) & $T_L$ (K) & $u_L$ ($\mathrm{m \, s^{-1}}$) \\
        \midrule
        1& RSR wave   & $2 \times 10^6$ & 300 & 652.4& 120 \\
        2& RS wave  & $2 \times 10^6$ & 400 & 617.9 & 80  \\
        3& R wave & $2 \times 10^6$ & 500 & 552.3& 80 \\
        \bottomrule
    \end{tabular}
\end{table}

\begin{figure}[htbp]
    \centering

    \begin{subfigure}[b]{0.32\textwidth}
        \centering
        \includegraphics[width=\linewidth]{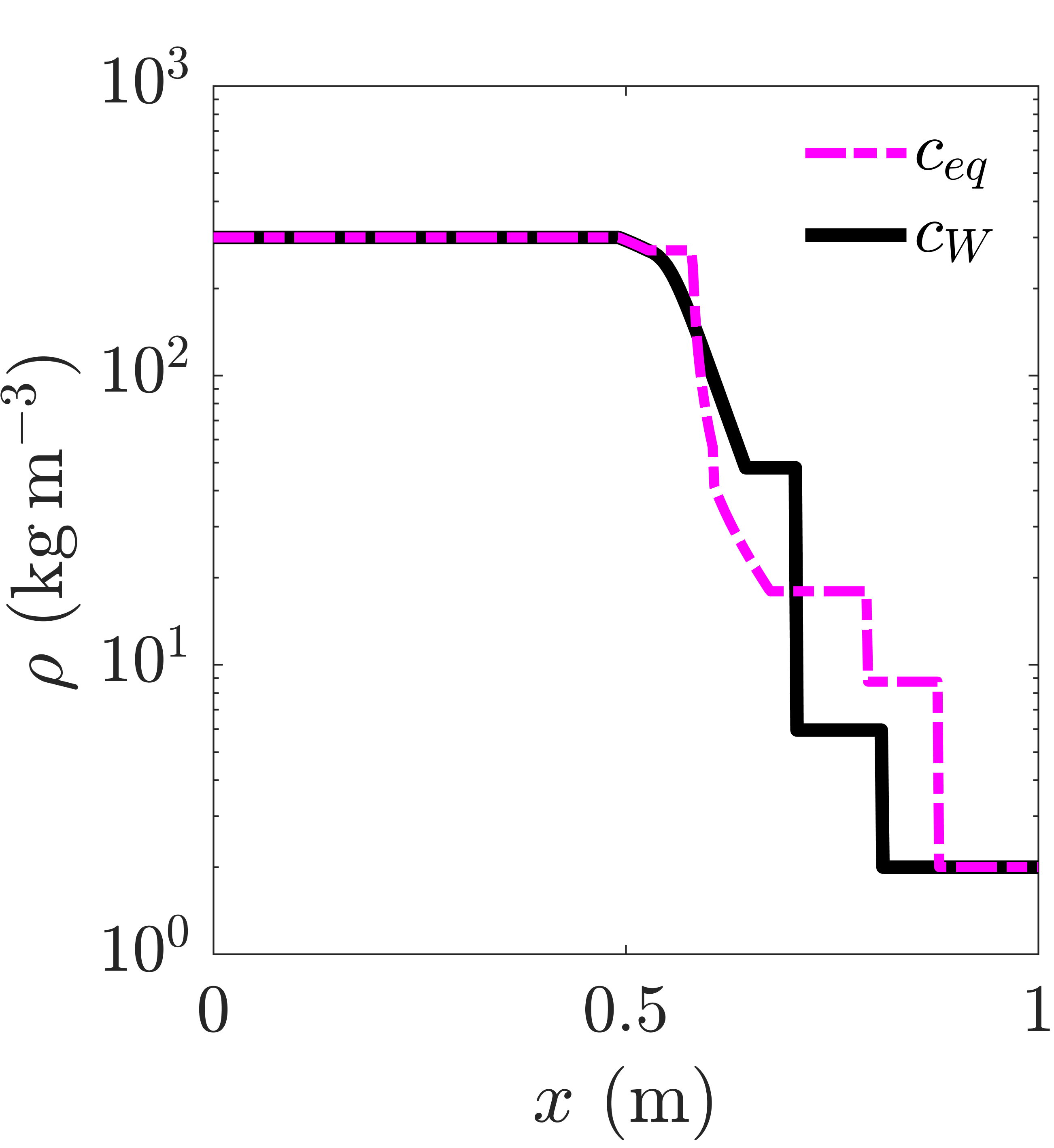}
        \caption{Density} 
        \label{fig9a}
    \end{subfigure}
    \hfill 
    \begin{subfigure}[b]{0.32\textwidth}
        \centering
        \includegraphics[width=\linewidth]{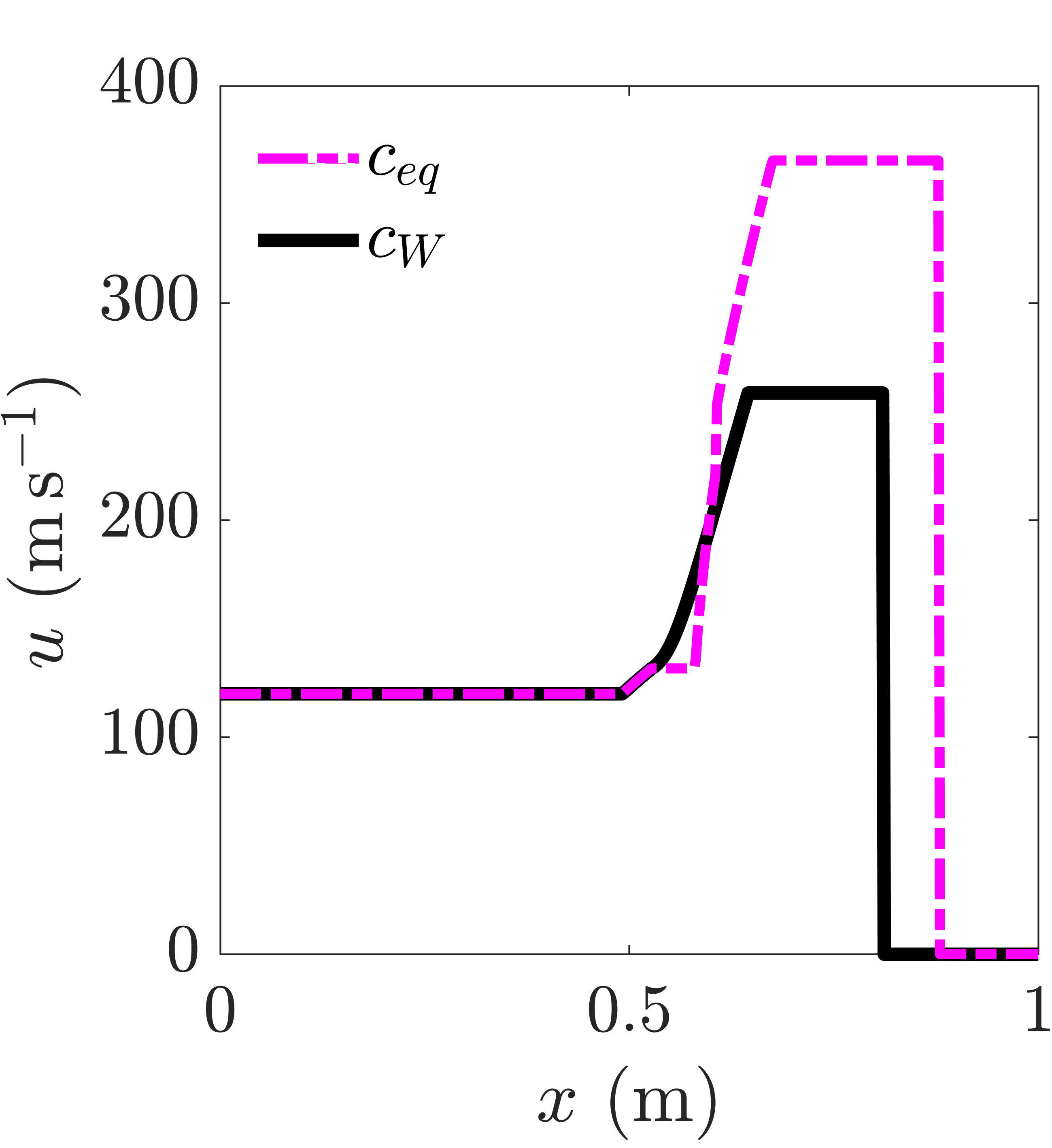}
        \caption{Velocity}
        \label{fig9b}
    \end{subfigure}
    \hfill
    \begin{subfigure}[b]{0.32\textwidth}
        \centering
        \includegraphics[width=\linewidth]{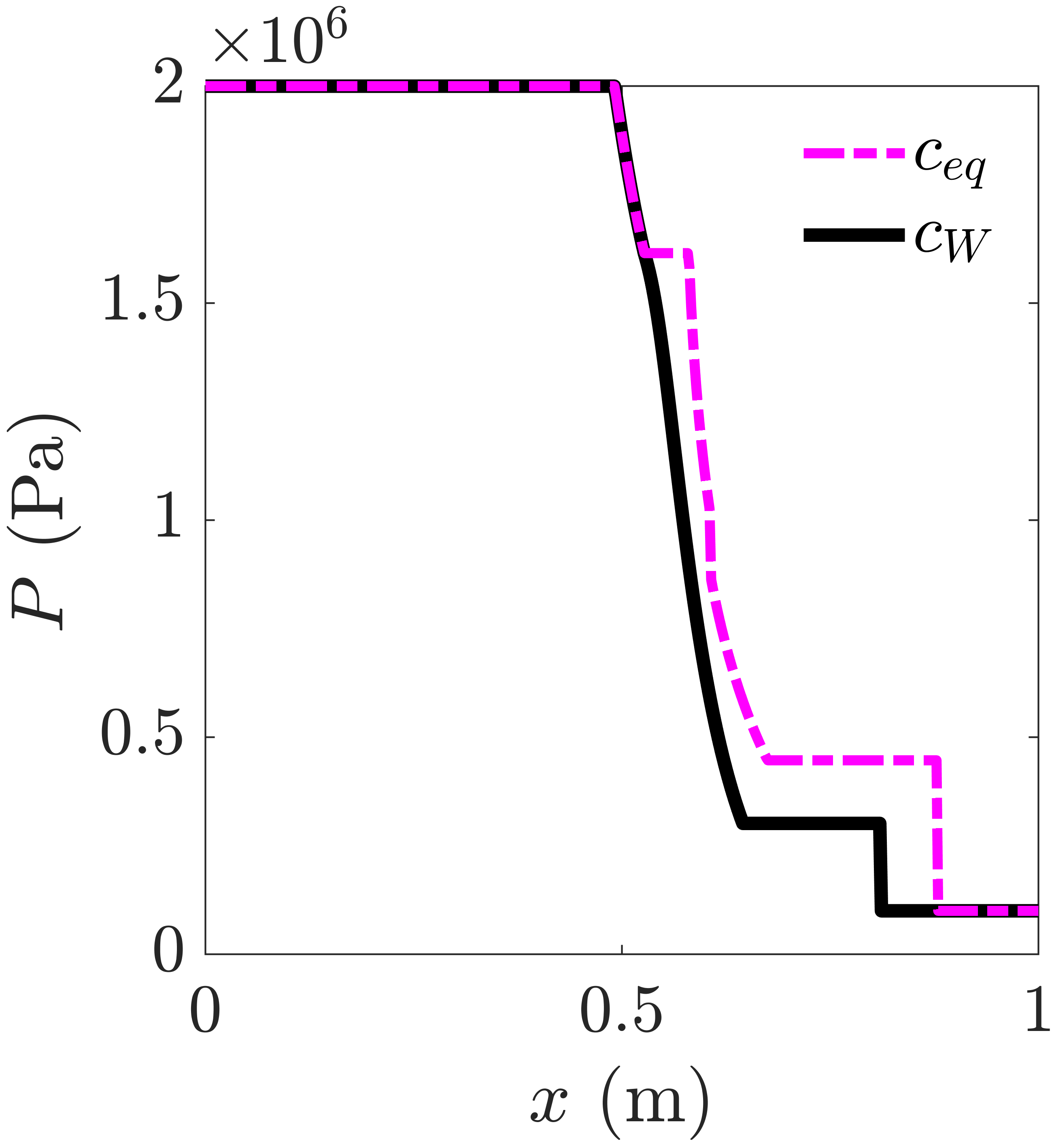}
        \caption{Pressure}
        \label{fig9c}
    \end{subfigure}
    
    \vspace{0.5em}
    
    \begin{subfigure}[b]{0.32\textwidth}
        \centering
        \includegraphics[width=\linewidth]{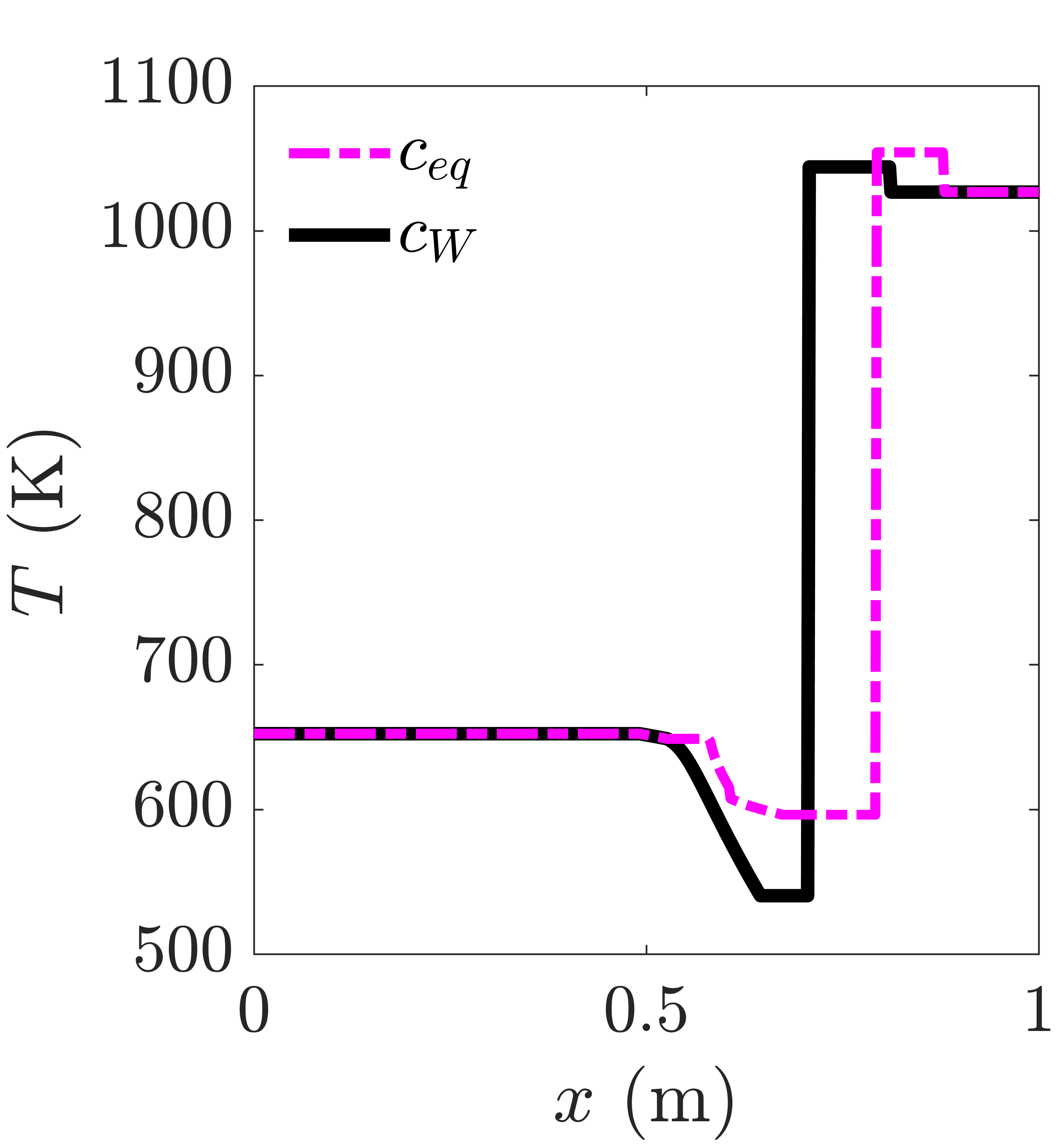}
        \caption{Temperature}
        \label{fig9d}
    \end{subfigure}
    \hfill
    \begin{subfigure}[b]{0.32\textwidth}
        \centering
        \includegraphics[width=\linewidth]{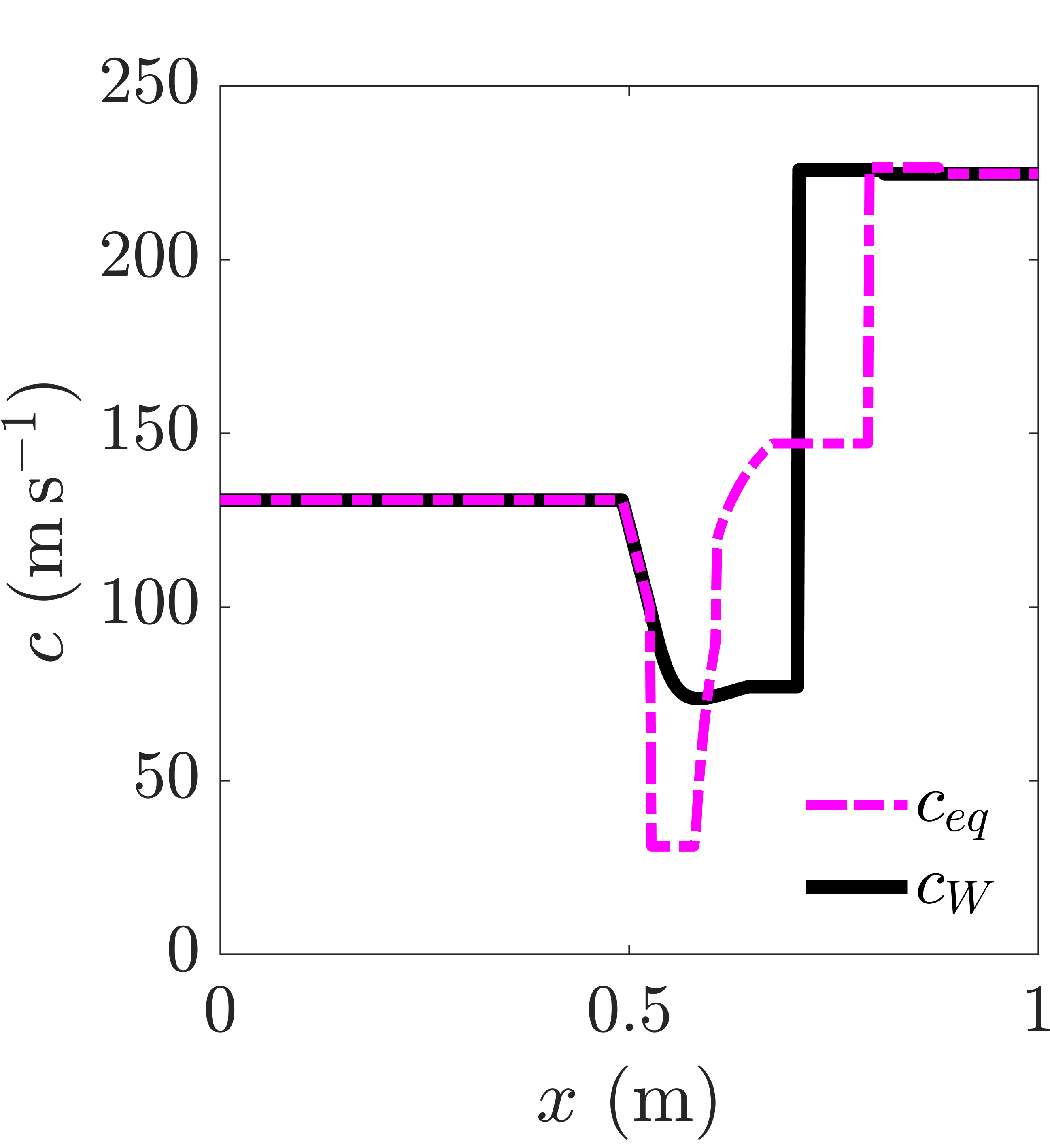}
        \caption{Speed of sound}
        \label{fig9e}
    \end{subfigure}
    \hfill
    \begin{subfigure}[b]{0.32\textwidth}
        \centering
        \includegraphics[width=\linewidth]{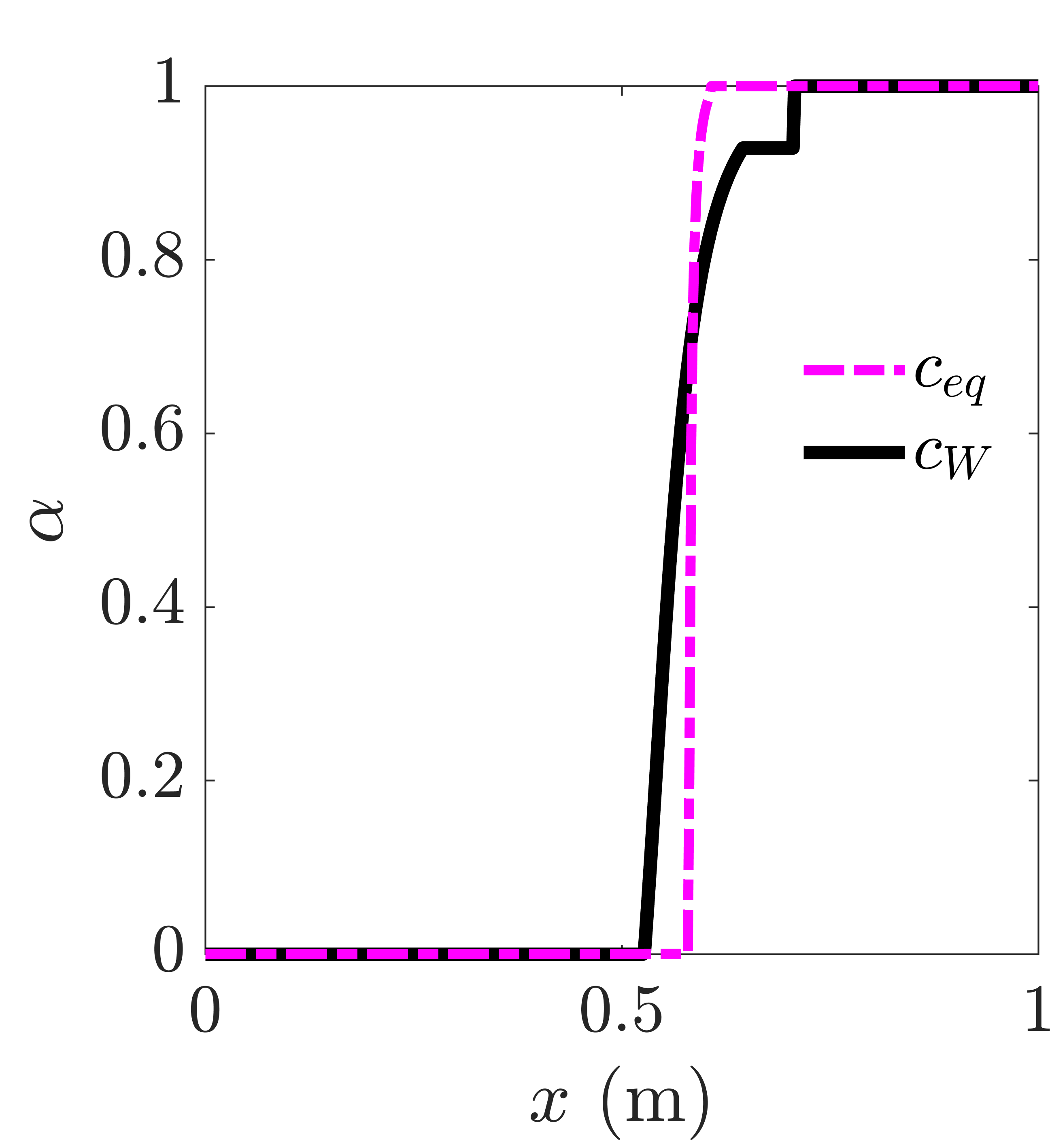}
        \caption{Vapor fraction}
        \label{fig9f}
    \end{subfigure}
    
    \vspace{0.5em} 
    
    \begin{subfigure}[b]{0.6\textwidth}
        \centering
        \includegraphics[width=\linewidth]{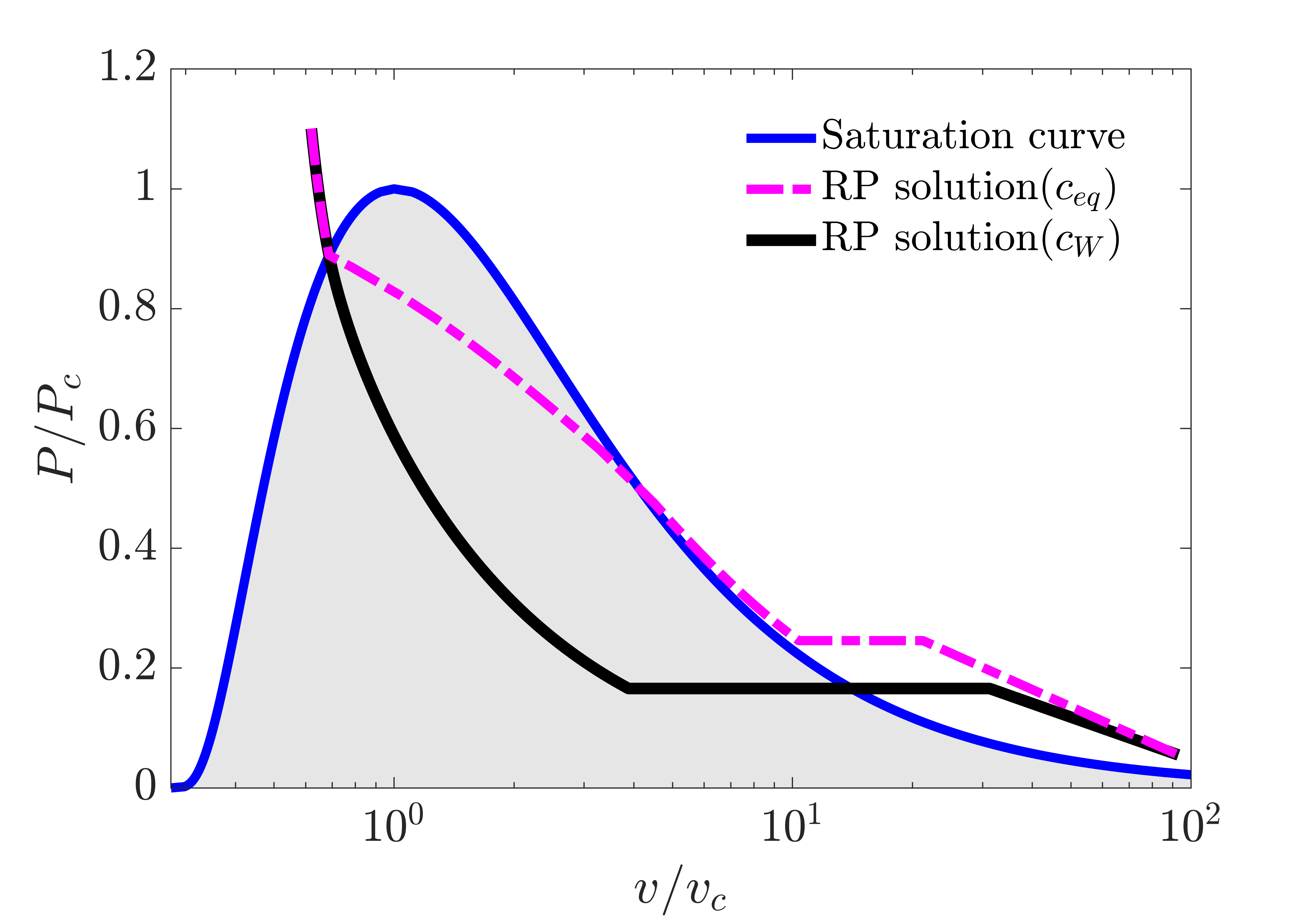}
        \caption{RP solution curve on $P-v$ coordinates}
        \label{fig9g}
    \end{subfigure}
    
    \caption{Exact solutions of the RP for Case 1, obtained with the $c_{eq}$ and $c_W$ models, respectively. For the $c_{eq}$ model, the expansion branch manifests as an RSR wave.}
      \label{fig9}
\end{figure}

Figure \ref{fig9} presents the simulation results for Case 1, which corresponds to the injection of n-dodecane into a low-pressure gas phase region at a high initial temperature. For the standard model ($c_{eq}$), the n-dodecane expands from the liquid phase, traverses the entire two-phase region, and completely transitions into the vapor phase. Consequently, the fluids on both sides of the contact discontinuity are in the vapor phase, as depicted in Figure \ref{fig9g}. In this scenario, the expansion branch evolves into a non-classical RSR wave, featuring an expansion shock sandwiched between two rarefaction waves. Conversely, for Wood's model, the rarefaction wave neither splits nor exhibits non-classical structures, which is consistent with our previous analysis regarding the continuity of the speed of sound. Furthermore, the intermediate pressure and velocity calculated by Wood's model are lower, and the extent of vaporization is significantly smaller.\par

\begin{figure}[htbp]
    \centering

    \begin{subfigure}[b]{0.32\textwidth}
        \centering
        \includegraphics[width=\linewidth]{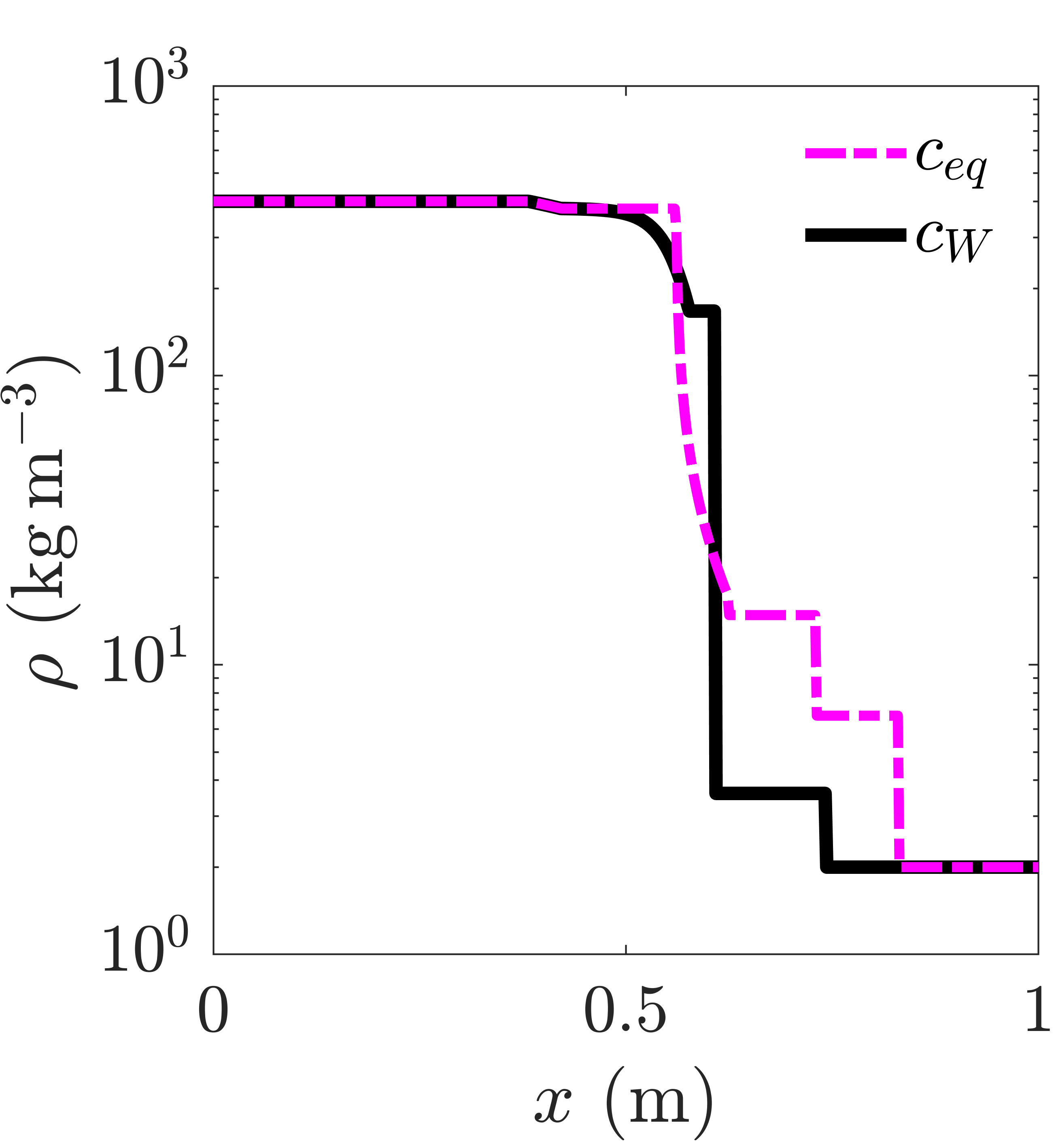}
        \caption{Density} 
        \label{fig10a}
    \end{subfigure}
    \hfill 
    \begin{subfigure}[b]{0.32\textwidth}
        \centering
        \includegraphics[width=\linewidth]{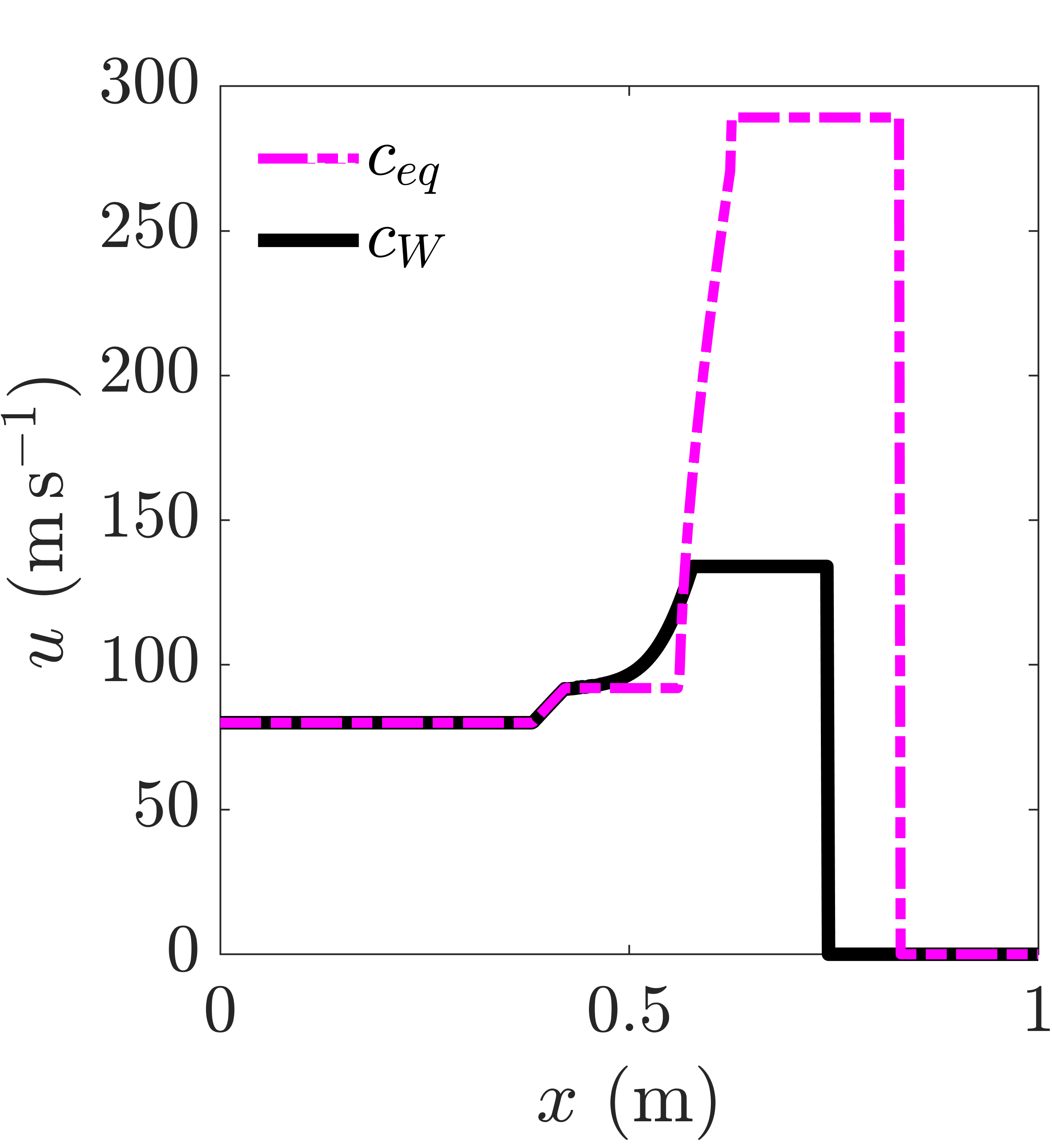}
        \caption{Velocity}
        \label{fig10b}
    \end{subfigure}
    \hfill
    \begin{subfigure}[b]{0.32\textwidth}
        \centering
        \includegraphics[width=\linewidth]{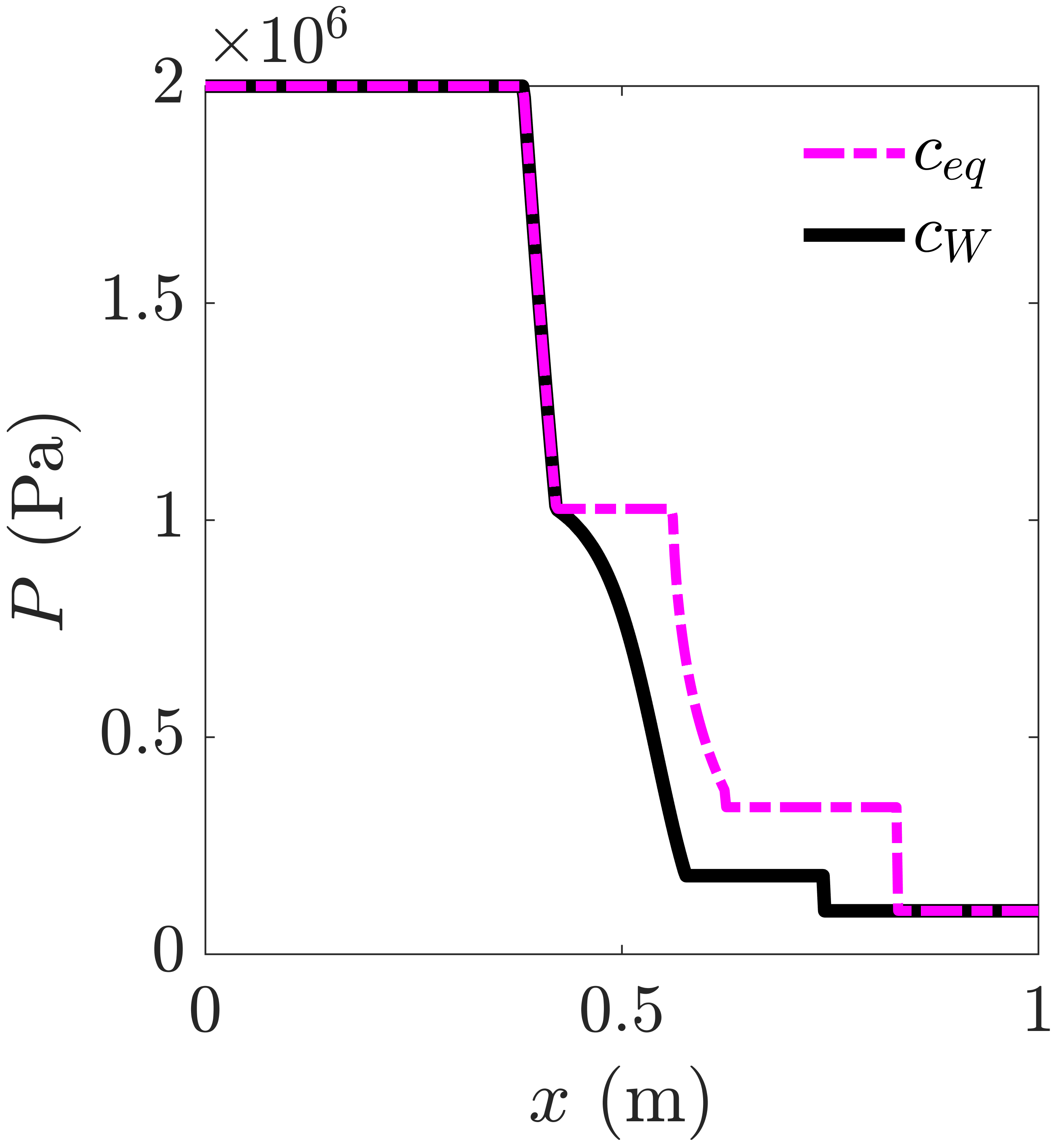}
        \caption{Pressure}
        \label{fig10c}
    \end{subfigure}
    
    \vspace{0.5em} 
    
    \begin{subfigure}[b]{0.32\textwidth}
        \centering
        \includegraphics[width=\linewidth]{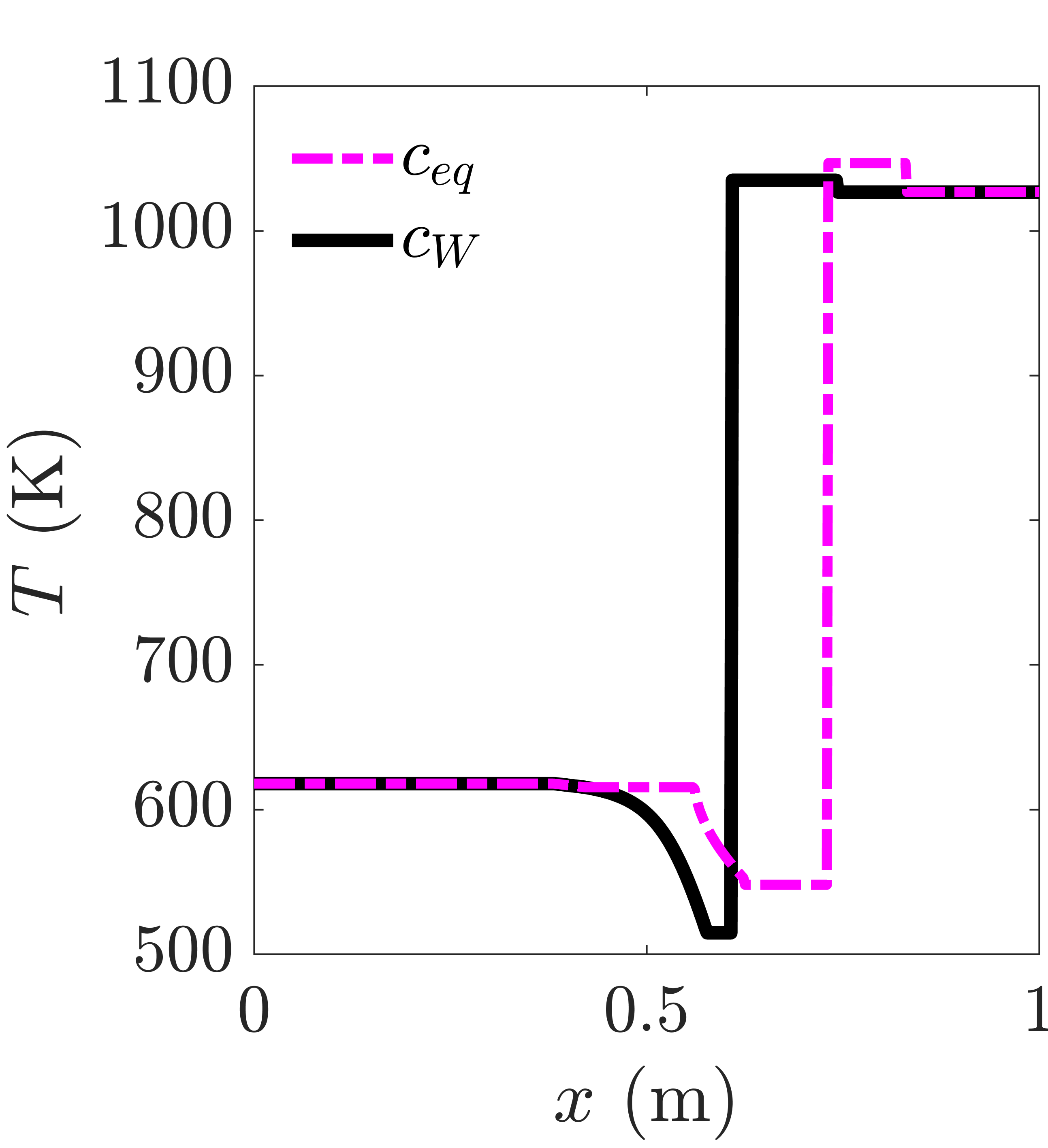}
        \caption{Temperature}
        \label{fig10d}
    \end{subfigure}
    \hfill
    \begin{subfigure}[b]{0.32\textwidth}
        \centering
        \includegraphics[width=\linewidth]{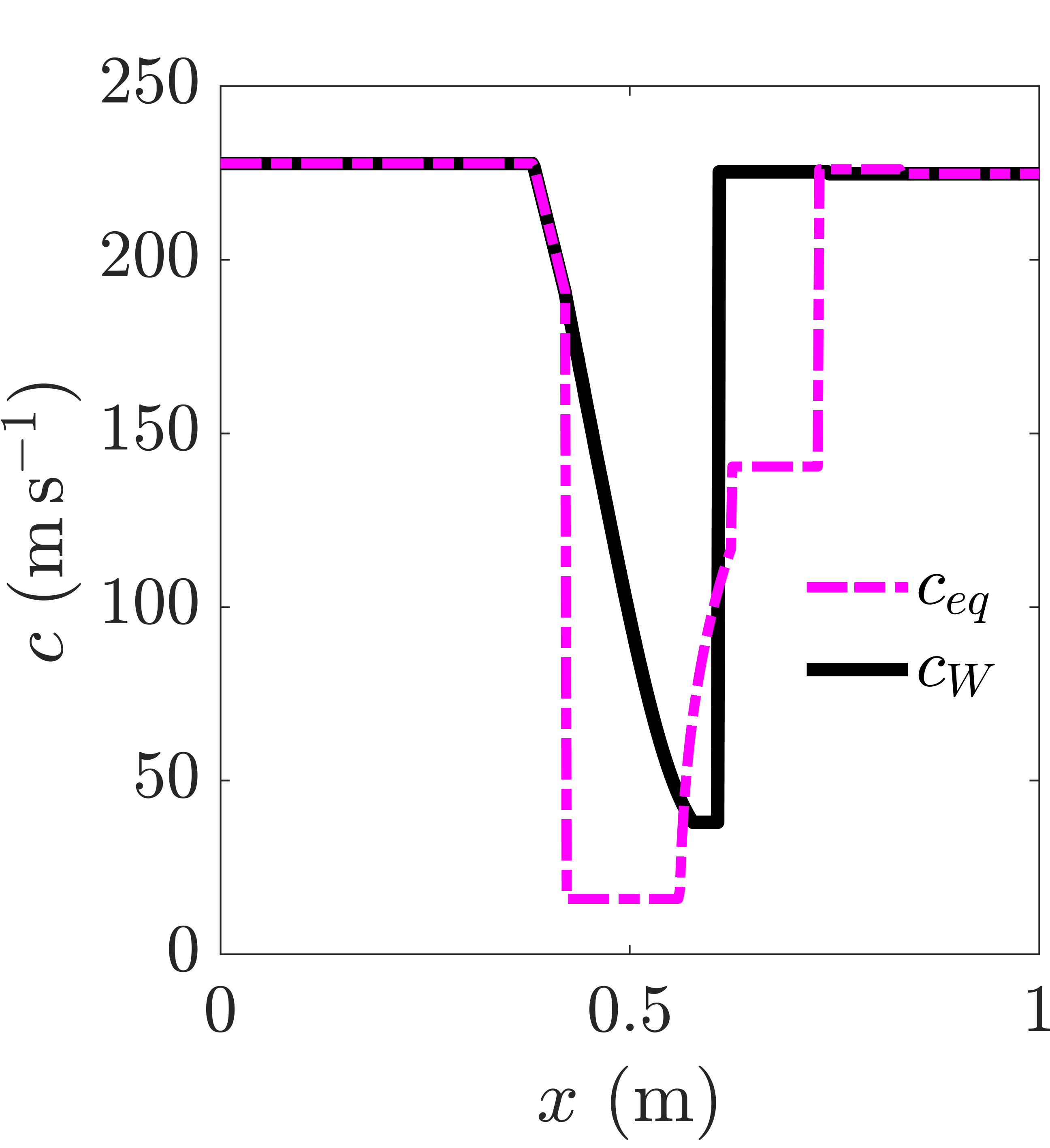}
        \caption{Speed of sound}
        \label{fig10e}
    \end{subfigure}
    \hfill
    \begin{subfigure}[b]{0.32\textwidth}
        \centering
        \includegraphics[width=\linewidth]{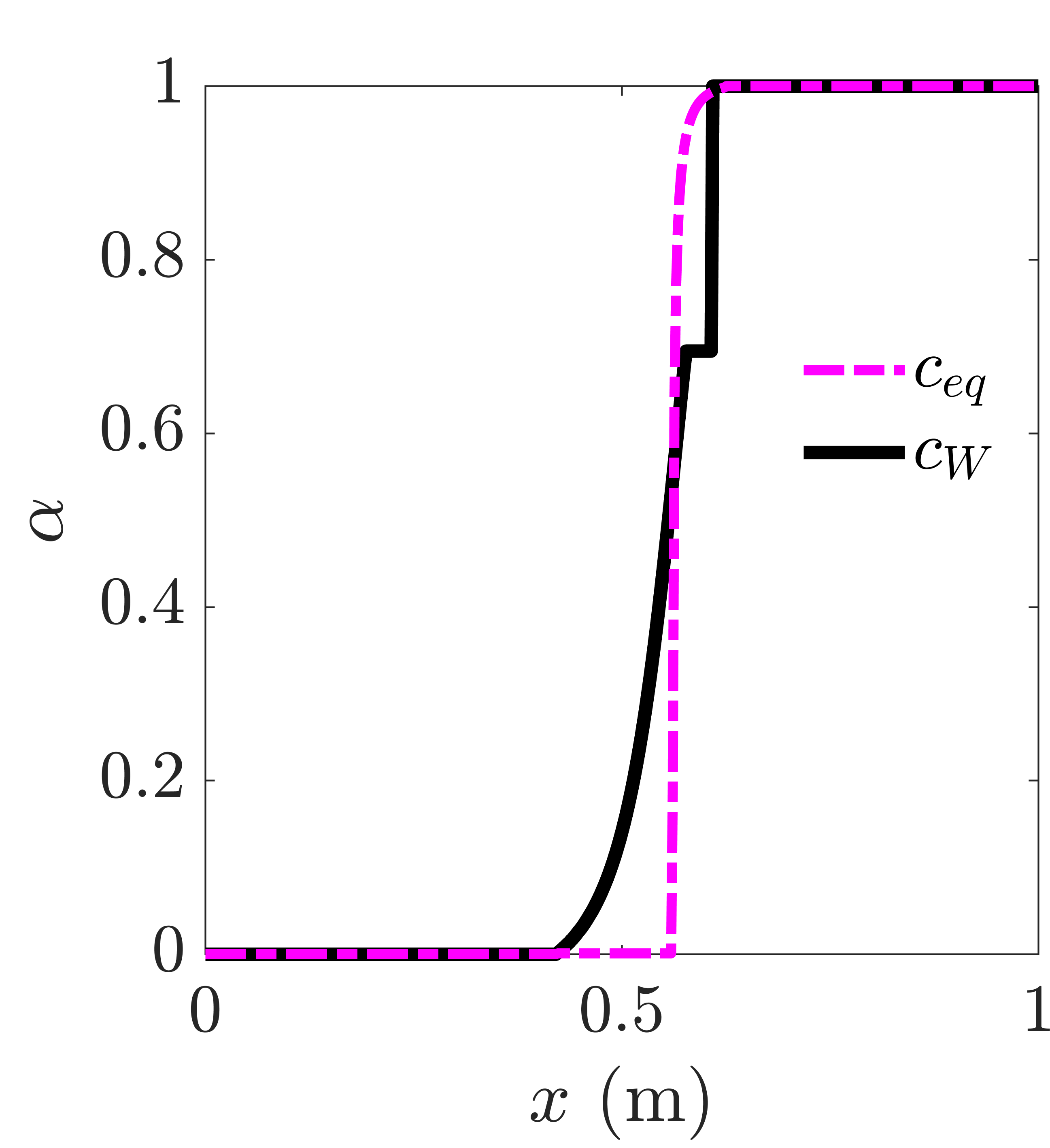}
        \caption{Vapor fraction}
        \label{fig10f}
    \end{subfigure}
    
    \vspace{0.5em} 
    
    \begin{subfigure}[b]{0.6\textwidth}
        \centering
        \includegraphics[width=\linewidth]{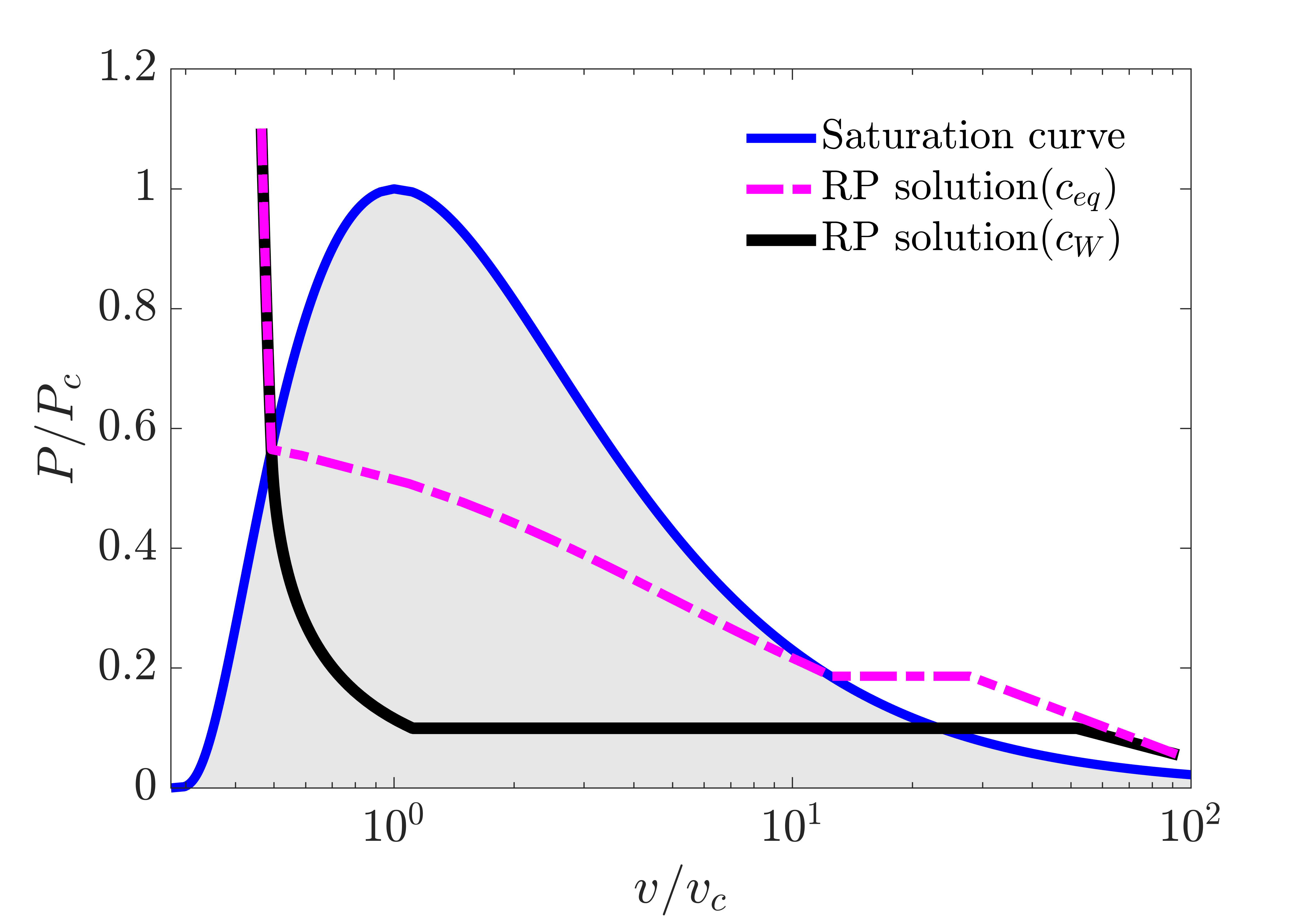}
        \caption{RP solution curve on $P-v$ coordinates}
        \label{fig10g}
    \end{subfigure}
    
    \caption{Exact solutions of the RP for Case 2, obtained with the $c_{eq}$ and $c_W$ models, respectively. For the $c_{eq}$ model, the expansion branch manifests as an RS wave.}
    \label{fig10}
\end{figure}

\begin{figure}[htbp]
    \centering

    \begin{subfigure}[b]{0.32\textwidth}
        \centering
        \includegraphics[width=\linewidth]{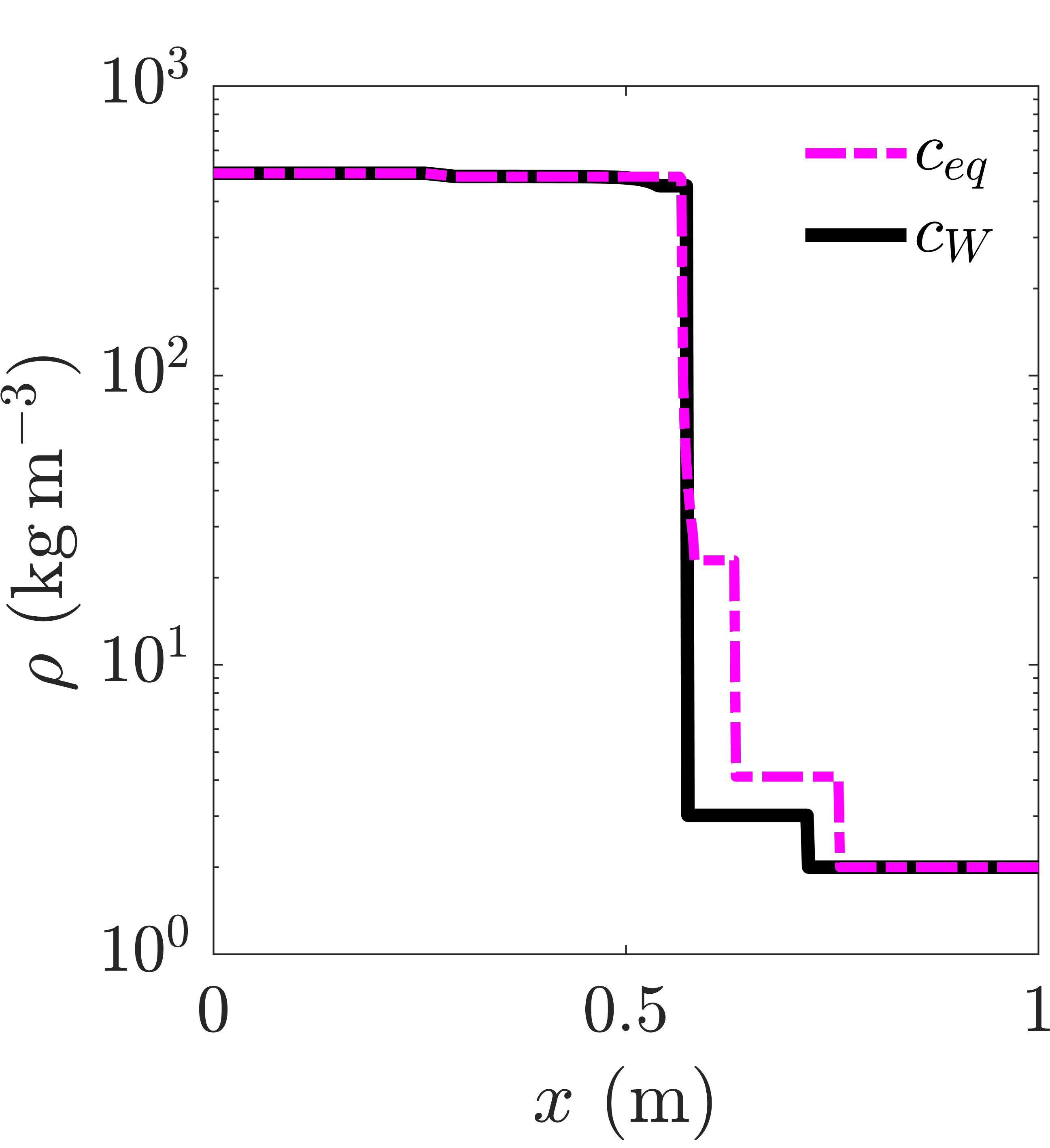}
        \caption{Density} 
        \label{fig11a}
    \end{subfigure}
    \hfill 
    \begin{subfigure}[b]{0.32\textwidth}
        \centering
        \includegraphics[width=\linewidth]{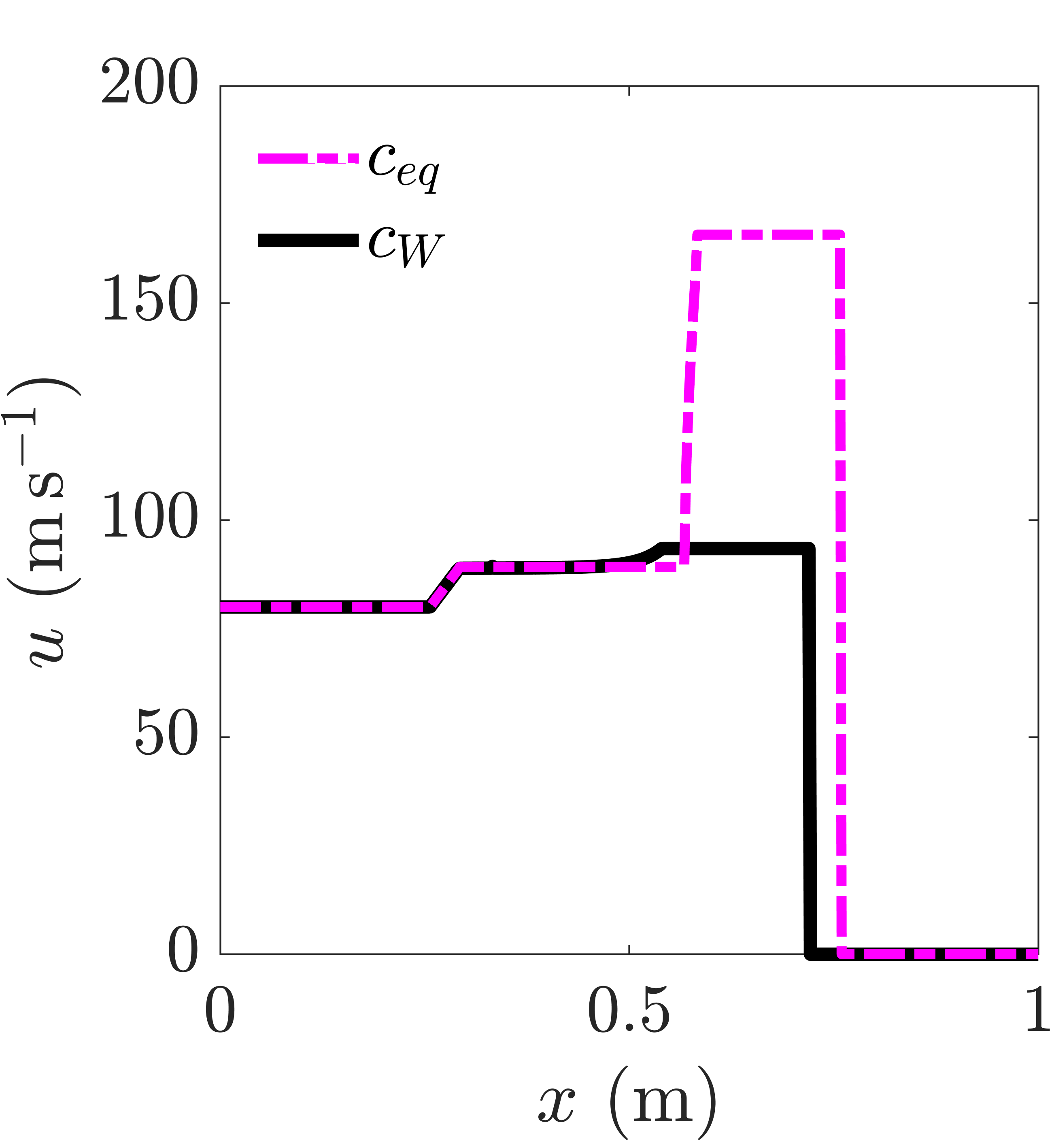}
        \caption{Velocity}
        \label{fig11b}
    \end{subfigure}
    \hfill
    \begin{subfigure}[b]{0.32\textwidth}
        \centering
        \includegraphics[width=\linewidth]{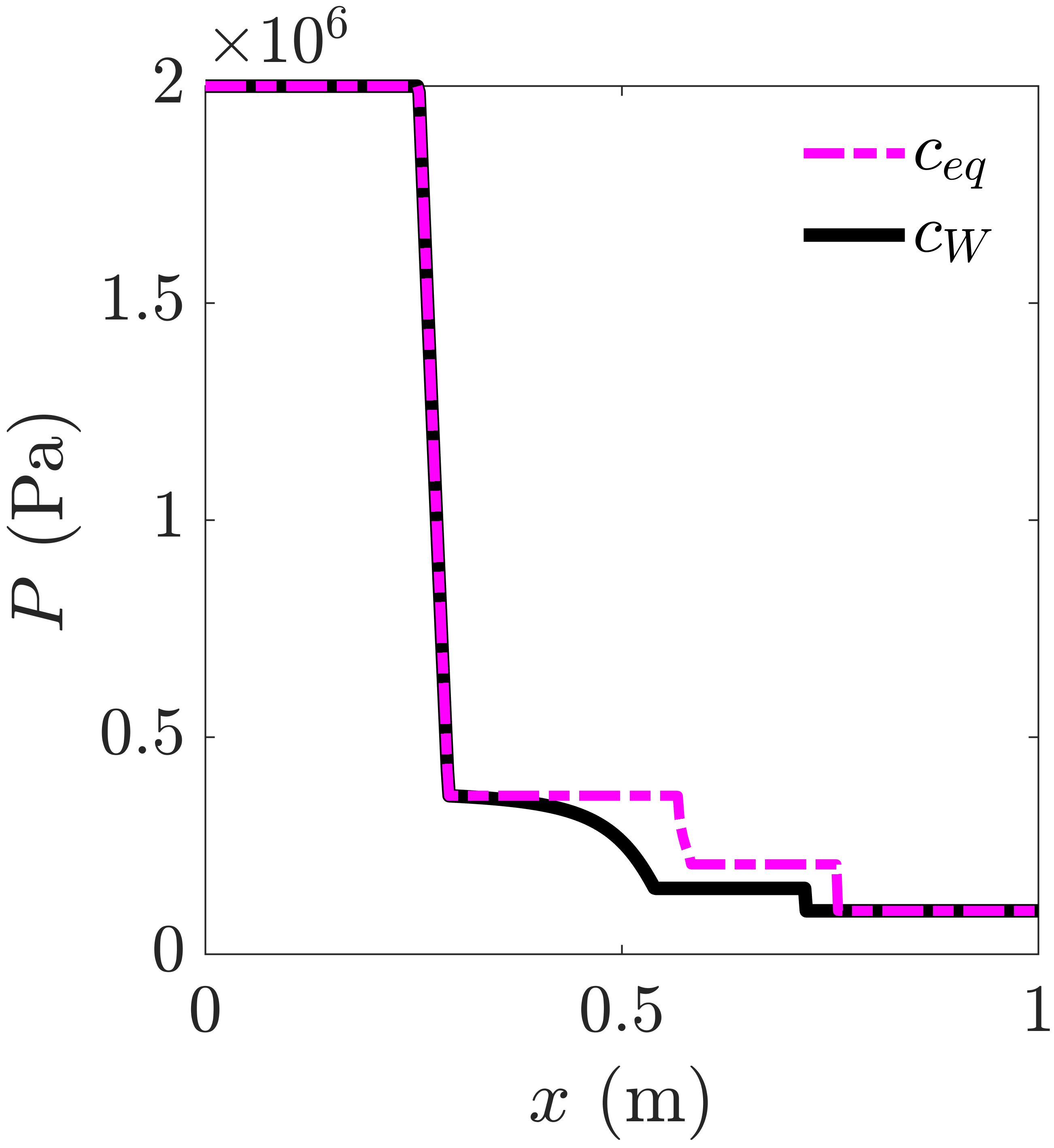}
        \caption{Pressure}
        \label{fig11c}
    \end{subfigure}
    
    \vspace{0.5em} 
    
    \begin{subfigure}[b]{0.32\textwidth}
        \centering
        \includegraphics[width=\linewidth]{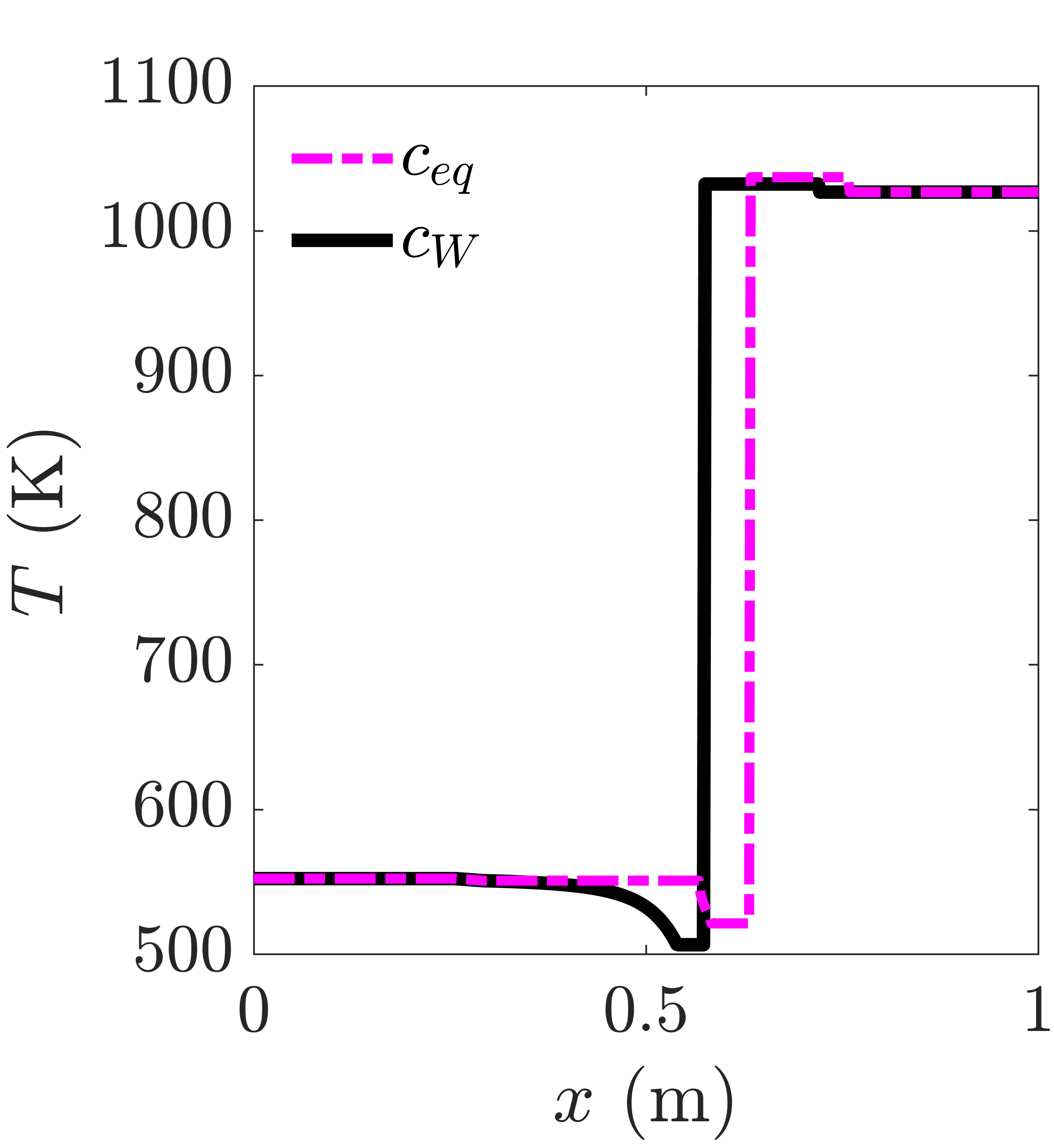}
        \caption{Temperature}
        \label{fig11d}
    \end{subfigure}
    \hfill
    \begin{subfigure}[b]{0.32\textwidth}
        \centering
        \includegraphics[width=\linewidth]{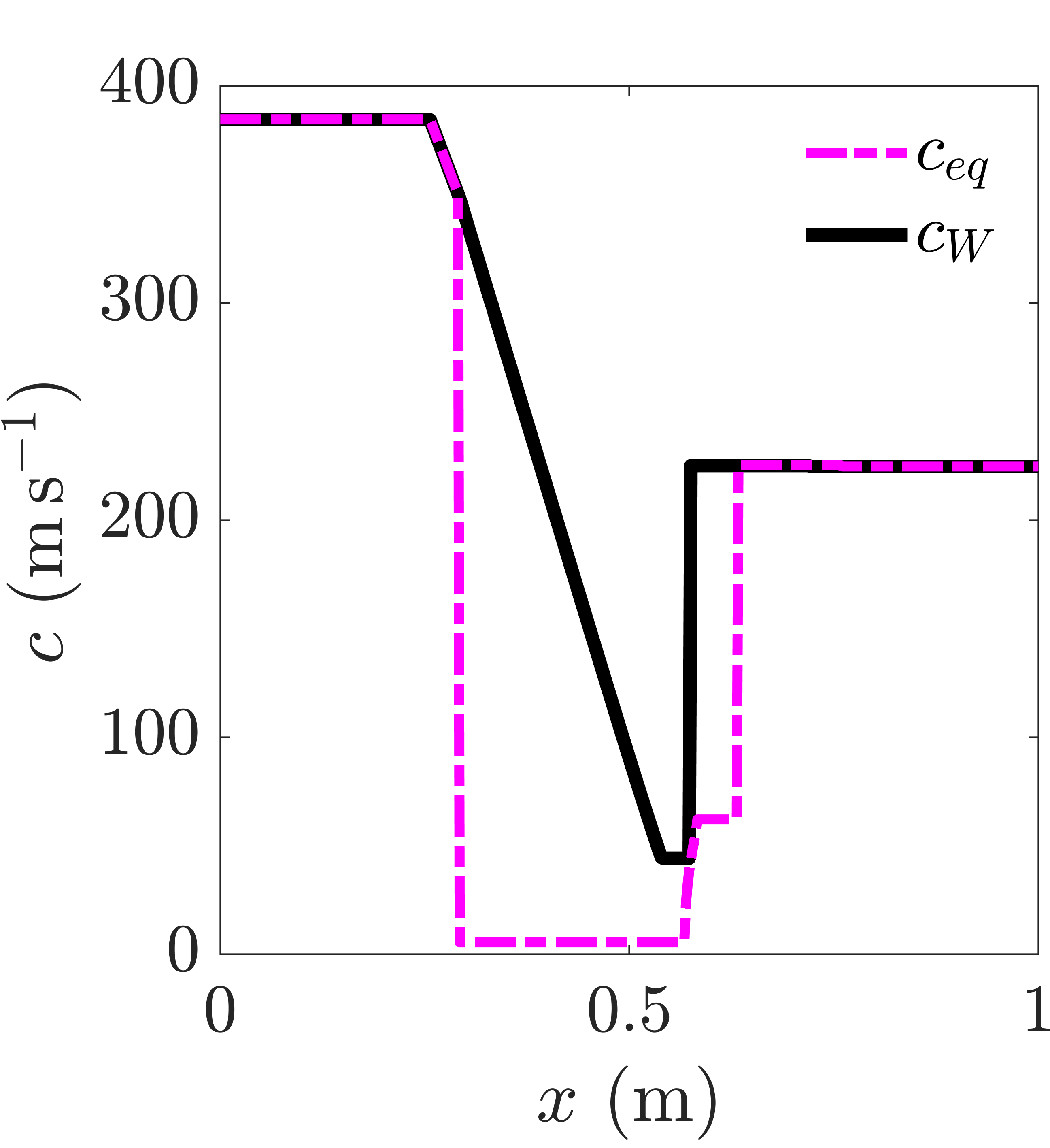}
        \caption{Speed of sound}
        \label{fig11e}
    \end{subfigure}
    \hfill
    \begin{subfigure}[b]{0.32\textwidth}
        \centering
        \includegraphics[width=\linewidth]{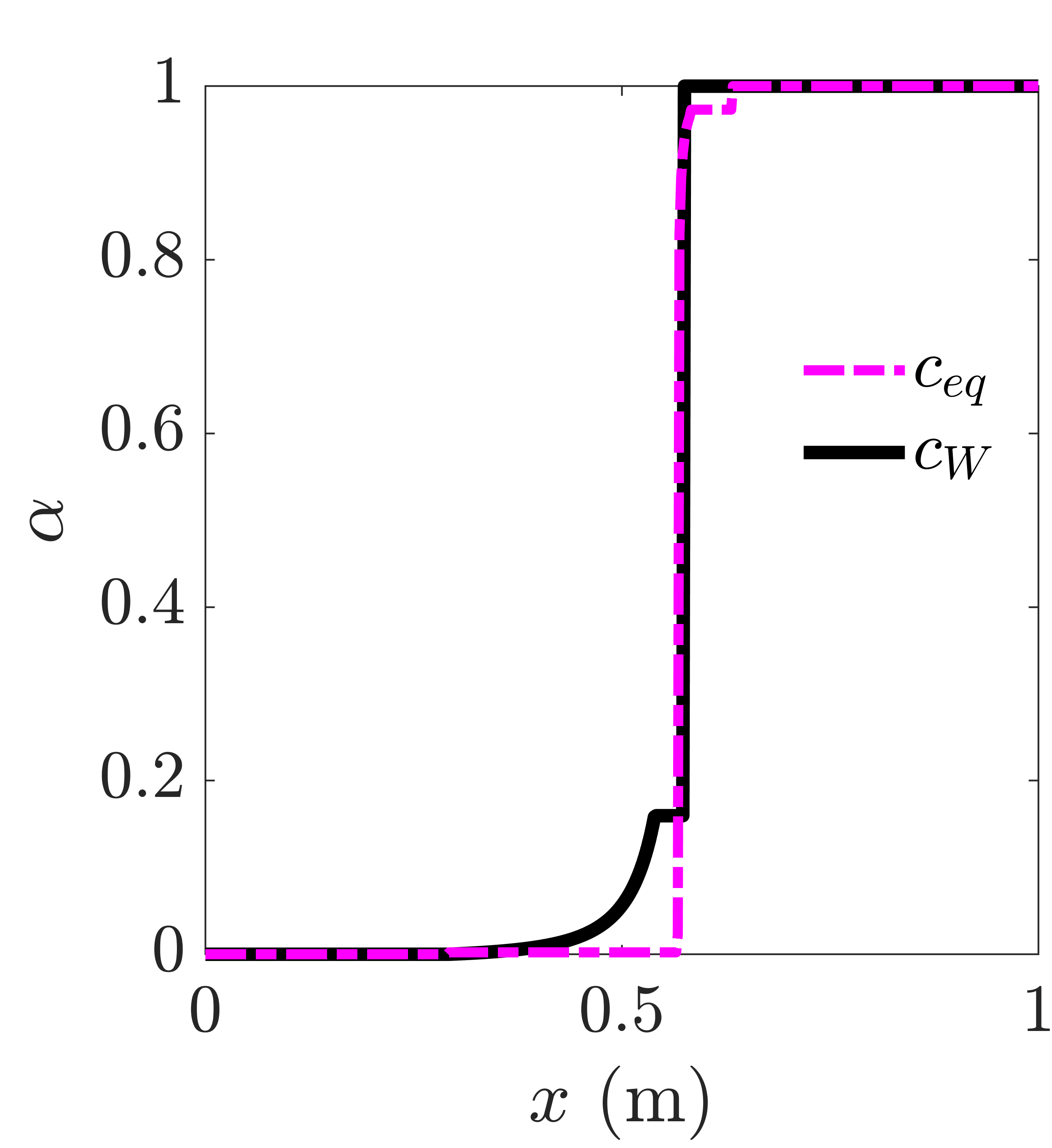}
        \caption{Vapor fraction}
        \label{fig11f}
    \end{subfigure}
    
    \vspace{0.5em} 
    
    \begin{subfigure}[b]{0.6\textwidth}
        \centering
        \includegraphics[width=\linewidth]{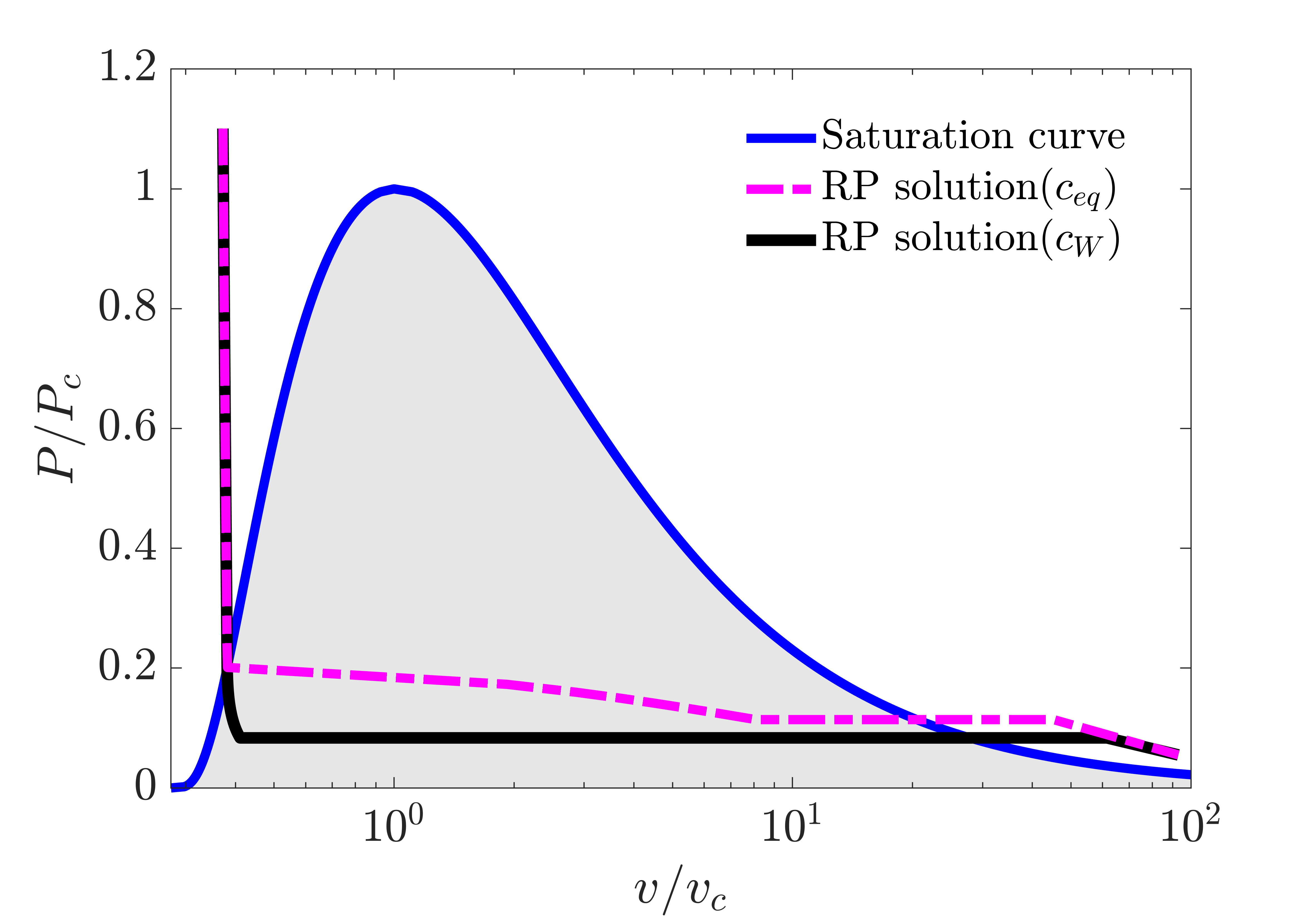}
        \caption{RP solution curve on $P-v$ coordinates}
        \label{fig11g}
    \end{subfigure}
    
    \caption{Exact solutions of the RP for Case 3, obtained with the $c_{eq}$ and $c_W$ models, respectively. For the $c_{eq}$ model, the expansion branch manifests as an R wave.}
    \label{fig11}
\end{figure}

Figure \ref{fig10} shows the results for Case 2, corresponding to n-dodecane injection at an intermediate temperature. In the standard model, the fluid expands from the liquid phase across the two-phase region into the vapor phase (Figure \ref{fig10g}). The expansion branch manifests as a non-classical RS wave, where the rarefaction wave is terminated by an expansion shock. This feature is most distinct in the velocity profile shown in Figure \ref{fig10b}. Figure \ref{fig11} displays the results for Case 3, representing the injection at a lower temperature. In the standard model, the n-dodecane expands from the liquid phase into the two-phase region, resulting in a contact discontinuity separating a two-phase mixture from the vapor phase (Figure \ref{fig11g}). In this case, the expansion branch manifests as a split classical rarefaction wave. Meanwhile, the rarefaction wave calculated by Wood's model exhibits a characteristic "pseudo-splitting" behavior, where a plateau with a minimal slope emerges in the pressure and density curves immediately after entering the two-phase region.

\clearpage

\clearpage

\clearpage

Compared with the standard model, the calculation results from Wood's model exhibit large deviations across all three conditions. Significant disparities are evident not only in numerical variables such as intermediate pressure and velocity but also in physical phenomena, including shock strength, the extent of vaporization, and the wave structure. Specifically, the isentropic path in Wood's model is characterized by a "density lag" effect(as previously discussed). This effect manifests as suppressed vaporization, underestimated intermediate pressure and velocity, and attenuated shock strength within the compression branch. Consequently, for CFD simulations of scramjet fuel sprays, the application of Wood's model is expected to yield a lower vaporization extent, reduced local pressure, and a diminished fuel dispersion region, thereby potentially affecting downstream mixing and combustion.

We deliberately characterize these discrepancies as "deviations" rather than "errors." This distinction is made because, based on our previous analysis, we do not regard Wood's speed of sound merely as a numerical approximation of the complete equilibrium speed of sound. Rather, Wood's model introduces a distinct (although thermodynamically less rigorous) entropy mixing model. Consequently, one cannot simply evaluate the superiority of either model based solely on the calculated results, as the validity of the standard model itself is strictly contingent upon the assumption of infinite inter-phase heat and mass transfer rates. The crucial insight offered by these findings is that the selection of the appropriate model should be guided by specific experimental validation and a profound physical assessment of relaxation timescales. As also concluded by \citet{Benjelloun2021}, it is essential to select the sound speed model based on the actual physical phenomena. However, in the absence of experimental data or clear physical mechanisms, we recommend the thermodynamically more consistent standard model.

\section{Conclusion}
\label{sec:conclusion}

This paper formalizes the "Flash evaporation Riemann Problem" (FeRP) within the framework of the Homogeneous Equilibrium Model and Vapor-Liquid Equilibrium assumptions. An exact FeRP solution framework based on the Newton iteration method has been established, in which all thermodynamic derivatives are rigorously resolved. The main conclusions are drawn as follows:

A non-classical gas dynamic behavior in the FeRP has been elucidated. A positive singularity of the Landau derivative $\Gamma$ arises on the liquid saturation line, inducing rarefaction wave splitting. On the vapor saturation line, the sign of the singularity is determined by the direction of isentropic traversal. Only for high-molecular-weight and high-heat-capacity fluids like $n$-dodecane can isentropes traverse the vapor saturation line retrogradely into the vapor phase, thereby inducing a negative singularity and generating two types of non-classical composite waves: the rarefaction-upstream sonic expansion shock (RS) and the rarefaction-double sonic expansion shock-rarefaction (RSR). Based on the unique thermodynamic properties of these non-classical waves, a Newton iteration method is established with the Chapman-Jouguet condition as the outer constraint, alongside the Rankine-Hugoniot and isentropic relations as inner constraints. This method enables the stable calculation of expansion shocks, avoiding convergence to trivial solutions.

While the formulation of FeRP constructs a theoretical bridge between the physical phenomenon of flash evaporation and the mathematical concept of the Riemann problem, the integrity of this bridge relies heavily on the assumption of thermal equilibrium. To assess the consequences of deviating from this ideal assumption, an exact solution for the FeRP incorporating Wood's mechanical equilibrium speed of sound is further established. Results demonstrate that Wood's speed of sound is not a simple approximation of the complete equilibrium speed of sound, but exhibits large deviations. Specifying Wood’s speed of sound as the characteristic propagation speed in the Euler equations implicitly defines a mechanical mixture entropy. This mixture entropy is inconsistent with the HEM framework, as it introduces dual temperature states within the two-phase region and an isentropic path characterized by a 'density lag' effect and non-physical entropy decrease. This effect leads to underestimated values for intermediate pressure, velocity, and vaporization extent. In scramjet fuel spray CFD simulations, Wood's model is expected to yield a diminished fuel dispersion region, reduced vaporization, and lowered local pressure relative to the complete equilibrium model, thereby potentially affecting downstream mixing and combustion.

Although Wood's model lacks rigor in thermodynamic closure, considering the complexity of non-equilibrium effects in practical scenarios, we maintain that the choice of speed of sound model for the FeRP should be guided by specific experimental data or physical mechanisms — particularly for complex flows sensitive to pressure fluctuations, such as scramjet internal flows. Otherwise, when such empirical constraints is lacking, the complete equilibrium model is suggested to be employed to ensure thermodynamic consistency.

\appendix

\section{PR Equation of State and its Derivatives}
\label{sec:PR_EOS}
In this work, all thermodynamic calculations employ the Peng-Robinson (PR) EoS, which is expressed as:
\begin{equation}
    p = \frac{\rho RT}{1 - b\rho} - \frac{a(T)\rho^2}{1 + 2b\rho - b^2\rho^2}.
\label{equ:PR_eos}
\end{equation}
The coefficients of the PR EoS are defined as:
\begin{align}
    &a(T) = a_c \cdot \alpha(T) = a_c \left[ 1 + c\left( 1 - \sqrt{T/T_c} \right) \right]^2 \\
    &a_c = 0.457236\frac{R^2 T_c^2}{P_c}\\
    &b = 0.077796 \frac{R T_c}{P_c} \\
    &c = 0.37464 + 1.54226\omega - 0.26992\omega^2, 
\end{align}
where $R$ is the gas constant and $\omega$ is the acentric factor. The derivative of coefficient $a(T)$ with respect to $T$ satisfies:
\begin{align}
    &\totalderiv{a}{T} = - \frac{c \sqrt{a \cdot a_c}}{\sqrt{T_c T}}\\
    &\frac{\mathrm{d}^2 a}{\mathrm{d} T^2} = \frac{a_c c (1+c)}{2 T \sqrt{T T_c}}.
\end{align}
Derivative of $P$ with respect to $T$:
    \begin{equation}
    \pderiv{P}{T}{\rho} = \frac{\rho R}{1 - b\rho} - \frac{\rho^2}{1 + 2b\rho - b^2\rho^2} \totalderiv{a}{T}.
\end{equation}
Derivative of $P$ with respect to $\rho$:
    \begin{equation}
    \pderiv{P}{\rho}{T} = \frac{RT}{(1 - b\rho)^2} - \frac{2a(T)\rho (1 + b\rho)}{(1 + 2b\rho - b^2\rho^2)^2}.
\end{equation}
Second derivative of $P$ with respect to $T$:
\begin{equation}
  \left( \frac{\partial^2 P}{\partial T^2} \right)_\rho =  - \frac{\rho^2}{1 + 2b\rho - b^2\rho^2} \frac{\mathrm{d}^2a}{\mathrm{d}T^2}.
\end{equation}
Mixed second derivative of $P$ with respect to $T$ and $\rho$:
\begin{equation}
    \frac{\partial^2 P}{\partial T \partial \rho} = \frac{R}{(1 - b\rho)^2} - \frac{2\rho (1 + b\rho)}{(1 + 2b\rho - b^2\rho^2)^2} \frac{\mathrm{d}a}{\mathrm{d}T}.
\end{equation}
Second derivative of $P$ with respect to $\rho$:
\begin{equation}
    \left( \frac{\partial^2 P}{\partial \rho^2} \right)_T = \frac{2bRT}{(1 - b\rho)^3} - \frac{2a(T) \left( 1 + 3b^2\rho^2 + 2b^3\rho^3 \right)}{(1 + 2b\rho - b^2\rho^2)^3}.
\end{equation}

\section{VLE Solver Based on PR Equation of State}
\label{sec:vle_solver}
Based on the PR EoS, this appendix provides a Newton iteration method for solving the saturation temperature of pure substances. For pure substances, the fugacity coefficients ($\phi$) of vapor and liquid phases are equal at saturation pressure and temperature. The fugacity coefficient is defined as:
\begin{equation}
    \phi_p = P \exp  \left((Z_p-1) - \ln(Z_p - B) - \frac{A}{2\sqrt{2}B} \ln E_p \right),
\end{equation}
where $Z_p$ is the compressibility factor, and subscript $p$ represents the phase. The dimensionless coefficients $Z_p, A, B, E_p$ satisfy:
\begin{equation}
\begin{aligned}
    A &= \frac{a(T) P}{R^2 T^2}; \quad B = \frac{b P}{R T}\\
    Z_p &=\frac{\rho_p P }{RT}; \quad  E_p = \frac{Z_p + (1+\sqrt{2})B}{Z_p + (1-\sqrt{2})B}.
\end{aligned}    
\end{equation}

Applying the condition of equal fugacity, a Newton iteration method is constructed with temperature $T$ as the iteration variable. The residual function is:
\begin{equation}
    F_\phi(T) = \ln \phi_l(T, P) - \ln \phi_v(T, P) = 0.
\end{equation}
Its derivative is:
\begin{equation}
    \pderiv{F_\phi(T)}{T}{P} = \pderiv{(\ln \phi_l)}{T}{P} - \pderiv{(\ln \phi_v)}{T}{P}.
\end{equation}
Computing these analytical derivatives is central to resolving the fugacity equilibrium. We decompose $\ln \phi$ into the form $f_1 + f_2 + f_3$:
\begin{equation}
    \pderiv{(\ln \phi_p)}{T}{P} = \pderiv{f_{1,p}}{T}{P} + \pderiv{f_{2,p}}{T}{P} + \pderiv{f_{3,p}}{T}{P},
\end{equation}
where:
\begin{align}
    \pderiv{f_{1,p}}{T}{P} &= \pderiv{Z_p}{T}{P}, \label{equ:df1} \\
    \pderiv{f_{2,p}}{T}{P} &= -\frac{1}{Z_p - B}\left(\pderiv{Z_p}{T}{P} + \frac{B}{T}\right), \label{equ:df2} \\
    \pderiv{f_{3,p}}{T}{P} &= -\frac{1}{2\sqrt{2}}\bigg[ \pderiv{}{T}{P}\left(\frac{A}{B}\right) \ln E_p + \frac{A}{B} \pderiv{\ln E_p}{T}{P} \bigg].
\label{equ:df3}
\end{align}

Evaluating Equations \ref{equ:df1}--\ref{equ:df3} requires determining the derivatives of $A, B, A/B, Z_p, E_p$ with respect to $T$, as well as the value of $Z_p$.

\textbf{1. Derivative of $A$:}
\begin{equation}
    \pderiv{A}{T}{P} = \frac{P}{R_u^2} \left[ \left(\totalderiv{a}{T}\right) T^{-2} + 
a(T) \cdot (-2 T^{-3}) \right] = \frac{A}{a}\totalderiv{a}{T} - \frac{2A}{T}.
\end{equation}

\textbf{2. Derivative of $B$:}
\begin{equation}
    \pderiv{B}{T}{P} = \frac{bP}{R} \pdv{}{T}\left(\frac{1}{T}\right) =  -\frac{B}{T}.
\end{equation}

\textbf{3. Derivative of $(A/B)$:}
\begin{equation}
    \pderiv{}{T}{P}\left(\frac{A}{B}\right)  = \frac{1}{B}\left( \frac{A}{a}\totalderiv{a}{T} - \frac{2A}{T} \right) - \frac{A}{B^2}\left( -\frac{B}{T} \right).
\end{equation}

\textbf{4. $Z_p$ and its derivative:}

$Z_p$ is the root of the cubic form of the PR equation. Solving the cubic equation:
\begin{equation}
    F_Z = Z^3 - (1 - B)Z^2 + (A - 2B - 3B^2)Z - (AB - B^2 - B^3) = 0.
\end{equation}
Three distinct roots will be obtained in the two-phase region. The largest real root corresponds to the vapor phase compressibility factor $Z_v = Z_{max}$, while the smallest real root corresponds to the liquid phase compressibility factor $Z_l = Z_{min}$. Implicit differentiation of $F_Z(Z, A, B) = 0$ yields:
\begin{equation}
    \pderiv{Z_p}{T}{P} = - \frac{\frac{\partial F}{\partial A} \pderiv{A}{T}{P} + \frac{\partial F}{\partial B} \pderiv{B}{T}{P}}{\frac{\partial F}{\partial Z_p}},
\end{equation}
where:
\begin{equation}
\begin{aligned}
    \frac{\partial F}{\partial Z_p} &= 3Z_p^2 - 2(1 - B)Z_p + (A 
- 2B - 3B^2), \\
    \frac{\partial F}{\partial A} &= Z_p - B, \\
    \frac{\partial F}{\partial B} &= Z_p^2 - 2Z_p - 6BZ_p - A + 2B + 3B^2.
\end{aligned}
\end{equation}

\textbf{5. Derivative of $\ln E_p$:}
\begin{equation}
    \pderiv{\ln E_p}{T}{P} = \frac{2\sqrt{2}\left(Z_p\pderiv{B}{T}{P}-B\pderiv{Z_p}{T}{P}\right)}{(Z_p + (1+\sqrt{2})B)(Z_p + (1-\sqrt{2})B)}.
\end{equation}

\section{Derivatives in the Single-Phase Region}
Calculating the intermediate pressure of the Riemann problem requires the partial derivatives of energy $e$ and entropy $s$. Calculating the Landau derivative and solving for the internal parameters of the rarefaction wave require the derivatives of the speed of sound $c$. This subsection first presents the analytical forms of these derivatives in the single-phase region.

\subsection{Derivatives of Energy and Entropy in the Single-Phase Region}
\label{sec:ds_de}
In the single-phase region, the derivatives of energy $e$ and entropy $s$ can be calculated by composite differentiation of Equations \ref{equ:e} and \ref{equ:s}. Derivatives of energy:
\begin{align}
    \pderiv{e}{\rho}{T} &= - \frac{1}{\rho^2}\left(T \pderiv{P}{T}{\rho} - P \right), \label{equ:dedrho_T}  \\ 
    \pderiv{e}{T}{\rho} &=  {C_v}.
\label{equ:dedT_rho}
\end{align}
Derivatives of entropy:
\begin{align}
    \pderiv{s}{\rho}{T} &= - \frac{1}{\rho^2}\pderiv{P}{T}{\rho}, \label{equ:dsdrho_T}\\
    \pderiv{s}{T}{\rho} &=  \frac{C_v}{T}.
\label{equ:dsdT_rho}
\end{align}
With $(P,\rho)$ selected as independent variables, the partial derivatives of $e(P,\rho)$ are:
\begin{align}
    \pderiv{e}{\rho}{P} &= -\frac{C_v \pderiv{P}{\rho}{T}}{\pderiv{P}{T}{\rho}} - \frac{1}{\rho^2}\left(T \pderiv{P}{T}{\rho} - P \right), \label{equ:dedrho_P}  \\ 
    \pderiv{e}{P}{\rho} &=  \frac{C_v}{\pderiv{P}{T}{\rho}}.
\label{equ:dedP_rho}
\end{align}
The partial derivatives of $s(P,\rho)$ are:
\begin{align}
    \pderiv{s}{\rho}{P} &= -\frac{C_v \pderiv{P}{\rho}{T}}{T \pderiv{P}{T}{\rho}} - \frac{1}{\rho^2}\pderiv{P}{T}{\rho}, \label{equ:dsdrho_P}\\
    \pderiv{s}{P}{\rho} &=  \frac{C_v}{T \pderiv{P}{T}{\rho}}.
\label{equ:dsdP_rho}
\end{align}

\subsection{Derivatives of Speed of Sound in the Single-Phase Region}
The speed of sound $c$ in the single-phase region is a function of $(\rho,T)$, and its derivatives are:
\begin{align}
    \pderiv{c}{\rho}{T} &= \frac{1}{2c} \left( 
        \pderiv{^2 P}{\rho^2}{T} 
        + \frac{2T}{C_v \rho^2} \pderiv{P}{T}{\rho} \frac{\partial^2 P}{\partial \rho \partial T} 
        - \frac{T}{C_v \rho^2} \left( \frac{2}{\rho} + \frac{1}{C_v} \pderiv{C_v}{\rho}{T} \right) \pderiv{P}{T}{\rho}^2 
    \right), \label{equ:dcdrho_T} \\
    \pderiv{c}{T}{\rho} &= \frac{1}{2c} 
\left( 
        \frac{\partial^2 P}{\partial T \partial \rho} 
        + \frac{2T}{C_v \rho^2} \pderiv{P}{T}{\rho} \pderiv{^2 P}{T^2}{\rho} 
        + \frac{1}{C_v \rho^2} \left( 1 - \frac{T}{C_v} \pderiv{C_v}{T}{\rho} \right) \pderiv{P}{T}{\rho}^2 
    \right), \label{equ:dcdT_rho}
\end{align}
where the derivative of $C_v$ with respect to $T$ is defined by the specific heat model. The derivative of $C_v$ with respect to $\rho$ satisfies the compatibility condition:
\begin{equation}
    \pderiv{C_v}{\rho}{T}  = -\frac{T}{\rho^2} \pderiv{^2 P}{T^2}{\rho}.
\end{equation}
Solving for the internal parameters of the rarefaction wave requires the derivatives of $c=c(P,\rho)$:
\begin{align}
\pderiv{c}{\rho}{P}&=\pderiv{c}{\rho}{T} - \pderiv{c}{T}{\rho}\frac{\pderiv{P}{\rho}{T}}{\pderiv{P}{T}{\rho}}, \label{equ:dcdrho_P} \\
\pderiv{c}{P}{\rho}&=\frac{\pderiv{c}{T}{\rho}}{\pderiv{P}{T}{\rho}}.
\label{equ:dcdP_rho}   
\end{align}

\section{Derivatives of Energy \texorpdfstring{$e$}{e} and Entropy \texorpdfstring{$s$}{s} in the Two-Phase Region}

Define the vapor volume fraction:
\begin{equation}
   \beta(\rho,T) = \frac{v-v_l(T)}{v_v-v_l(T)}
   = \frac{\alpha \rho_v}{\rho}.
\end{equation}
The mixture internal energy \ref{equ:mix_e} and mixture entropy \ref{equ:mix_s} are rewritten as:
\begin{equation}
   e(\rho,T)=\beta e_v(T) +(1-\beta)e_l(T),
    \label{equ:mix2_e} 
\end{equation}
\begin{equation}
   s(\rho,T)=\beta s_v(T) +(1-\beta)s_l(T).
\label{equ:mix2_s} 
\end{equation}
Since the mixing rules for entropy $s$ and energy $e$ are identical, their derivative forms in the two-phase region are analogous. Differentiating Equation \ref{equ:mix2_e} with respect to $\rho$ and $T$, respectively, we get:
\begin{align}
    \pderiv{e}{\rho}{P} &= (e_v-e_l)\pderiv{\beta}{\rho}{P},\\
     \pderiv{e}{P}{\rho} &= \beta \totalderiv{e_v}{P}+(1-\beta)\totalderiv{e_l}{P}+(e_v-e_l)\pderiv{\beta}{P}{\rho},
\end{align}

Differentiating Equation \ref{equ:mix2_s} with respect to $\rho$ and $T$, respectively, yields:
\begin{align}
    \pderiv{s}{\rho}{P} &= (s_v-s_l)\pderiv{\beta}{\rho}{P},\\
     \pderiv{s}{P}{\rho} &= \beta \totalderiv{s_v}{P}+(1-\beta)\totalderiv{s_l}{P}+(s_v-s_l)\pderiv{\beta}{P}{\rho},
\end{align}
where the derivatives of $\beta$ are:
\begin{align}
    \pderiv{\beta}{\rho}{P} &= -\frac{1}{(v_v-v_l)\rho^2}, \label{equ:dbdrho}\\
    \pderiv{\beta}{P}{\rho} &= \frac{1}{\rho (\rho_l - \rho_v)^2} \left( \rho_v (\rho - \rho_v) \totalderiv{\rho_l}{P} + \rho_l (\rho_l - \rho) \totalderiv{\rho_v}{P}  \right).
\label{equ:dbdP}
\end{align}

Derivatives of saturated internal energy $e_v, e_l$:
\begin{equation}
    \totalderiv{e_p}{P} = \pderiv{e}{\rho}{T} \totalderiv{\rho}{P} + \pderiv{e_p}{T}{\rho} \totalderiv{T_{\text{sat}}}{P},
    \label{equ:dedP_t}
\end{equation}

Derivatives of saturated internal entropy $s_v, s_l$:
\begin{equation}
    \totalderiv{s_p}{P} = \pderiv{s}{\rho}{T} \totalderiv{\rho}{P} + \pderiv{s_p}{T}{\rho} \totalderiv{T_{\text{sat}}}{P},
    \label{equ:dsdP_t}
\end{equation}
where the partial derivatives of $e_p$ and $s_p$ with respect to $\rho$ and $T$ are the single-phase region derivatives (Equations \ref{equ:dedrho_T}--\ref{equ:dedT_rho} and Equations \ref{equ:dsdrho_T}--\ref{equ:dsdT_rho}). Equations \ref{equ:dbdP}, \ref{equ:dsdP_t} and \ref{equ:dedP_t} both use the derivative of saturated density:
\begin{equation}
    \totalderiv{\rho_p}{P} = \frac{1 - \pderiv{P}{T}{\rho_p} \totalderiv{T_{\text{sat}}}{P}}{ \pderiv{P}{\rho_p}{T} },
    \label{equ:drhodP_t}
\end{equation}
where the partial derivatives of $P$ with respect to $\rho$ and $T$ are the partial derivatives of the EOS in Appendix \ref{sec:PR_EOS}. The derivative of saturation temperature ${\mathrm{d}T_\text{sat}}/{\mathrm{d}P}$ satisfies the Clausius-Clapeyron equation:
\begin{equation}
    \totalderiv{T_{\text{sat}}}{P} = \frac{T_{\text{sat}} (v_v - v_l)}{h_v - h_l} = \frac{T_{\text{sat}} (v_v - v_l)}{e_v-e_l+P(v_v-v_l)}.
\label{equ:dTdP_t}
\end{equation}

\section{Derivatives of Wood's Speed of Sound (\texorpdfstring{$c_W$}{cW})}
\label{sec:dc_w}
The derivatives of Wood's speed of sound $c_W$ are mathematically less complex than those of the complete thermodynamic equilibrium speed of sound $c_{eq}$. Furthermore, the derivatives of $c_W$ are required for deriving $c_{eq}$, so $c_W$ is addressed first. Additionally, in the exact solution based on Wood's model, these derivatives are essential for resolving the internal parameters of the rarefaction wave. In Wood's compressibility $K_W$ (Equation \ref{equ:Kw}), define $D_v = 1/(\rho_v c_v^2)$ and $D_l = 1/(\rho_l c_l^2)$. Thus:
\begin{equation}
    K_W = \rho ( \alpha (D_v - D_l) + D_l).
\end{equation}
Note that $D_v$ and $D_l$ are functions of $P$ only.

\subsection{Derivative \texorpdfstring{$\pderiv{c_W}{\rho}{P}$}{(dcW/d(rho))P}}
\begin{equation}
    \pderiv{c_W}{\rho}{P} = -\frac{c_W^3}{2} \pderiv{K_W}{\rho}{P}.
\end{equation}
Since $P=\text{const}$, $D_v, D_l, \rho_v, \rho_l$ are all constants. Thus:
\begin{equation}
    \pderiv{K_W}{\rho}{P} = \frac{K_W}{\rho}+ \rho (D_v - D_l) \pderiv{\alpha}{\rho}{P},
\end{equation}
where:
\begin{equation}
    \pderiv{\alpha}{\rho}{P} = \frac{1}{\rho_v - \rho_l}.
\label{equ:dadrho}
\end{equation}
Finally:
\begin{equation}
    \pderiv{c_W}{\rho}{P} =  -\frac{c_W}{2\rho} - \frac{c_W^3 \rho}{2} \left( \frac{D_v - D_l}{\rho_v - \rho_l} \right).
\end{equation}

\subsection{Derivative \texorpdfstring{$\pderiv{c_W}{P}{\rho}$}{(dcW/dP)rho}}
\begin{equation}
    \pderiv{c_W}{P}{\rho} = -\frac{c_W^3}{2} \pderiv{K_W}{P}{\rho}.
\end{equation}
When $\rho=\text{const}$, $D_v, D_l, \alpha$ are all functions of $P$.
\begin{equation}
    \pderiv{K_W}{P}{\rho} = \rho \left( (D_v - D_l)\pderiv{\alpha}{P}{\rho}  + \alpha \left(\totalderiv{D_v}{P} - \totalderiv{D_l}{P}\right) + \totalderiv{D_l}{P} \right),
\end{equation}
where:
\begin{equation}
    \left( \frac{\partial \alpha}{\partial P} \right)_{\rho} = \frac{\rho_l \totalderiv{\rho_v}{P} - \rho_v \totalderiv{\rho_l}{P} + \rho (\totalderiv{\rho_l}{P}-\totalderiv{\rho_v}{P})}{(\rho_l - \rho_v)^2},
    \label{equ:dadP}
\end{equation}
\begin{equation}
    \totalderiv{D_p}{P} = \totalderiv{}{P} \left( (\rho_p c_p^2)^{-1} \right) = -D_p^2 \left[ \rho_p' c_p^2 + \rho_p \totalderiv{(c_p^2)}{P} \right].
\end{equation}
The derivative of $c_l^2$ is the total derivative along the saturation line:
\begin{equation}
    \totalderiv{(c_p^2)}{P} = 2c_p\left(\pderiv{c_p}{\rho}{T} \totalderiv{\rho_p}{P} + \pderiv{c_p}{T}{\rho} \totalderiv{T_\text{sat}}{P}\right),
\end{equation}
where the partial derivatives of saturated speed of sound $c_p$ with respect to $\rho$ and $T$ are calculated using Equations \ref{equ:dcdrho_T} and \ref{equ:dcdT_rho}.

\section{ Derivatives of Complete Thermodynamic Equilibrium Speed of Sound (\texorpdfstring{$c_{eq}$}{ceq})}
\label{sec:dc_eq}
The compressibility $K_{eq}$ of the complete equilibrium speed of sound consists of Wood's compressibility $K_W$ and the acoustic compressibility $K_H$. The derivative of $K_W$ has been obtained in Appendix \ref{sec:dc_w}. The key to solving for the derivative of $c_{eq}$ is to solve for the derivative of $K_H$. Defining two auxiliary functions $f(P), g(P)$ which depend only on $P$, the acoustic compressibility $K_H$ is written as:
\begin{equation}
     K_H(\rho, P) = \rho \cdot \left( 
\alpha(\rho, P) \cdot (f(P) - g(P)) + g(P) \right),
\end{equation}
where the auxiliary functions $f(P)$ and $g(P)$ are defined along the saturation line:
\begin{align}
        f(P) &= T_{\text{sat}} \cdot \rho_v \cdot C_{P,v}^{-1} \cdot \left(\totalderiv{s_v}{P}\right)^2, \\
    g(P) &= T_{\text{sat}} \cdot \rho_l \cdot C_{P,l}^{-1} \cdot \left(\totalderiv{s_l}{P}\right)^2.
\end{align}
For brevity, we use the superscript $'$ to denote the total derivative with respect to $P$ along the saturation line, and $''$ to denote the second derivative (e.g., $T' = \totalderiv{T_{\text{sat}}}{P}; T'' = \totalderiv{^2T_{\text{sat}}}{P^2}$).

\subsection{Derivative \texorpdfstring{$\pderiv{c_{eq}}{\rho}{P}$}{(dceq/d(rho))P}}
We first need $\pderiv{K_{eq}}{\rho}{P}$.
\begin{align}
    \pderiv{c_{eq}}{\rho}{P} &= -\frac{1}{2} (K_W + K_H)^{-3/2} \left( \pderiv{K_W}{\rho}{P} + \pderiv{K_H}{\rho}{P} \right) \notag \\
    &= -\frac{c_{eq}^3}{2} \left( \pderiv{K_W}{\rho}{P} + \pderiv{K_H}{\rho}{P} \right),
\end{align}
where $\pderiv{K_W}{\rho}{P}$ has been obtained in Appendix \ref{sec:dc_w}. Now we derive $\pderiv{K_H}{\rho}{P}$.
Since $f(P)$ and $g(P)$ are constants when $P=\text{const}$:
\begin{equation}
    \pderiv{K_H}{\rho}{P} = \frac{K_H}{\rho} + \rho (f - g) \pderiv{\alpha}{\rho}{P}.
\end{equation}
Substituting Equation \ref{equ:dadrho}:
\begin{equation}
    \pderiv{K_H}{\rho}{P} 
= \frac{K_H}{\rho} + \frac{\rho (f(P) - g(P))}{\rho_v(P) - \rho_l(P)}.
\label{equ:dKH_drh}
\end{equation}

\subsection{Derivative \texorpdfstring{$\pderiv{c_{eq}}{P}{\rho}$}{(dceq/dP)rho}}
\label{sec:dceqdP_rho}
\begin{align}
    \pderiv{c_{eq}}{P}{\rho} &=  -\frac{1}{2} (K_W + K_H)^{-3/2} \left( \pderiv{K_W}{P}{\rho} + \pderiv{K_H}{P}{\rho} \right) \notag \\
   & = -\frac{c_{eq}^3}{2} \left( \pderiv{K_W}{P}{\rho} + \pderiv{K_H}{P}{\rho} \right),
\end{align}
where the derivation of $\pderiv{K_W}{P}{\rho}$ is given in Appendix \ref{sec:dc_w}. Below we focus on deriving $\pderiv{K_H}{P}{\rho}$. This derivative is quite complex and requires the following five steps:

\textbf{Step 1. Main derivative $\pderiv{K_H}{P}{\rho}$}

Using the product rule:
\begin{equation}
    \left( \frac{\partial K_H}{\partial P} \right)_{\rho} = \rho \cdot \left( (f - g) \pderiv{\alpha}{P}{\rho} + \alpha f' + (1 - \alpha) g' \right),
    \label{equ:dKH_dP}
\end{equation}
where $\pderiv{\alpha}{P}{\rho}$ has been obtained in Equation \ref{equ:dadP}. We need to derive $f'$ and $g'$.

\textbf{Step 2: Sub-derivatives}

We use logarithmic differentiation to derive $f'$.
\begin{equation}
    \ln(f) = \ln(T_{\text{sat}}) + \ln(\rho_v) - \ln(C_{p,v}) + 2 \ln(s_v').
\end{equation}
Differentiating with respect to $P$:
\begin{equation}
    f' = f \cdot \left( \frac{T'}{T_{\text{sat}}} + \frac{\rho_v'}{\rho_v} - \frac{C_{P,v}'}{C_{P,v}} + \frac{2 s_v''}{s_v'} \right).
\label{equ:df}
\end{equation}
The derivation of $g'$ is similar:
\begin{equation}
    g' = g \cdot \left( \frac{T'}{T_{\text{sat}}} + \frac{\rho_l'}{\rho_l} - \frac{C_{P,l}'}{C_{P,l}} + \frac{2 s_l''}{s_l'} \right).
\label{equ:dg}
\end{equation}
This introduces new derivatives to be solved: $C_{P,p}'$ and $s_p''$, which are total derivatives along the saturation line.

\textbf{Step 3: Second-order saturation derivatives}

Derivative of specific heat at constant pressure:
\begin{equation}
    C_{P,p}' = \left( \frac{\partial C_P}{\partial \rho} \right)_T \rho_p' + \left( \frac{\partial C_P}{\partial T} \right)_\rho T'.
\label{equ:dCp}
\end{equation}
The partial derivatives $\left( \frac{\partial C_P}{\partial \rho} \right)_T$ and $\left( \frac{\partial C_P}{\partial T} \right)_\rho$ can be derived directly from the $C_P$ model, or from the $C_v$ model and Equation \ref{equ:CPCv}.
Solving for $s_p''$ requires the first derivative $s_p'$ (which we have obtained in Equation \ref{equ:dsdP_t}):
\begin{equation}
    s_p' =\pderiv{s}{\rho}{T} \rho_p' + \pderiv{s}{T}{\rho} T',
\end{equation}
where the partial derivatives of saturated entropy $s_p$ with respect to $\rho$ and $T$ are calculated using Equations \ref{equ:dsdrho_T} and \ref{equ:dsdT_rho}. Now, we differentiate $s_p'$ with respect to $P$, applying the product rule to both terms:
\begin{equation}
    s_p'' = \left( \totalderiv{}{P}\left[\pderiv{s}{\rho}{T}\right] \rho_p' + \pderiv{s}{\rho}{T} 
\rho_p'' \right) + \left( \totalderiv{}{P}\left[\pderiv{s}{T}{\rho}\right] T' + \pderiv{s}{T}{\rho} T'' \right).
\end{equation}
Expanding and simplifying the derivative $\frac{\mathrm{d}}{\mathrm{d}P}[\dots]$, we finally get:
\begin{equation}
    s_p'' = \left( \frac{\partial^2 s}{\partial \rho^2} \right)_T (\rho_p')^2 + 2 \left( \frac{\partial^2 s}{\partial T \partial \rho} \right) T' \rho_p' + \left( \frac{\partial^2 s}{\partial T^2} \right)_\rho (T')^2 + \left( \frac{\partial s}{\partial \rho} \right)_T \rho_p'' + \left( \frac{\partial s}{\partial T} \right)_\rho T''.
\label{eq:svpp_deriv}
\end{equation}
This expression depends on the second-order derivatives of entropy.

\textbf{Step 4: Second-order entropy derivatives from EoS}

We differentiate the first-order entropy derivative expressions (Equations \ref{equ:dsdT_rho}, \ref{equ:dsdrho_T}):
\begin{equation}
    \left( \frac{\partial^2 s}{\partial T^2} \right)_\rho = \frac{\partial}{\partial T} \left( \frac{C_v}{T} \right)_\rho = \frac{1}{T} \pderiv{C_v}{T}{\rho} - \frac{C_v}{T^2},
    \label{eq:d2sdT2}
\end{equation}
\begin{equation}
    \left( \frac{\partial^2 s}{\partial \rho^2} \right)_T = \frac{\partial}{\partial \rho} \left( -\frac{1}{\rho^2} \pderiv{P}{T}{\rho} \right)_T = \frac{2}{\rho^3} \pderiv{P}{T}{\rho} - \frac{1}{\rho^2} \left( \frac{\partial^2 p}{\partial \rho \partial T} \right),
    \label{eq:d2sdrho2}
\end{equation}
\begin{equation}
    \left( \frac{\partial^2 s}{\partial T \partial \rho} \right) = \frac{\partial}{\partial T} \left( -\frac{1}{\rho^2} \pderiv{P}{T}{\rho} \right)_\rho = -\frac{1}{\rho^2} \left( \frac{\partial^2 P}{\partial T^2} \right)_\rho.
    \label{eq:d2sdTdr}
\end{equation}

\textbf{Step 5: Second-order saturation line derivatives}

The remaining unknown terms are the second-order total derivatives $T''$ and $\rho_v''$.\par
\textbf{1. Derivative $T''$}

Starting from the Clausius-Clapeyron relation (Equation \ref{equ:dTdP_t}), let the latent heat of phase transition be $h_v-h_l=L_v$. Applying the quotient rule:
\begin{equation}
    T'' = \totalderiv{(T')}{P} = \frac{ \left( \totalderiv{(T_{\text{sat}} \Delta v)}{P} \right) L_v - (T_{\text{sat}} \Delta v) \left( \totalderiv{L_v}{P} \right) }{L_v^2}.
\end{equation}
Applying the product rule on the first term and expanding $L_v' = h_v' - h_l'$:
\begin{equation}
    T'' = \frac{ \left[ T' \Delta v 
+ T_{\text{sat}} (v_v' - v_l') \right] L_v - (T_{\text{sat}} \Delta v) (h_v' - h_l') }{L_v^2},
    \label{eq:Tpp_deriv}
\end{equation}
where:
\begin{align}
    v_p'&=-\frac{\rho_p'}{\rho_p^2},\\
    h_p'&=e_p'+v_p+Pv_p'.
\end{align}

\textbf{2. Derivative $\rho_v''$}

Starting from the expression for $\rho_p'$ (Equation \ref{equ:drhodP_t}):
\begin{equation}
    \rho_p' = \frac{1 - \pderiv{P}{T}{\rho} T'}{ \pderiv{P}{\rho}{T} }.
\end{equation}
We apply the quotient rule $\left(\frac{U}{V}\right)' = \frac{U'V - UV'}{V^2}$. Let:
\begin{equation}
U = 1 - \pderiv{P}{T}{\rho} T'; \quad V = \pderiv{P}{\rho}{T}.
\end{equation}
Their derivatives $U'$ and $V'$ are:
\begin{align}
    U' & = - \totalderiv{}{P} \left( \pderiv{P}{T}{\rho} T' \right) =  -  \left( \left( \frac{\partial^2 P}{\partial \rho \partial T} \right) \rho_v' + \left( \frac{\partial^2 P}{\partial T^2} \right)_\rho T' \right) T' - \pderiv{P}{T}{\rho} T'', \\
    V' &= \totalderiv{V}{P} = \totalderiv{}{P} \left( \pderiv{P}{\rho}{T} \right) = \left( \frac{\partial^2 P}{\partial \rho^2} \right)_T \rho_v' + \left( \frac{\partial^2 P}{\partial T \partial \rho} \right) T'.
\end{align}
Substituting $U, V, U', V'$ into the quotient rule yields the final expression for $\rho_p''$.

\section{Thermophysical Parameters of n-dodecane}
\label{sec:thermo_dodecane}
The parameters required to determine the saturation properties of n-dodecane via the PR EoS are listed in Table \ref{tab:dodecane1}.
\begin{table}
    \centering
    \caption{Thermodynamic parameters of n-dodecane}
    \label{tab:dodecane1}
    \begin{tabular}{ccccc}
        \toprule
         $W$ ($\mathrm{kg\,kmol^{-1}}$) & $\omega$ & $T_c$ (K) & $\rho_c$ ($\mathrm{kg \,m^{-3}}$) & $P_c$ (MPa) \\
        \midrule
        170.33& 0.574 & 658.1 & 186 & 1.82\\
        \bottomrule
    \end{tabular}
\end{table}
The parameters required for computing internal energy, entropy, and speed of sound via exact integration of the PR EoS are provided in Table \ref{tab:dodecane2}.

\begin{table}
    \centering
    \caption{The parameters of the energy and entropy model}
    \label{tab:dodecane2}
    \begin{tabular}{cccc}
        \toprule
          $C_{v,\infty}^c$ 
($\mathrm{J \, kg^{-1} \,K^{-1}}$) & $n $ & $e_\text{ref}$ ($\mathrm{J \, kg^{-1}}$) & $s_\text{ref}$ ($\mathrm{J \, kg^{-1} \,K^{-1}}$)  \\
        \midrule
        $2.970 \times 10^3$& 0.613 & $6.948 \times 10^5$ & $1.401 \times 10^3$\\
        \bottomrule
    \end{tabular}
\end{table}

\begin{figure}[H]
\centering
\begin{subfigure}{0.49\textwidth}
    \includegraphics[width=\linewidth]{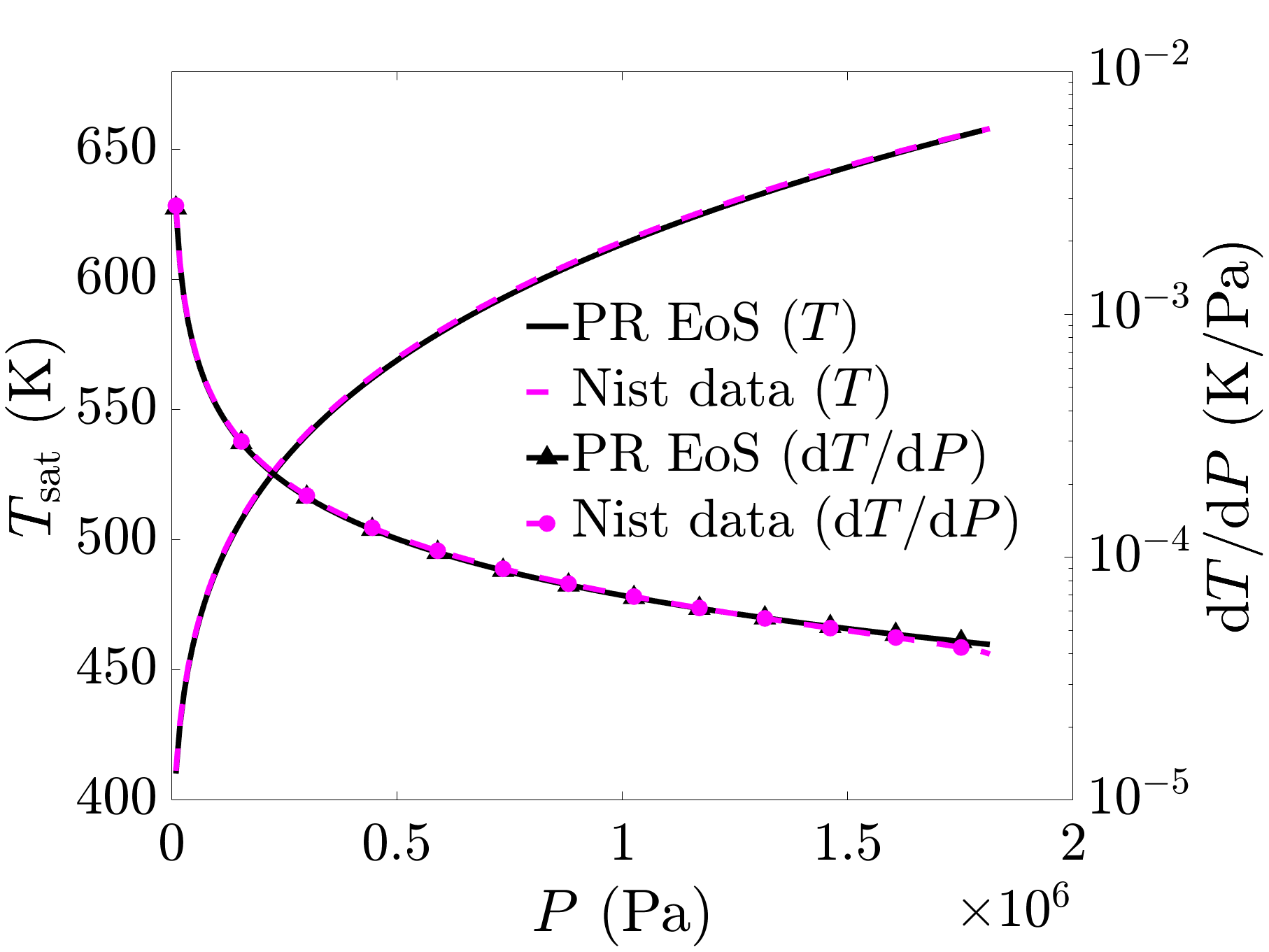}
    \caption{Saturation lines and its derivatives}
    \label{fig12a}
\end{subfigure}
\hfill
\begin{subfigure}{0.49\textwidth}
    \includegraphics[width=\linewidth]{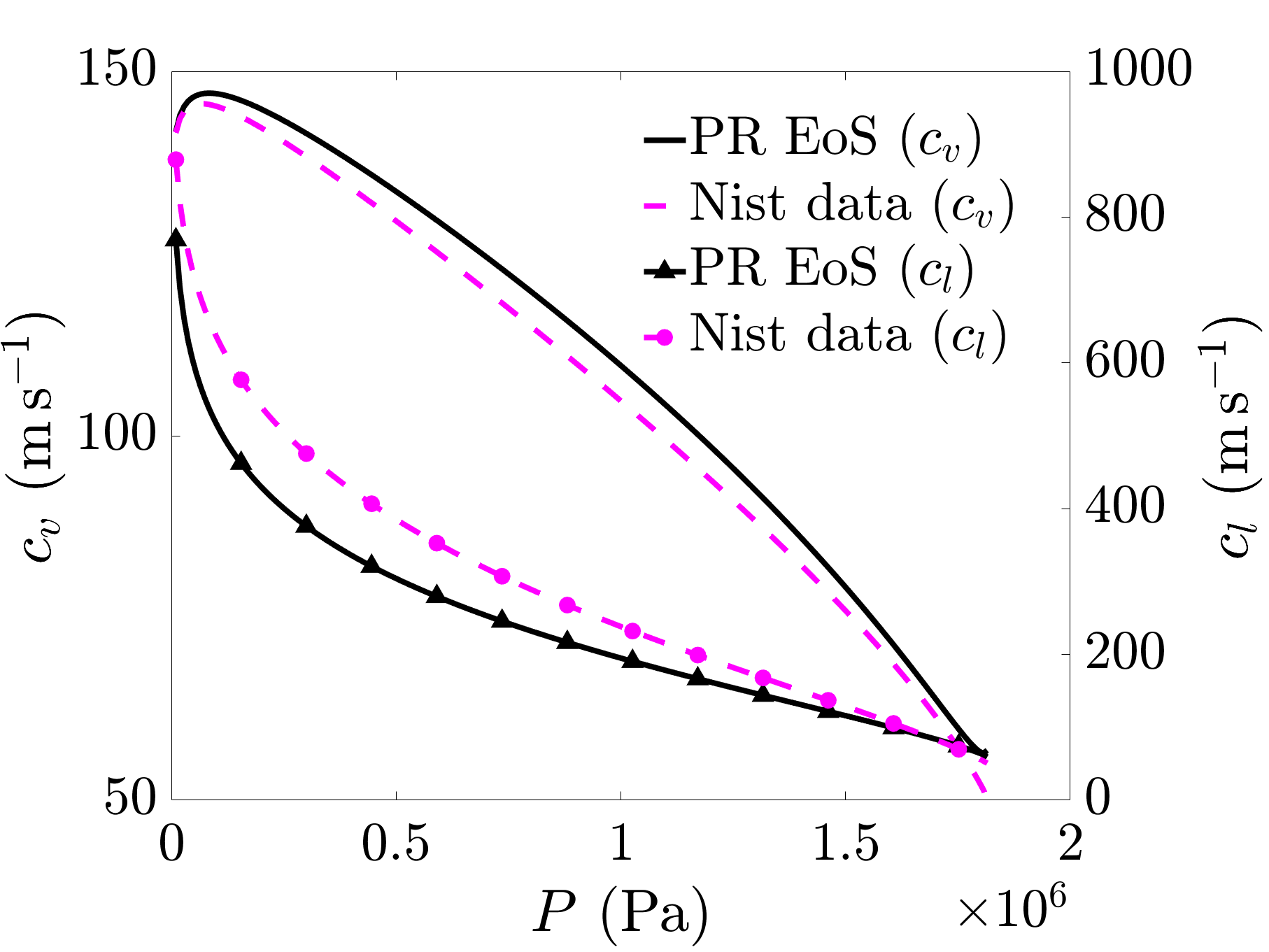}
    \caption{Speed of sound along saturation lines}
    \label{fig12b}
\end{subfigure}
\caption{Comparison of PR EoS results with NIST data}
\label{fig12}
\end{figure}

Detailed definitions of the parameters above are available in reference \citep{GuardoneArgrow2005}. Among them, $\rho_c$ is derived from the PR EoS based on $T_C$ and $P_c$, which is smaller than the NIST data ($\rho_c = 226\mathrm{kg \,m^{-3}}$). The calculated key thermodynamic variables, such as saturation line properties and saturated speed of sound, exhibit excellent agreement with NIST data, as shown in Figure \ref{fig12}.

\section{MATLAB Program for Solving the Exact Solution of the Flash evaporation Riemann Problem}
\label{sec:matlab}
This paper provides a MATLAB program for solving the exact solution of the FeRP, provided as supplementary material to readers, with the URL https://github.com/boss-monkey/FeRP. Detailed comments are provided inside each function for reader reference.

\bibliographystyle{plainnat}

\end{document}